\documentclass[11pt]{article}

\usepackage[utf8]{inputenc}
\usepackage{bm}
\usepackage{amsmath}
\usepackage{amssymb}
\usepackage[labelfont=bf,textfont=md]{caption} % bold caption
\usepackage[usenames]{color}
\usepackage{hyperref}
\usepackage{epsfig,wrapfig,amsthm,dsfont,bbm}
\usepackage{mathrsfs}
\usepackage{amsfonts}
\usepackage{authblk}

\usepackage{graphicx}
\usepackage{caption}
\usepackage{subcaption}
\usepackage{natbib}

\usepackage[margin=1in]{geometry}
\usepackage{setspace}

\usepackage{threeparttable}
\usepackage{float} % force figure/table's position

\usepackage{multicol}
\usepackage[export]{adjustbox}

\usepackage{authblk} % one-line affiliations
\makeatletter
\renewcommand\AB@affilsepx{, \protect\Affilfont}
\makeatother

\title{\textbf{An Adaptive and Robust Method for Multi-trait Analysis of Genome-wide Association Studies Using Summary Statistics}}
\author[1,2]{Qiaolan Deng}
\author[2]{Chi Song}
\author[1,3]{Shili Lin}

\affil[1]{Interdisciplinary Ph.D. Program in Biostatistics}
\affil[2]{Division of Biostatistics, College of Public Health}
\affil[3]{Department of Statistics, The Ohio State University, Columbus, Ohio}

\date{}

\begin{document}

%\doublespacing

\maketitle

%\newpage

\begin{abstract}

 Genome-wide association studies (GWAS) have identified thousands of genetic variants associated with human traits or diseases in the past decade. Nevertheless, much of the heritability of many traits is still unaccounted for. Commonly used single-trait analysis methods are conservative, while multi-trait methods improve statistical power by integrating association evidence across multiple traits. In contrast to individual-level data, GWAS summary statistics are usually publicly available, and thus methods using only summary statistics have greater usage. Although many methods have been developed for joint analysis of multiple traits using summary statistics, there are many issues, including inconsistent performance, computational inefficiency, and numerical problems when considering lots of traits. To address these challenges, we propose a multi-trait adaptive Fisher method for summary statistics (MTAFS), a computationally efficient method with robust power performance. We applied MTAFS to two sets of brain imaging derived phenotypes (IDPs) from the UK Biobank, including a set of 58 \textit{Volumetric} IDPs and a set of 212 \textit{Area} IDPs. Together with results from a simulation study, MTAFS shows its advantage over existing multi-trait methods, with robust performance across a range of underlying settings. It controls type 1 error well, and can efficiently handle a large number of traits.

\end{abstract}


%\newpage
%\tableofcontents
%\newpage

\section{Introduction}

Genome-wide association studies (GWAS) have identified thousands of genetic variants associated with complex diseases \citep{visscher201710}. However, for many complex traits, the heritability attributed to the genetic variants identified is still quite limited and a large proportion of the heritability remains unexplained \citep{manolio2009finding,visscher201710}. In GWAS, it is typical to test the association between a single trait and a single variant one at a time, the so called single-trait analysis. In reality, a common phenomenon is pleiotropy, in which a genetic variant is associated with multiple traits \citep{solovieff2013pleiotropy}. As such, conducting single-trait analyses may lose statistical power when genetic variants are truly associated with multiple traits. Therefore, there is an increasing need for methods that jointly analyze multiple traits together.

Although there are numerous existing multi-trait methods, many require individual-level genotype data \citep{o2012multiphen, wu2016sequence, zhang2014testing}. Due to privacy concern and data logistics, individual-level genotype data require permissions for access, limiting the applicability of methods relying on such data. In contrast, GWAS summary statistics such as effect sizes, standard errors, z-scores, and p-values, are publicly available for most published studies. With increasing availability of GWAS summary statistics, methods that only require such information for multi-trait analysis undoubtedly will see greater usage.

Although relatively limited compared to other types of data and methods, a number of multi-trait methods using only summary statistics have been proposed. We categorize them into two groups. The first group consists of non-adaptive method. The method using the sum of squared z scores, denoted as SSU, is a special case of SPU proposed by Pan \citep{pan2009asymptotic,kim2015adaptive}. He et al. \citep{he2013general} proposed a method using the sum of z scores, called SUM. Zhu et al. \citep{zhu2015meta} proposed a meta analysis method, called HOM, that is particularly suited for testing homogeneous effects across all traits. A chi-squared test, metaMANOVA, was proposed by Xu et al. \citep{xu2003combining} . Cauchy's method proposed by Liu and Xie \citep{liu2020cauchy} provides a general way to combine dependent p-values after appropriate transformation. The second group contains adaptive methods, which evaluate evidence adaptively and are particularly suited for heterogeneous situations where not all traits are associated, nor in the same directions or with the same effect size. An early example is TATES \citep{van2013tates} which combines the p-value of each trait to obtain an overall p-value and it has similar power performance to the SSU \citep{liu2019geometric}. aSPU adaptively combines powered score test statistics and requires permutation to obtain p-values \citep{kim2015adaptive}. Another method, metaUSAT, adaptively combines SSU and metaMANOVA \citep{ray2018methods}. MixAda \citep{liu2018multiple} adaptively combines SUM and a squared score test statistic, which is less powerful than metaUSAT under many scenarios \citep{ray2018methods}. On the other hand, MTAR is an adaptive principal component (PC)-based association test \citep{guo2019integrate}. Wu recently proposed aMAT, claimed to be feasible for any number of traits \citep{wu2020multi}.

Increased availability of GWAS summary statistics in recent years further points to the need for considering many traits simultaneously without accessing raw data. For example, in recent years, the UK Biobank has made thousands of functional and structural brain imaging phenotypes available, thus, a joint analysis of a large number of such traits may help better understand the biological mechanism of complex brain functions and diseases \citep{bycroft2018uk, elliott2018genome}. However, many previous methods using summary statistics as discussed above have only explored settings with a small number of traits \citep{zhu2015meta, ray2018methods, guo2019integrate}, rendering their performance of analyzing a large number of traits unknown. Our preliminary simulation study indicates that methods such as SSU and aMAT are sensitive to sparsity of signals and underlying correlation structures, whereas metaUSAT and SSU may not control type 1 error well at small significance levels. Computational issues also exist in some methods: aSPU is extremely time-consuming when the significance level is small due to its use of permutations; metaUSAT also becomes time-consuming when the p-values are extremely small, and it may return invalid values when the number of traits is large (e.g. over 200). Therefore, there is a need for robust and computationally efficient methods for settings where the number of traits is large, in the hundreds.

In this paper, we take up this challenge and propose a Multi-Trait Adaptive Fisher method for Summary statistics (MTAFS), a computationally efficient and statistically powerful method. In particular, MTAFS has three advantages over many existing methods. First, it controls type 1 error well compared to SSU and metaUSAT, regardless of the number of traits and significance levels. Second, it is robust, with good statistical power under different sparse and dense scenarios. Third, it is computationally efficient compared to metaUSAT and aSPU by avoiding permutations and is feasible for settings with a large number of traits. 

\section{Method}

\subsection{Setup}

Let $\bm{Z}=(z_1,\cdots,z_T)'_{(1 \times T)}$ be the GWAS summary statistics, the z scores, across $T$ traits for a given SNP. Our goal is to test whether the SNP is associated with at least one of the $T$ traits. Under the null hypothesis of no association between the SNP and any of the traits, we assume $\bm{Z} \sim \mathcal{N}(0,\bm{R})$. Here, $\bm{R}$ is referred to as the trait correlation matrix, and can be estimated by the sample correlation of $\bm{Z}$ based on the independent and identically distributed assumption across SNPs \citep{zhu2015meta} . Linkage disequilibrium score regression (LDSC) is another option \citep{turley2018multi, guo2019integrate}. We denote the estimated correlation matrix by either method as $\bm{\hat{R}}$. For computational efficiency, we used sample correlation estimates in the simulation studies and LDSC in the real data applications.

First, we use eigen-decompostion to decorrelate the z scores. Let $\bm{\hat{R}}=\bm{Q}\bm{\Lambda}\bm{Q}'$, where the columns of $\bm Q$ are eigenvectors in decreasing order of their corresponding eigenvalues given in the corresponding diagonal elements of $\bm{\Lambda}$. We denote the proportion of variance explained by the first two eigenvalues as $v_0\%$. Then let $v_1\%$, $v_2\%$, and $v_3\%$ be the three percentages evenly distributed between $v_0\%$ and $100\%$, with $q_1$, $q_2$, and $q_3$ denoting the corresponding number of eigenvalues achieving the percent of variance explained for the first time.

For each of the 5 levels of percentage of variance explained, we use the corresponding $E$ eigenvalues, $E \in \{2,q_1,q_2,q_3,T\}$, along with their eigenvectors to construct the transformed z score vector $\bm{U}_E$: $\bm{U}_E'=\bm{Z}'\bm{Q}_E\bm{\Lambda}_E^{-\frac{1}{2}}$, where $\bm{Q}_{E_{(T \times E)}}$ consists of the first $E$ columns of $\bm{Q}$ and $\bm{\Lambda}_{E_{(E \times E)}}$ is a submatrix of $\bm{\Lambda}$ containing only the first $E$ eigenvalues. As a result, $\bm{U}_E$ is a column vector of length $E$, and $\bm{U}_E \sim \mathcal{N}(0,\bm{I})$ under the null hypothesis. We then propose an adaptive method, as described in the following, in the spirit of the adaptive Fisher's method \citep{song2016screening}  for each of the five levels of variance explained. The resulting five p-values are then combined to construct an omnibus test statistic based on our proposed MTAFS method; the steps are depicted in a flow chart (supplementary Figure \ref{fig:v0}).

\subsection{Adaptive Method} \label{sec:adaptive}

Unlike the traditional Fisher's method which directly combines the ($-\log$)-transformed p-values, the adaptive Fisher's method considers ordered p-values and combines them adaptively \citep{song2016screening}. The method we are proposing here also considered ordered p-values, but they are combined adaptively using a different strategy for computational efficiency. Specifically, based on an $\bm{U}_E$, we obtain a vector of independent (two-sided) p-values, denoted as $\bm{p}_E = (p_1,\cdots,p_E)$, such that $\bm{p}_E = 2[\bm 1-\Phi(|\bm{U}_E|)]$, where $\Phi(\cdot)$ is the cumulative distribution of standard normal distribution, and is a component-wise operation. We calculate the sum of the ordered negative log p-values and let $s_k = \sum_{j=1}^k -(\log p_{(j)})$, where $p_{(j)}$ is the $j^{th}$ smallest p-value and $k \in \{1,\cdots,E\}$. We can rewrite $s_k$ as a weighted sum of independent $\chi^2$ variables \citep{david2004order,nagaraja2006order}, for which Davies method (R package CompQuadForm) or the saddlepoint approximation method (R package Survey) can efficiently approximate its p-value \citep{wu2016efficient}, denoted as $p_{s_k}$. We define the test statistic of our adaptive method for level $E$ as follows:

\begin{equation} 
    AF(E) = Cauchy(p_{s_k}; k=1,\ldots,E) = \sum_{k=1}^E \omega_k \tan\{(0.5-p_{s_k})\pi \},
\end{equation}
where $\omega_k = \frac{1}{E}$ for all $k$'s. This way of combining the evidence from p-values follows what was referred to as the Cauchy's method in the literature \citep{liu2020cauchy}, and the p-value of the test statistic can be calculated analytically:

\begin{equation} \label{eq:af}
    p_{AF(E)} = 0.5 - \frac{\arctan(AF(E))}{\pi}.
\end{equation}
We note that Cauchy's method is similar to the minP method because only a few of the smallest p-values would typically dominate the overall significance \citep{liu2020cauchy}. Nevertheless, since the p-values are calculated analytically, Cauchy's method is much more computationally efficient than minP.

\subsection{MTAFS}

From the literature \citep{aschard2014maximizing} and our own preliminary study (Figure \ref{fig:s2}-\ref{fig:s4}), it is shown that using either the first few or all eigenvectors would lead to unstable power performance. Therefore, we propose MTAFS, which integrates evidence from five levels of variance explained, for robust consideration. Specifically, MTAFS constructs a test statistic that combines the $\{p_{AF(E)}, E \in \{2,q_1,q_2,q_3,T\}\}$ obtained from Equation (\ref{eq:af}) for each of the 5 levels of variance explained. We define the test statistics of MTAFS as

\begin{equation} 
    T_{MTAFS} = Cauchy(p_{AF(E)}; E \in \{2,q1,q2,q3,T\}) = \sum_{E \in \{2,q1,q2,q3,T\}} \omega_E \tan\{(0.5-p_{AF(E)})\pi \},
\end{equation}
where $\omega_E = \frac{1}{5}$ for all $E$'s. As described above, the p-value of $T_{MTAFS}$ is 

\begin{equation*} 
    p_{MTAFS} = 0.5 - \frac{\arctan(T_{MTAFS})}{\pi}.
\end{equation*}
Since we have vectorized the R function of MTAFS, it can simultaneously analyze a large number of SNPs without using the ``for'' loop, which further increases its computational efficiency. MTAFS is implemented in an R package available at \url{http://www.github.com/Qiaolan/MTAFS}.

\section{Simulations and results}

\subsection{Simulation Setup}

We simulated z scores from $\mathcal{N}(\bm{\mu},\bm{R})$ following previous studies \citep{guo2019integrate, liu2020cauchy, wu2020multi}. Various scenarios were constructed by setting different correlation matrices, association models and strengths, and levels of signal sparsity. For $\bm R$, we considered two realistic correlation matrices estimated from real data and two commonly used structures. Specifically, we used the UK Biobank brain image-derived phenotypes (IDPs): the set of 58 volumetric IDPs, with the resulting estimated correlation matrix referred to as UKCOR1 (Figure \ref{fig:cor_volume}); and the T1 FAST region of interests containing 139 IDPs, denoted as UKCOR2 its correlation matrix (Figure \ref{fig:cor_t1fast}). Moreover, we also examined two commonly-used correlation structures, compound symmetry (CS) and autocorrelation structure of order 1 (AR), each with two levels of correlation --- weak (0.3) or strong (0.7). This leads to a total of 6 correlation matrices (Table \ref{tab:scenarios}). For analyzing data simulated, we re-estimated the correlation matrix instead of using the one for simulating the data.

We considered using two association models. In model 1, denoted as M1, we generated $\bm{\mu} = \sum_{j=1}^J c \lambda_j \bm{u}_j$, where $c$ is the parameter denoting the effect size, $\lambda_j$ and $\bm{u}_j$ are the $j^{th}$ eigenvalue and eigenvector of $\bm{R}$ respectively, and $J$ represents the top $J$ eigenvectors. We simulated different level of sparsity by varing $J$ (Table \ref{tab:scenarios}). This association model was also simulated in other studies \citep{guo2019integrate, wu2020multi}. In the second association model (M2), we generated scenarios by directly setting some elements of $\bm{\mu}$ to be nonzero, with fewer non-zeros denoting greater sparsity. We note that when $c=0$ in M1 or all elements of $\bm \mu$ are 0 in M2, we are in fact investigating the type 1 error.

Finally, we considered three levels of sparsity, high, intermediate, and low. For the highly sparse scenarios, in M1, only the top 2\% or 5\% of the eigenvectors had nonzero effect sizes, depending on the correlation structures; in M2, either 2\%, 4\%, or 5\% of the traits had nonzero effect sizes, also depending on the correlation structures. The proportion of nonzero effect sizes was 20\% in both models for the intermediate level of sparsity. The low sparsity scenarios had the proportion equal to 50\% in both models. The specific eigenvectors (for M1) or the specific traits (for M2) that corresponds to an nonzero effect size are given in Table \ref{tab:scenarios}.

%In addition to normal distributions, we generated z scores from non-normal t distributions to briefly evaluate how sensitive the methods were to the normality assumption of z scores. We had noncentrality parameters as $\bm \mu$ and the degree of freedom 20.

We include eight competing methods in our simulation study for comparison with MTAFS: SUM, SSU, metaUSAT, metaMANOVA, Cauchy, HOM, MTAR, and aMAT. These are representatives of currently available multi-trait methods using summary statistics.

\subsection{Type 1 errors}

We first evaluated the type 1 error of MTAFS and the comparison methods at various significance levels, from $5 \times 10^{-2}$ to $1 \times 10^{-5}$, for multiple correlation matrices. Table \ref{tab:type1} shows the results for UKCOR1. The estimated correlation matrix is obtained using the sample correlation over $10^5$ replicates. One can see that all the methods except SSU and metaUSAT controlled type 1 error well. For SSU and metaUSAT, their type 1 errors are inflated when the significance levels are smaller (bolded values in Table \ref{tab:scenarios}). Because metaUSAT adaptively combines metaMANOVA and SSU, the inflation of metaUSAT could be caused by SSU. On the other hand, we see that Cauchy and MTAFS controlled type 1 error better with increasing significance level, consistent with a previous study \citep{liu2020cauchy}. Since MTAFS used Cauchy to combine p-values, MTAFS naturally shared the characteristics of Cauchy. We evaluated the type 1 error with UKCOR2 and observed similar findings (Table \ref{tab:type1_t1fast}). We still observed type 1 error inflation for SUM and metaUSAT with the CS and AR correlation structure under either weak or strong correlation, although the magnitude were not as severe (Table \ref{tab:type1_cs03}-\ref{tab:type1_ar07_100}).

\subsection{Power comparisons}

\begin{figure}[htbp]
    \begin{multicols}{2}
     \centering
      \begin{subfigure}{0.4\textwidth}
        \includegraphics[width=\textwidth, height=6cm]{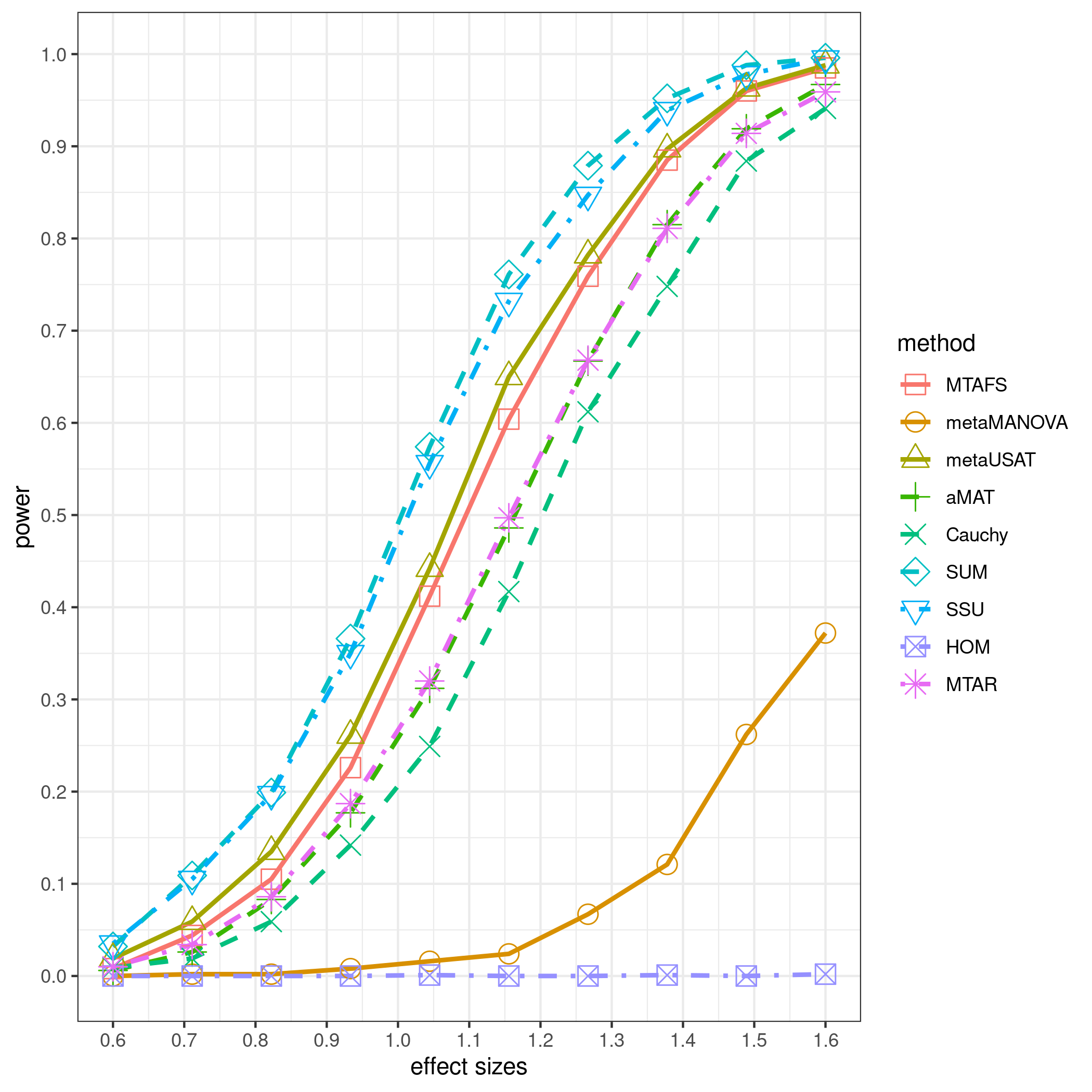}
        \caption{}
        \label{fig:v3a}
      \end{subfigure}
      \par
      \begin{subfigure}{0.4\textwidth}
        \includegraphics[width=\textwidth, height=6cm]{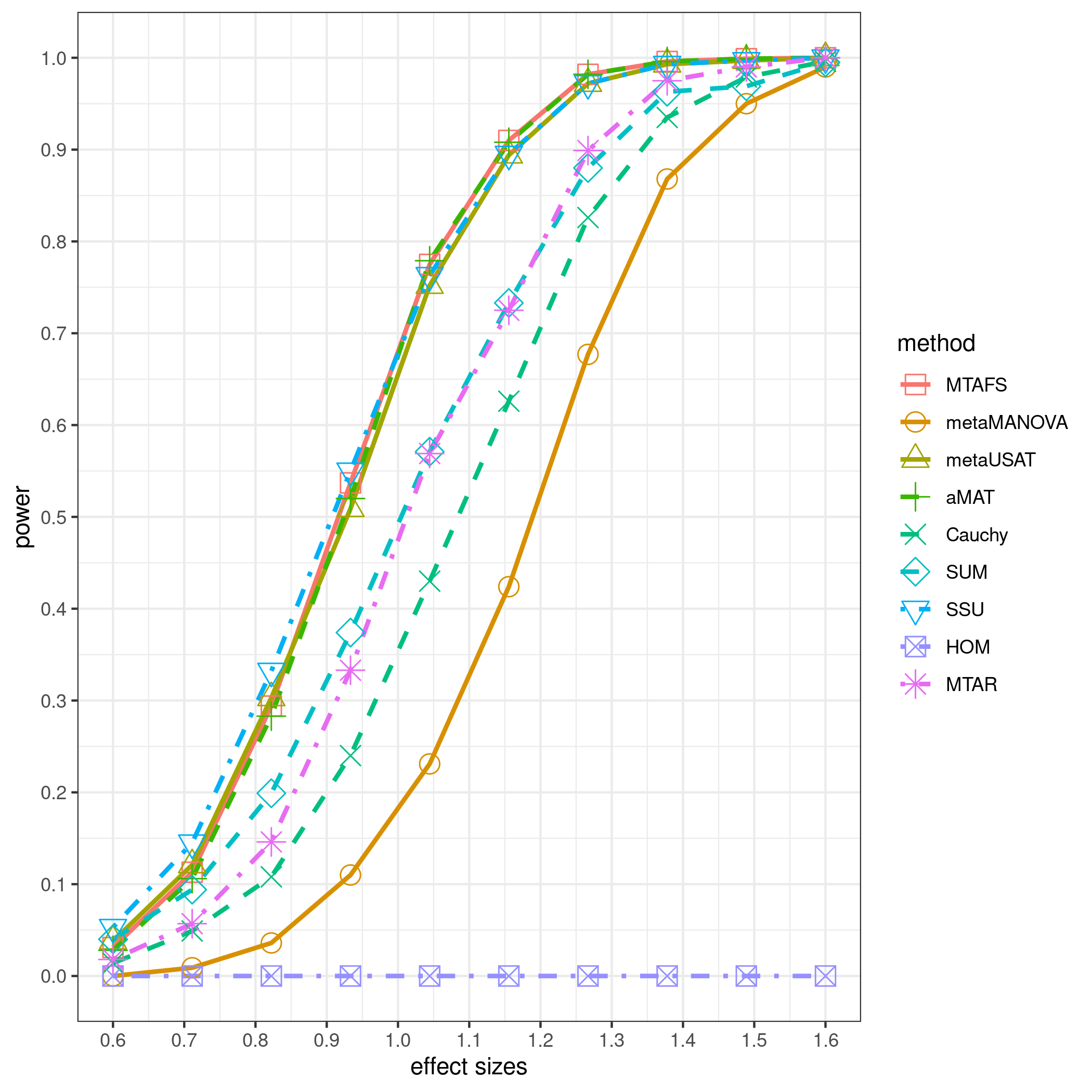}
        \caption{}
        \label{fig:v3b}
      \end{subfigure}
      \par
      \begin{subfigure}{0.4\textwidth}
        \includegraphics[width=\textwidth, height=6cm]{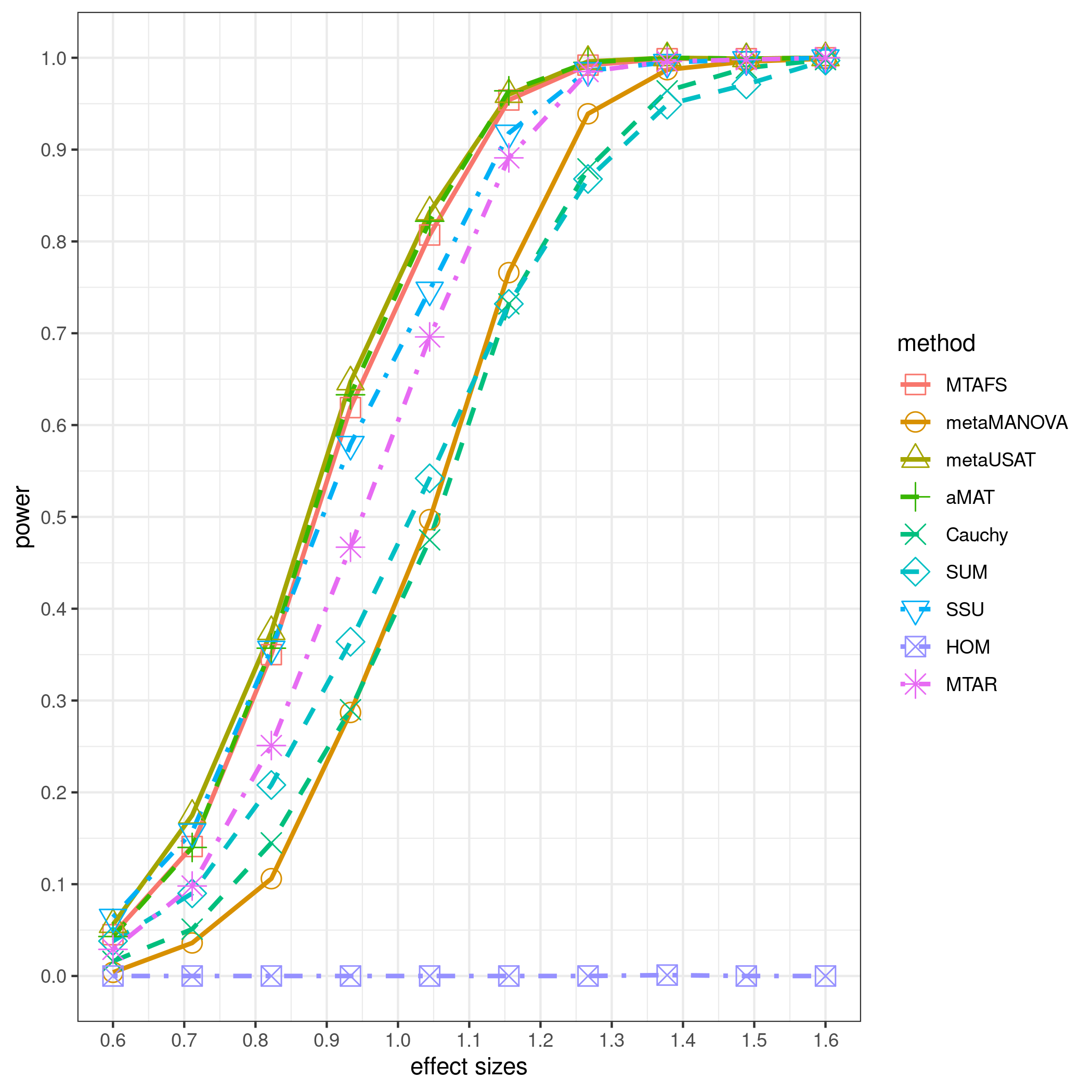}
        \caption{}
        \label{fig:v3c}
      \end{subfigure}
      \par
      \begin{subfigure}{0.4\textwidth}
        \includegraphics[width=\textwidth, height=6cm]{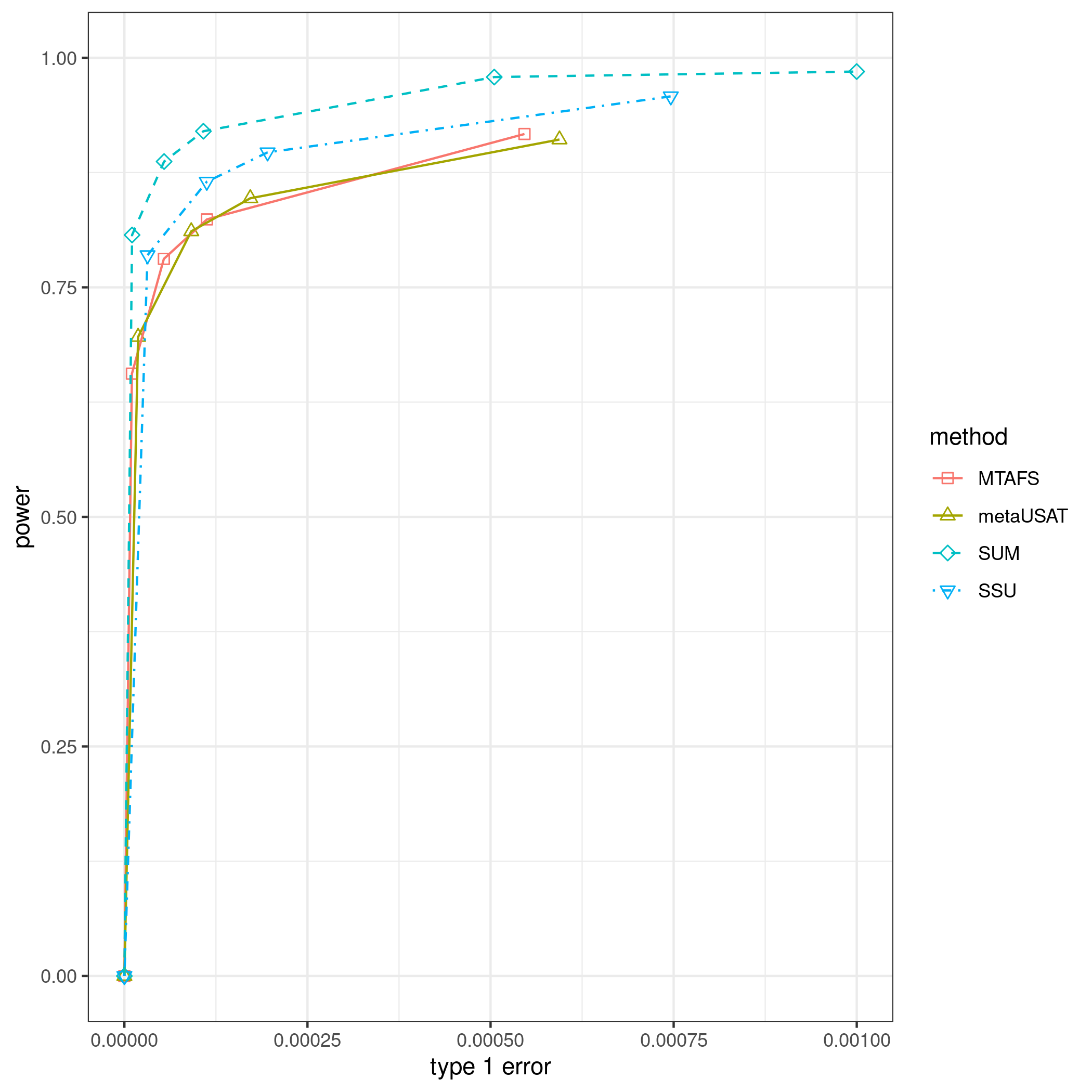}
        \caption{}
        \label{fig:v3d}
      \end{subfigure}
      \par
      \begin{subfigure}{0.4\textwidth}
        \includegraphics[width=\textwidth, height=6cm]{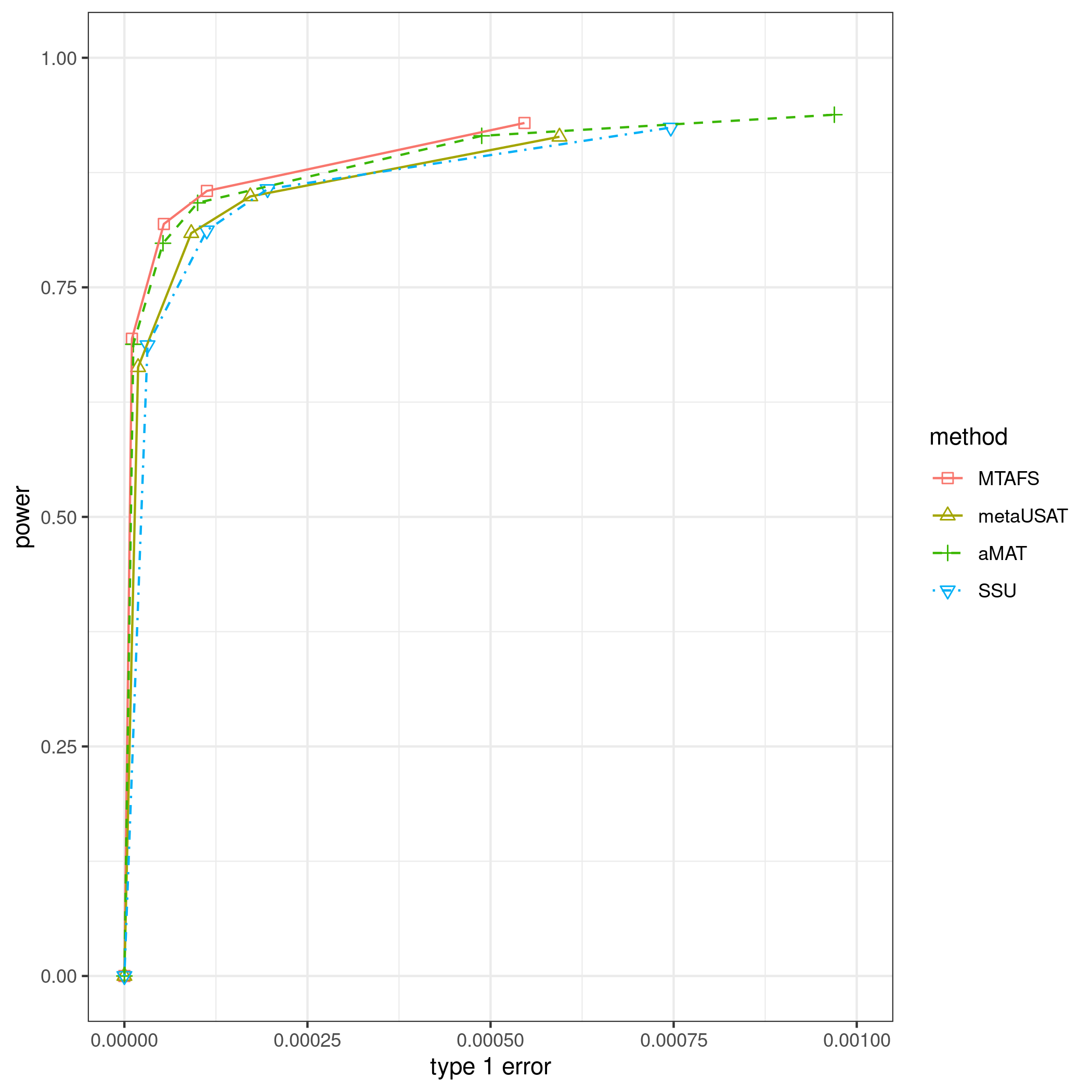}
        \caption{}
        \label{fig:v3e}
      \end{subfigure}
      \par
      \begin{subfigure}{0.4\textwidth}
        \includegraphics[width=\textwidth, height=6cm]{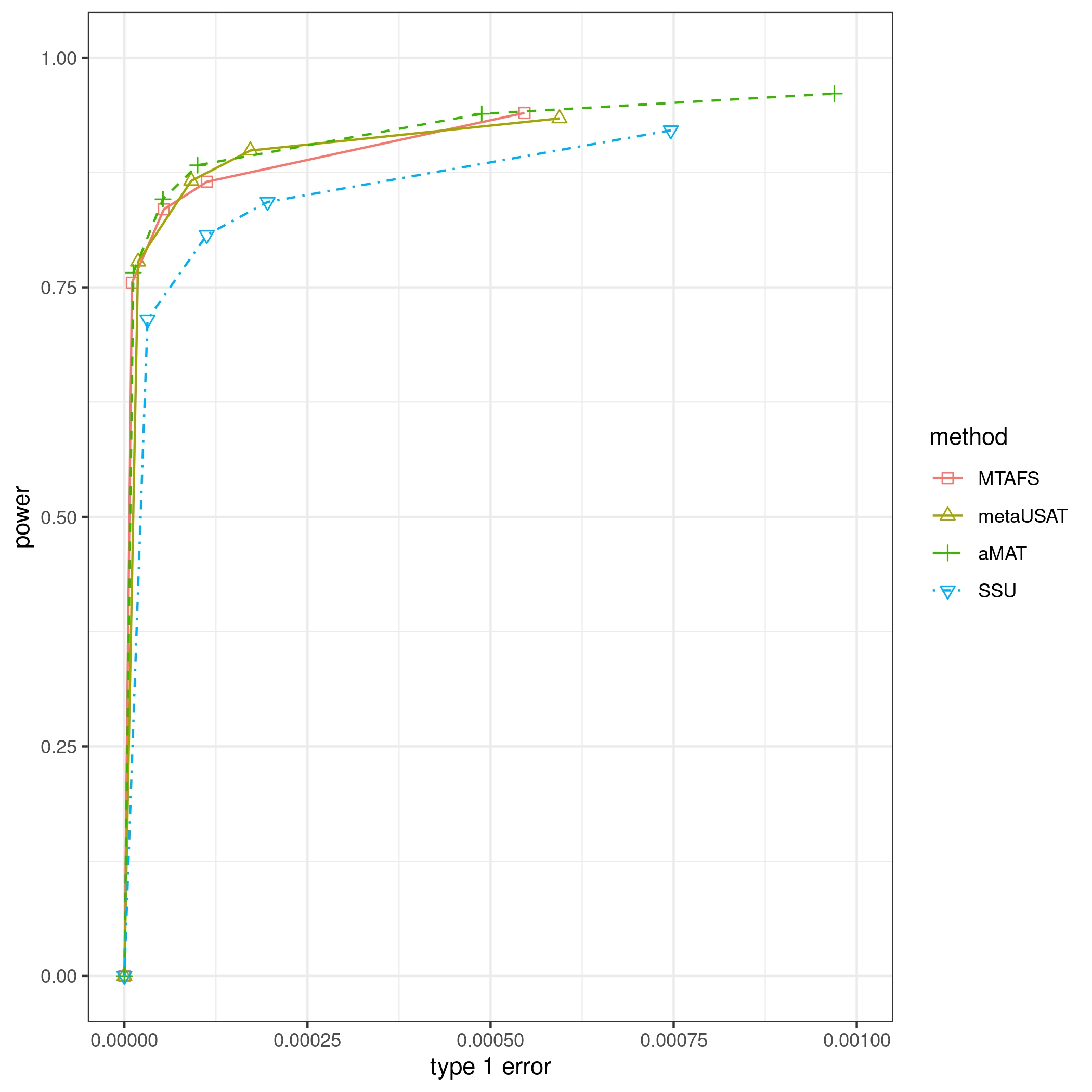}
        \caption{}
        \label{fig:v3f}
      \end{subfigure}
    \end{multicols}
  \caption{Comparison of methods for model M1 using the UKCOR1 correlation matrix. (a) high sparsity, with only top 2 eigenvectors informative; (b) intermediate sparsity, with top 11 eigenvectors informative; (c) low sparsity, with top 25 eigenvectors informative; (d) partial ROC curves for the four best methods with comparable power in (a); (e) partial ROC curves for the four best methods with comparable power in (b); (f) partial ROC curves for the four best methods with comparable power in (c).}
  \label{fig:v3}
\end{figure}

For power comparisons, we simulated 1000 z scores and the significance level was set to be $5 \times 10 ^{-5}$. First, we evaluated the power of the different methods with UKCOR1. For the association model M1 (Figure \ref{fig:v3}), when only the top two eigenvectors were informative, SUM and SSU were the most powerful methods, followed by metaUSAT and MTAFS (Figure \ref{fig:v3a}). As more eigenvectors become informative, the power of SUM decreased, while SSU, metaUSAT, and MTAFS continue to perform well, and aMAT also joined this group for the less sparse scenarios (Figure \ref{fig:v3}(b,c)). Considering the type 1 error inflation of SSU and metaUSAT, receiver operating characteristic (ROC) curves (with a particular effect size for each of the three sparsity settings) restricted to a small type 1 error range were used to measure the performance of the top 4 methods in each sparsity level, for a fairer comparison of power (Figure \ref{fig:v3}(d-f)). Due to the inflated type 1 error of SSU and metaUSAT, they in fact have smaller power compared to SUM and MTAFS when the empirical type 1 errors are the same at a very small level, especially with less sparse scenarios (Figure \ref{fig:v3}(e-f)). We note that HOM had no power at all three sparsity levels, an observation consistent with previous studies \citep{wu2020multi}.

\begin{figure}[htbp]
    \begin{multicols}{2}
     \centering
      \begin{subfigure}{0.4\textwidth}
        \includegraphics[width=\textwidth, height=6cm]{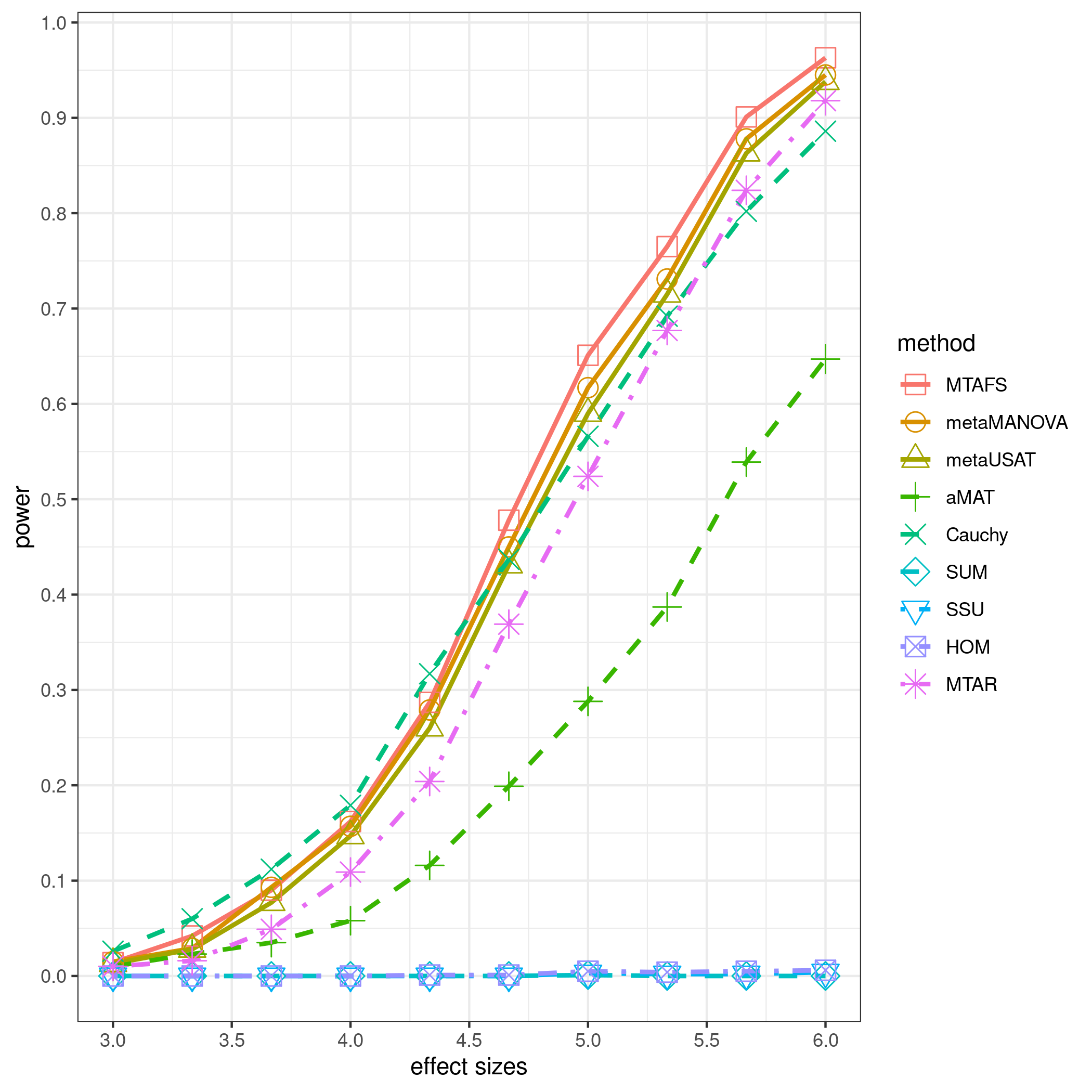}
        \caption{}
        \label{fig:v4a}
      \end{subfigure}
      \par
      \begin{subfigure}{0.4\textwidth}
        \includegraphics[width=\textwidth, height=6cm]{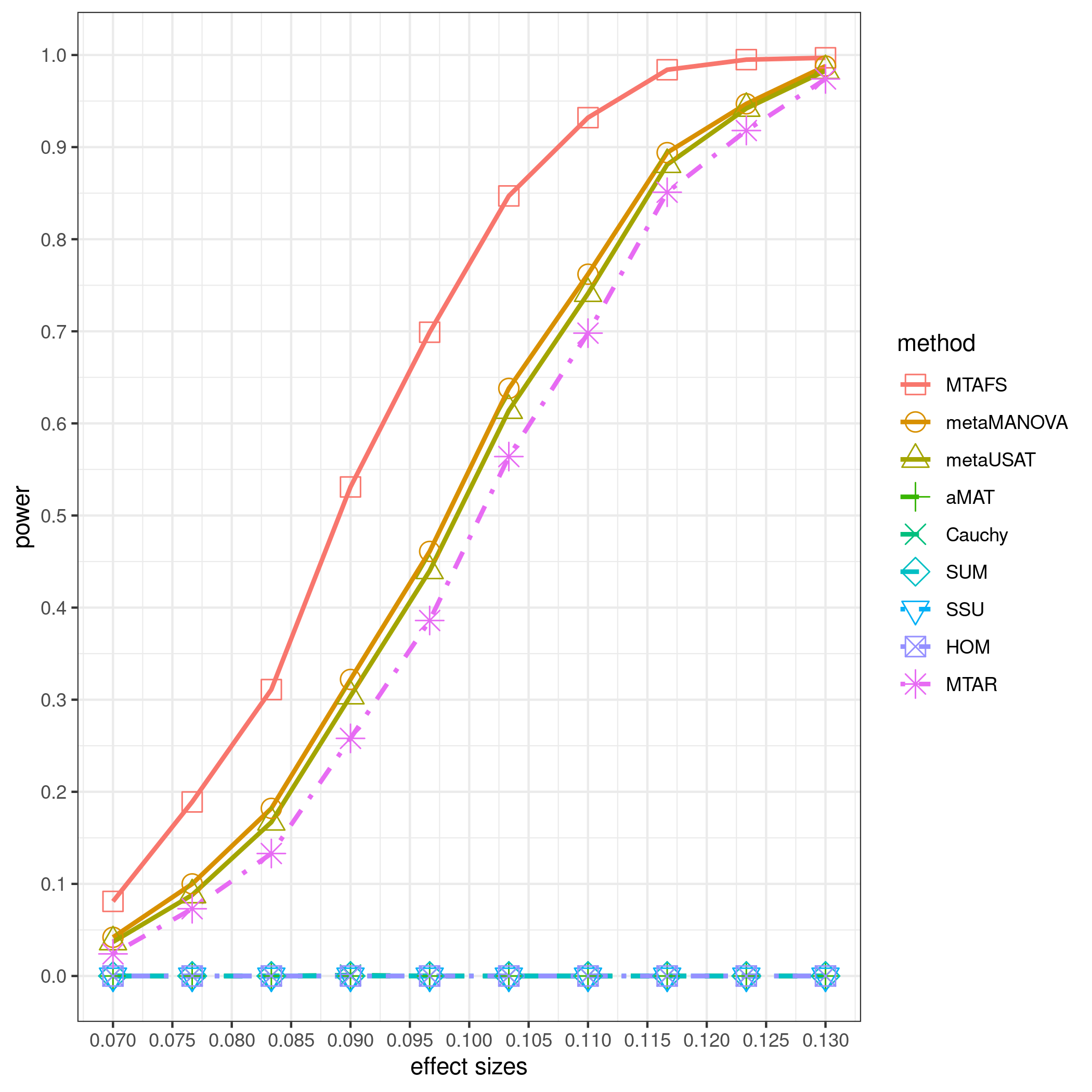}
        \caption{}
        \label{fig:v4b}
      \end{subfigure}
      \par
      \begin{subfigure}{0.4\textwidth}
        \includegraphics[width=\textwidth, height=6cm]{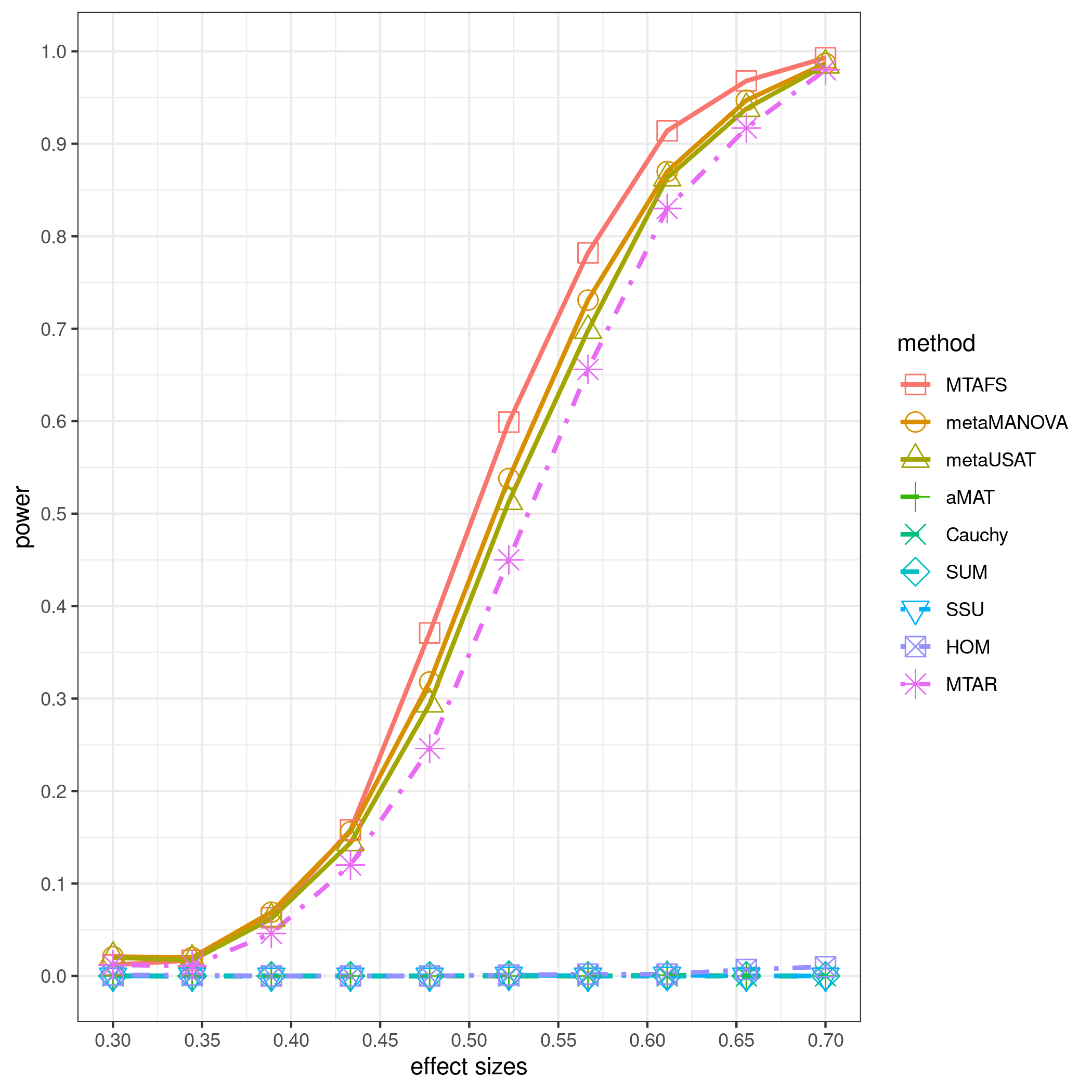}
        \caption{}
        \label{fig:v4c}
      \end{subfigure}
      \par
      \begin{subfigure}{0.4\textwidth}
        \includegraphics[width=\textwidth, height=6cm]{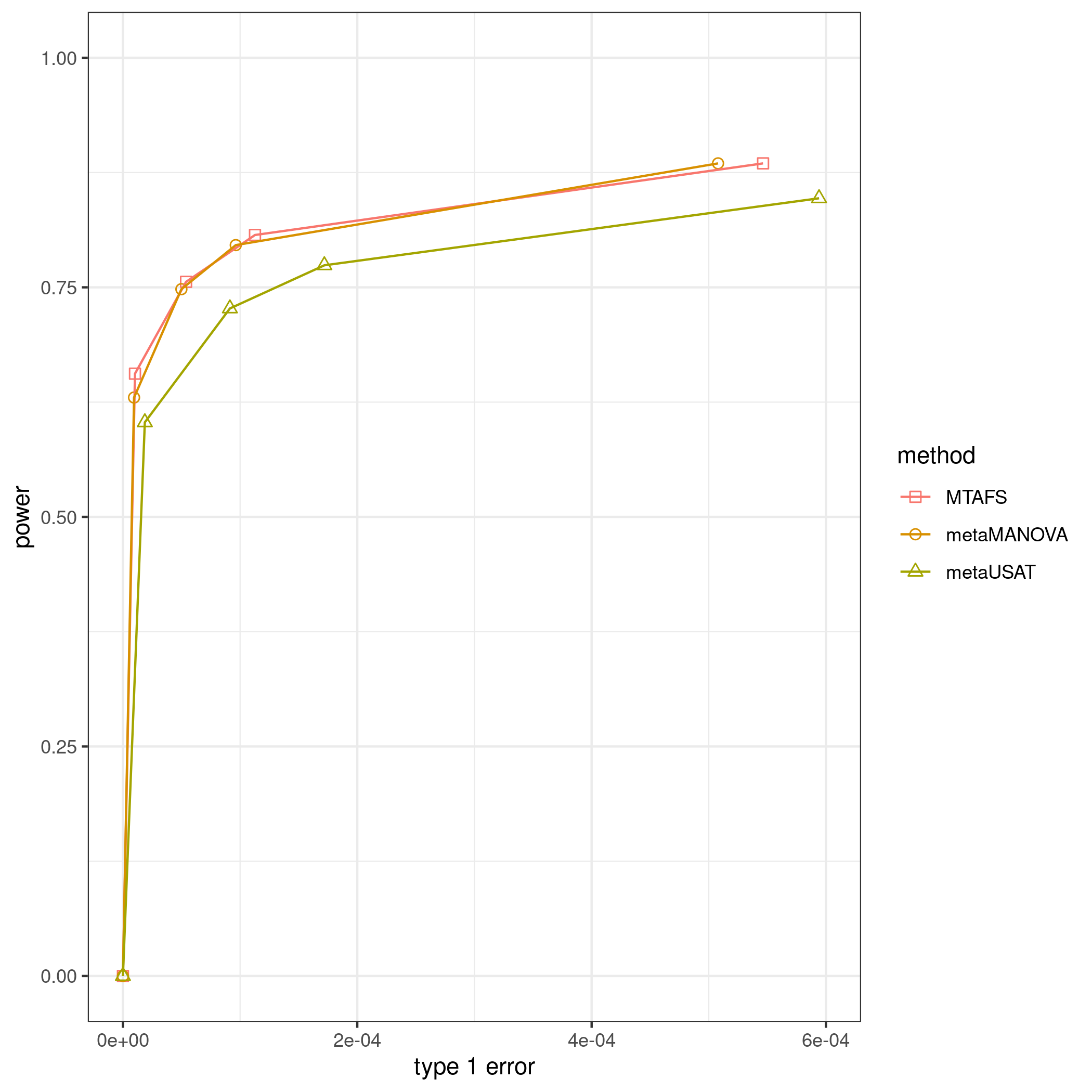}
        \caption{}
        \label{fig:v4d}
      \end{subfigure}
      \par
      \begin{subfigure}{0.4\textwidth}
        \includegraphics[width=\textwidth, height=6cm]{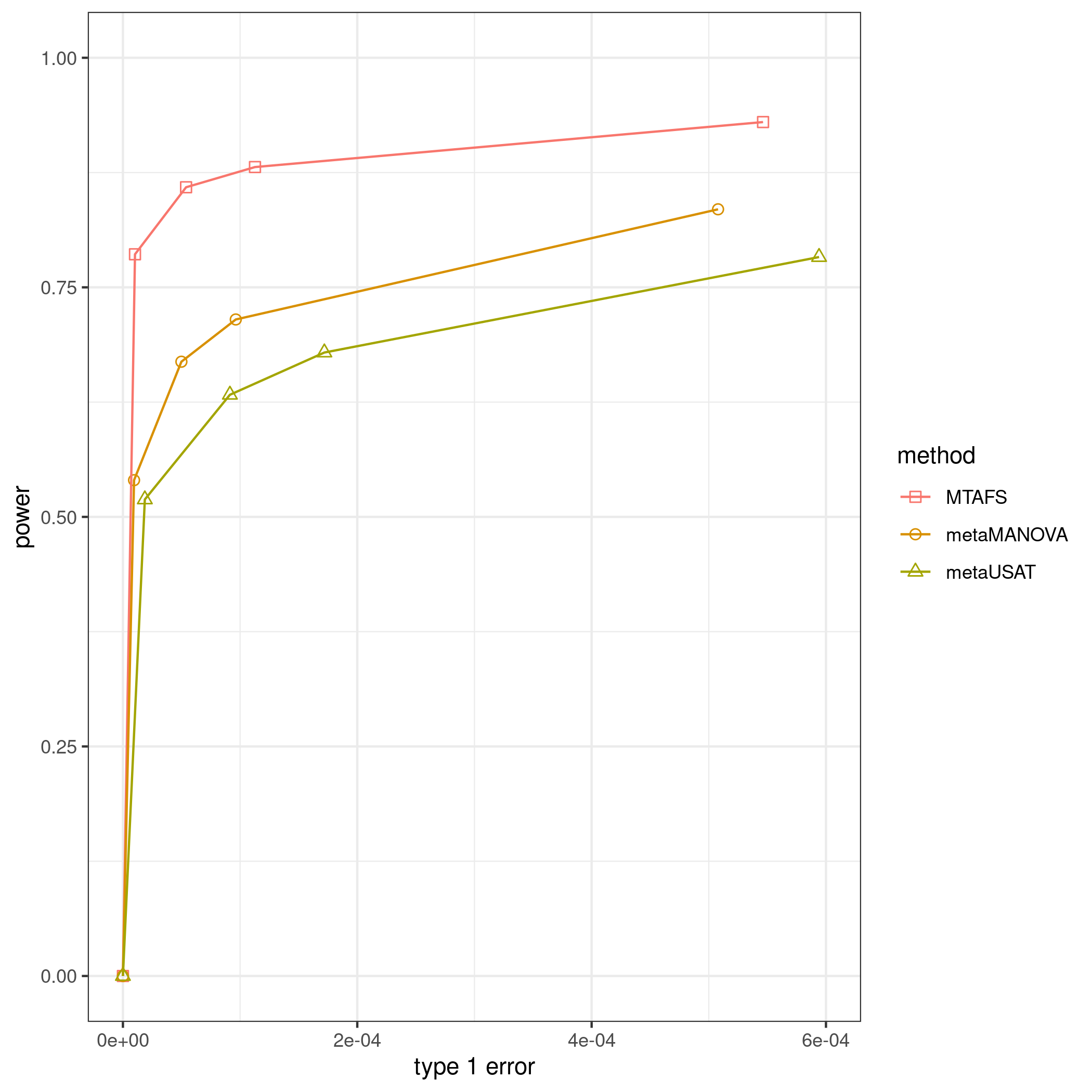}
        \caption{}
        \label{fig:v4e}
      \end{subfigure}
      \par
      \begin{subfigure}{0.4\textwidth}
        \includegraphics[width=\textwidth, height=6cm]{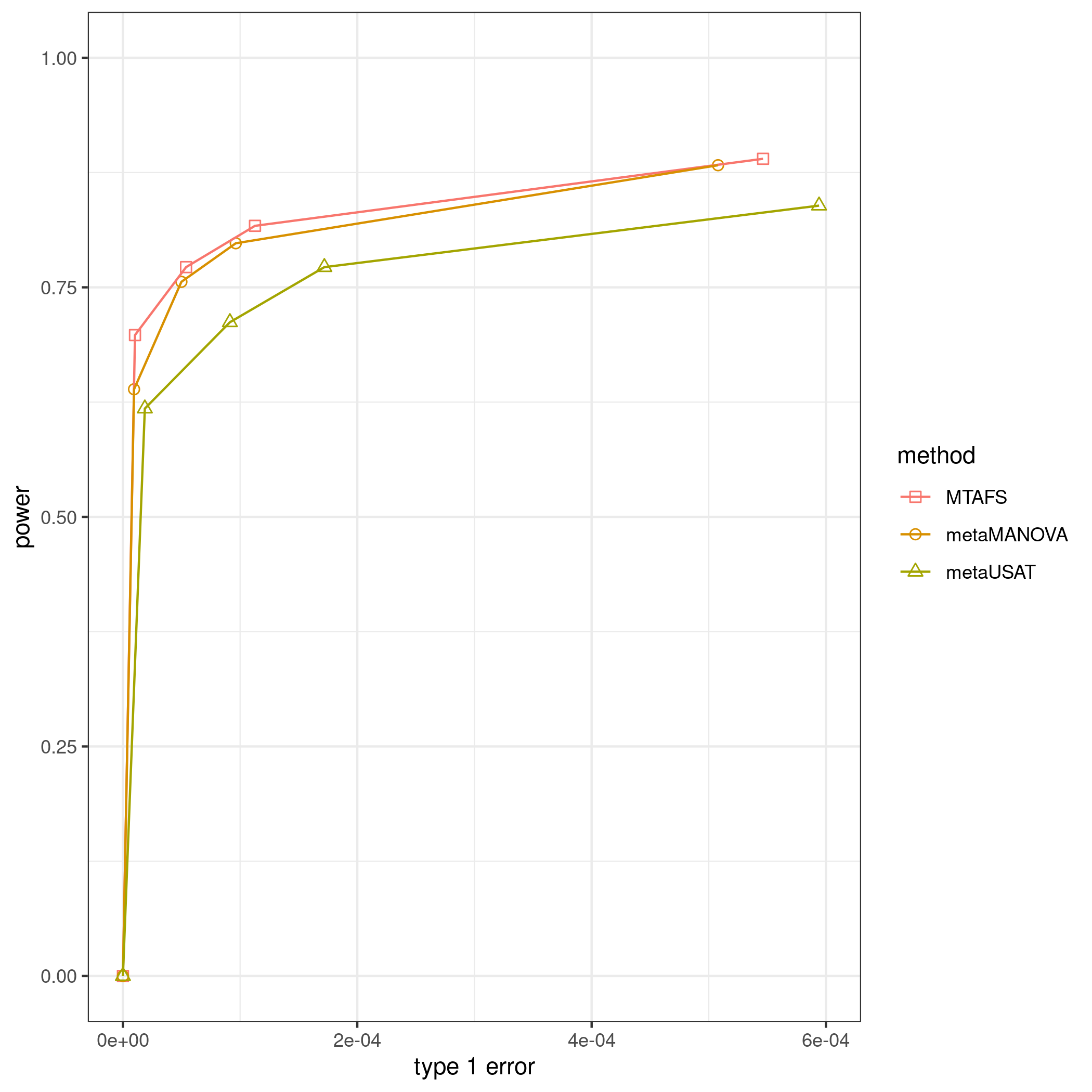}
        \caption{}
        \label{fig:v4f}
      \end{subfigure}
    \end{multicols}
  \caption{Comparison of methods for model M2 using the UKCOR1 correlation matrix. (a) high sparsity, with only 3 nonzero components of $\bm \mu$; (b) intermediate sparsity, with 13 nonzero components of $\bm \mu$; (c) low sparsity, with 30 nonzero components of $\bm \mu$ out of a total of 58; (d) partial ROC curves for the three best methods with comparable power in (a); (e) partial ROC curves for the three best methods with comparable power in (b); (f) partial ROC curves for the three best methods with comparable power in (c).}
  \label{fig:v4}
\end{figure}

For M2 with UKCOR1, MTAFS was seen to be the most powerful methods at all three sparsity levels (Figure \ref{fig:v4}). It is interesting to see that, other than MTAFS, the other methods have unstable performance, depending on the sparsity levels. For example, Cauchy was competitive in the high sparsity setting, but its power dropped down to zero at intermediate and low sparsity levels. Comparing across models M1 and M2, we see that SSU was among the powerful for M1, but its power dropped down to zero for M2. Whereas MTAFS performs well consistently across the association models, effect sizes, and sparsity levels. 

Next, we compare the results when using the correlation matrix UKCOR2 (Figure \ref{fig:power_t1fast_m1},\ref{fig:power_t1fast_m2}). For both M1 and M2, the results were similar to those for UKCOR1. Considering all the results together, the main qualitative observation for UKCOR1 remains the same for the UKCOR2 correlation matrix: the performance of the other methods are unstable, and MTAFS is extremely consistent across all settings and was always among the top performers, whereas all the other methods are less stable. 

For the CS covariance matrix with model M2 (Figure \ref{fig:power_cs50},\ref{fig:power_cs100}), MTAFS remains among the group of most powerful methods. This is also true for M2 with the AR structure (Figure \ref{fig:power_ar50},\ref{fig:power_ar100}), except that in the high sparsity setting, Cauchy outperformed all other methods by a large margin.

Considering all results from the simulation study with two different association models, effect sizes, sparsity levels, and covariance structures. It is clear that MTAFS is the most robust method. Although metaUSAT is also among the leaders in all settings in terms of power, we would argue that MTAFS is preferred since its type 1 error is well controlled while metaUSAT has been seen to have severely inflated type 1 error in some settings. Further, MTAFS may outperform metaUSAT in some scenarios (Figure \ref{fig:power_t1fast_m2}), whereas MTAFS was never greatly outperformed by metaUSAT.

%\subsection{Sensitivity analysis}

%We first checked type 1 error rates given the t distribution (Table \ref{tab:s2}). Compared with normal distributions, the t distribution had a heavier tail. Thus, all methods had inflated type 1 error rates. However, the extents of inflation were different among the methods. Although SSU was the most inflated method given normal distributions, its inflation was much slighter than many other methods under the t distribution. The t distribution resulted in severe inflation of MTAFS, metaMANOVA, metaUSAT, and especially MTAR, suggesting a strong dependence on the normality assumption. Next, we evaluated the power performance under the weighted PCs with the Volume trait correlation matrix. The power of the methods increased given the small effect sizes, because t distributions had thicker tails than normal distributions leading to more large z-scores. We also used ROC curves to further compare the methods (Figure \ref{fig:s12}). It showed that considering the type 1 error inflation problem, Cauchy had the best performance. 

\section{Real data application}

\subsection{Data and pre-processing}

Regional brain morphology such as surface area and thickness of the cerebral cortex, and volume of subcortical structures has a complex genetic architecture involving many common genetic variants with small effect sizes and the strongly overlapped genetic architectures of sets of regional brain features \citep{van2020understanding}. \cite{van2020understanding} applied an multi-trait method to 171 regional brain morphology measures and identified much more significant SNPs than the single-trait analysis, suggesting that multi-trait analysis of regional measures can be powerful to discover genetic variants. Also, brain imaging data (e.g., functional magnetic resonance imaging (fMRI) data) have been proved useful for investigating connections between brain function and genetics \citep{liu2009combining}.

UK Biobank is a rich and long-term prospective epidemiological study of 500,000 volunteers \citep{sudlow2015uk}. Participants were 40–69 years old at recruitment, with one aim being to acquire as rich data as possible before disease onset. \cite{elliott2018genome} investigated the genetic architecture of brain structure and function by conducting GWAS of 3,144 functional and structural brain imaging phenotypes from the UK Biobank (\url{http://big.stats.ox.ac.uk/}), which cover the entire brain and including multimodal information on grey matter volume, area and thickness, white matter connections and functional connectivity. The single-trait analyses were mainly applied in the study, thus we would like to apply our multi-trait method to potentially discovery more genetic variants. We carried out two multi-trait analyses, one with a moderate number of traits: 58 \textit{Volumetric} IDPs, and one with a large number of traits: 212 \textit{Area} IDPs of grey matter (Figure \ref{fig:cor_area}).

The summary statistics included the z scores from measuring the associations between each of the 11,734,353 SNPs and each of the 58 or 212 IDPs. LDSC was applied to estimate the volume and the area IDP correlation matrices. To obtain independent SNPs to satisfy the assumption of our method, we applied LD clumping to remove SNPs whose correlation with index SNPs were above 0.2 in each window of 250kb, leading to 593,416 SNPs remaining. MTAFS and a subset of the competing methods (those performed well in some settings in the simulation study) were applied to identify significant SNPs that are associated with at least one IDP in each of the two sets of traits. We used a genome-wide significance threshold of $5 \times 10^{-8}$ for each of the multi-trait analysis methods. The genes corresponding to the significant SNPs were identified using NCBI dbSNP \citep{sayers2021database}. To investigate gene annotations, we used Functional Mapping and Annotation (FUMA) \citep{watanabe2017functional} to show tissue specific expression patterns of genes identified by MTAFS and other methods.

\subsection{Results of 58 \textit{Volumetric} IDPs}

\begin{figure}
\centering

\tabskip=0pt
\valign{#\cr
  \hbox{%
    \begin{subfigure}[b]{.48\textwidth}
    \centering
    \includegraphics[width=\textwidth]{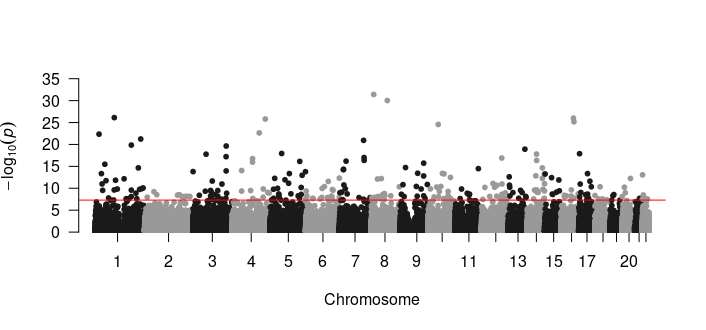}
    \caption{}
    \label{fig:volume_man}
    \end{subfigure}%
  }\vfill
  \hbox{%
    \begin{subfigure}{.48\textwidth}
    \centering
    \includegraphics[width=\textwidth]{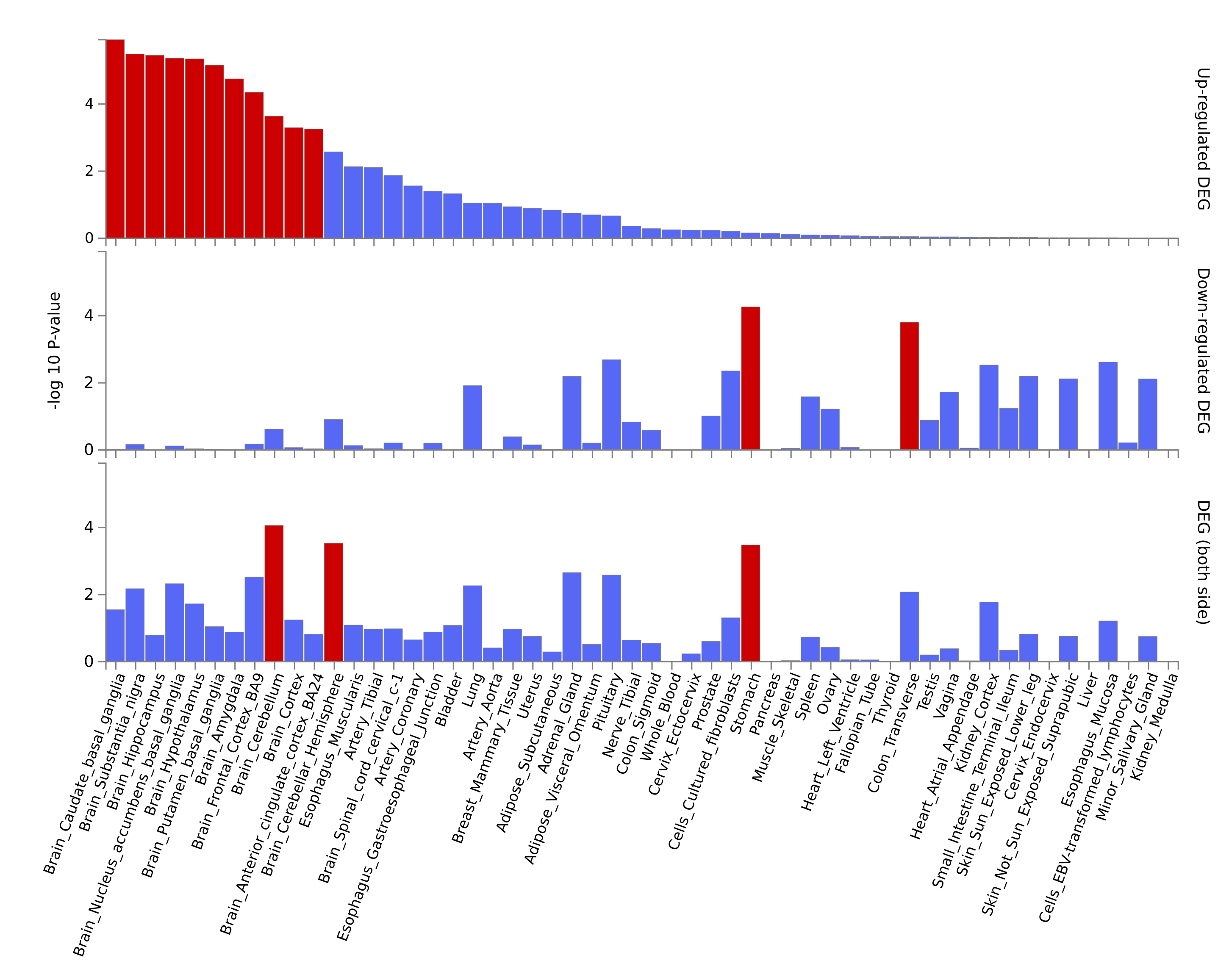}
    \caption{}
    \label{fig:volume_deg}
    \end{subfigure}%
  }\cr
  \noalign{\hfill}
  \hbox{%
    \begin{subfigure}[]{.48\textwidth}
    \centering
    \includegraphics[width=\textwidth]{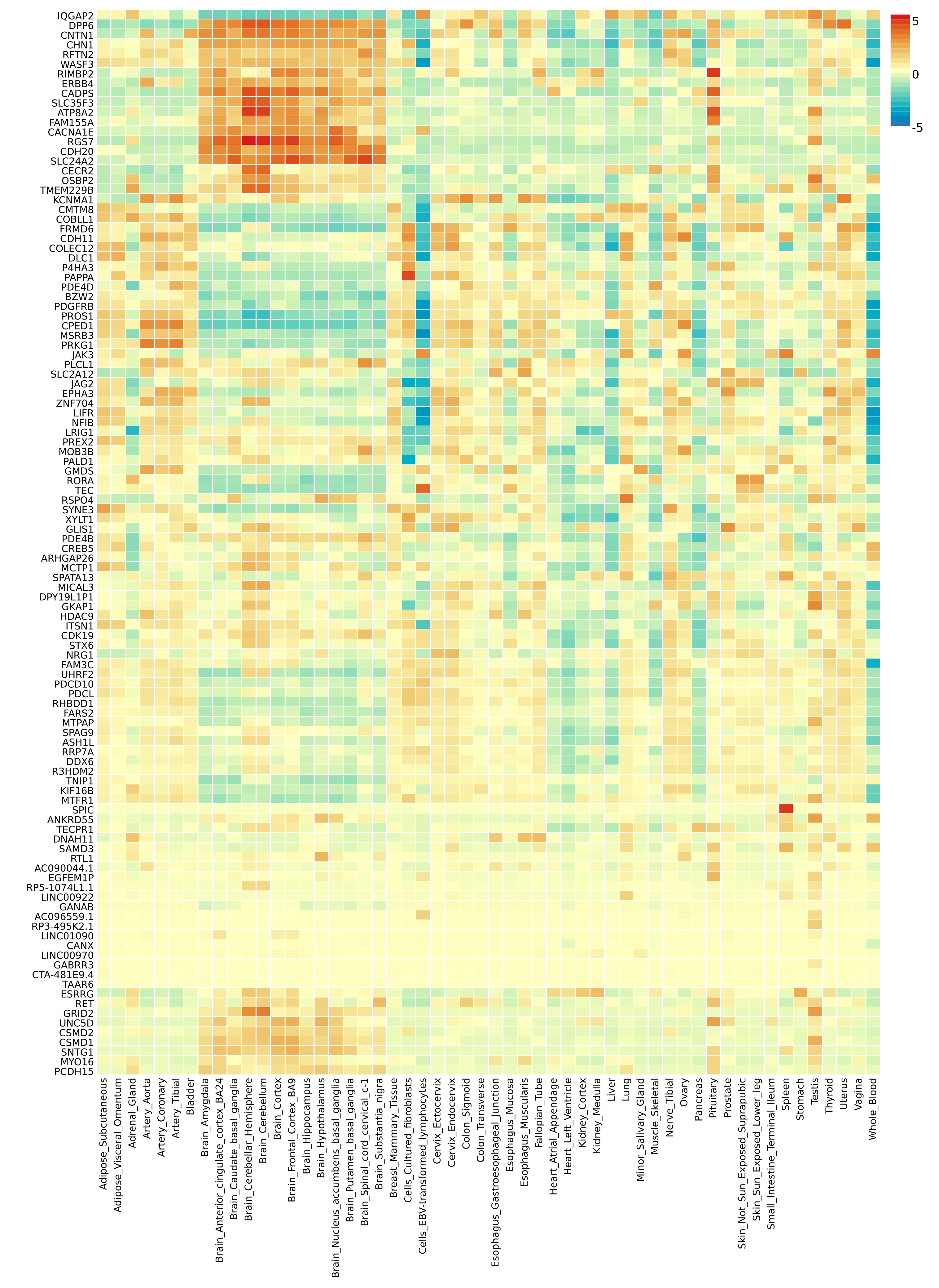}
    \caption{}
    \label{fig:volume_heat}
    \end{subfigure}%
  }\cr
}

\caption{Analysis results of the 58 \textit{Volumetric} IDPs. (a) Manhattan plot of the SNPs identified by MTAFS. For (b) and (c), we use the GTEx data over 54 tissue types. (b) Tissue expression analysis for genes uniquely identified by MTAFS for volume. Significant enrichment are in red with p-values less than 0.05 after Bonferroni correction; (c) The expression heatmap of all genes identified by MTAFS for volume. The red clusters at the top of the figure close to the left have higher relative expression.}
\label{fig:volume_realdata}
\end{figure}

MTAFS identified 264 SNPs with p-values less than $5 \times 10^{-8}$ (Figure \ref{fig:volume_man}), followed by metaMANOVA with 90 SNPs (Table \ref{tab:sigsnp}). The rest of the methods (metaUSAT, aMAT, MTAR) identified even fewer SNPs (Figure \ref{fig:volume_venn}). We also carried out single-trait analysis as a comparison, which identified only 6 significant SNPs (at the significance level of $5 \times 10^{-8}/58$), all of which were also identified by each of the multi-trait methods. 

Many of the unique genes found by MTAFS, including \textit{ATP8A2}, \textit{DPP6}, \textit{ERBB4}, and \textit{GRID2}, have been reported previously to be associated with brain structure and function. Several studies showed that \textit{ATP8A2} was closely related to cerebellum, and its mutation could cause cognitive impairment and intellectual disability \citep{martin2016new, mcmillan2018recessive, onat2013missense}. \textit{DPP6} has been reported to be associated with human neural diseases\citep{cacace2019loss, clark2008dpp6} and thalamus volume \citep{alliey2019nrxn1}. A knockout of it in mice led to impaired hippocampal-dependent learning and memory and smaller brain size \citep{lin2020novel}. \textit{ERBB4} is a candidate risk gene for schizophrenia \citep{silberberg2006involvement, law2007disease} and an essential regulator of central neural system \citep{gassmann1995aberrant} . It was also reported to be associated with total intracranial volume \citep{alliey2019nrxn1}. \textit{GRID2} is known to be differentially expressed in Purkinje cells in the cerebellum and the deletion of it causes cerebellar ataxia \citep{hills2013deletions, van2015early}. \textit{PAPPA} was also found by both single-trait analysis and MTAFS. It was reported to be associated with brain region volumes in previous studies \citep{elliott2018genome,zhao2019genome}.	

To further investigate the biological mechanism, we used FUMA to annotate the genes identified in terms of biological context. Figure \ref{fig:volume_heat} shows the gene expression heatmap of significant genes found by MTAFS. The expression value depends on the genotype-tissue expression (GTEx) project \citep{lonsdale2013genotype} including 54 human tissues. There were 14 tissues specifically related to brain such as amygdala and caudate basal ganglia. There was a cluster of genes close to the top left with higher relative expression; this cluster includes 13 of the 14 brain-related tissues. In FUMA, we also tested if the gene set was significantly enriched in tissues. Especially, we were interested in whether the set of genes uniquely identified by MTAFS are biologically relevant. Figure \ref{fig:volume_deg} shows that those genes were enriched significantly in most brain-related tissues (red bars in the top plot showing up-regulation). In contrast, we found the gene set consisting of genes identified by the comparison methods was not significantly enriched in any of the brain-related tissues (Figure \ref{fig:volume_other_deg}).

\subsection{Results of 212 \textit{Area} IDPs}

\begin{figure}
\centering

\tabskip=0pt
\valign{#\cr
  \hbox{%
    \begin{subfigure}[b]{.48\textwidth}
    \centering
    \includegraphics[width=\textwidth]{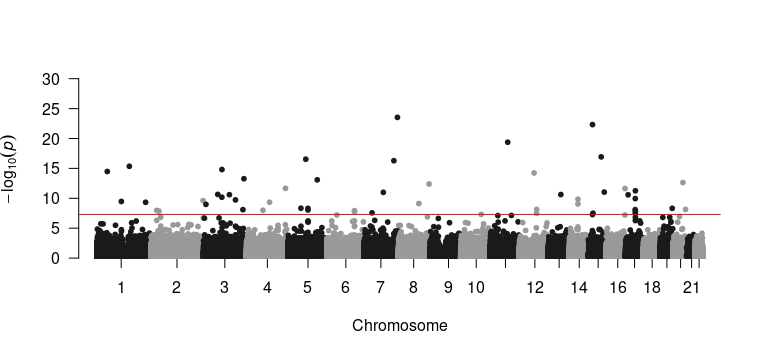}
    \caption{}
    \label{fig:area_man}
    \end{subfigure}%
  }\vfill
  \hbox{%
    \begin{subfigure}{.48\textwidth}
    \centering
    \includegraphics[width=\textwidth]{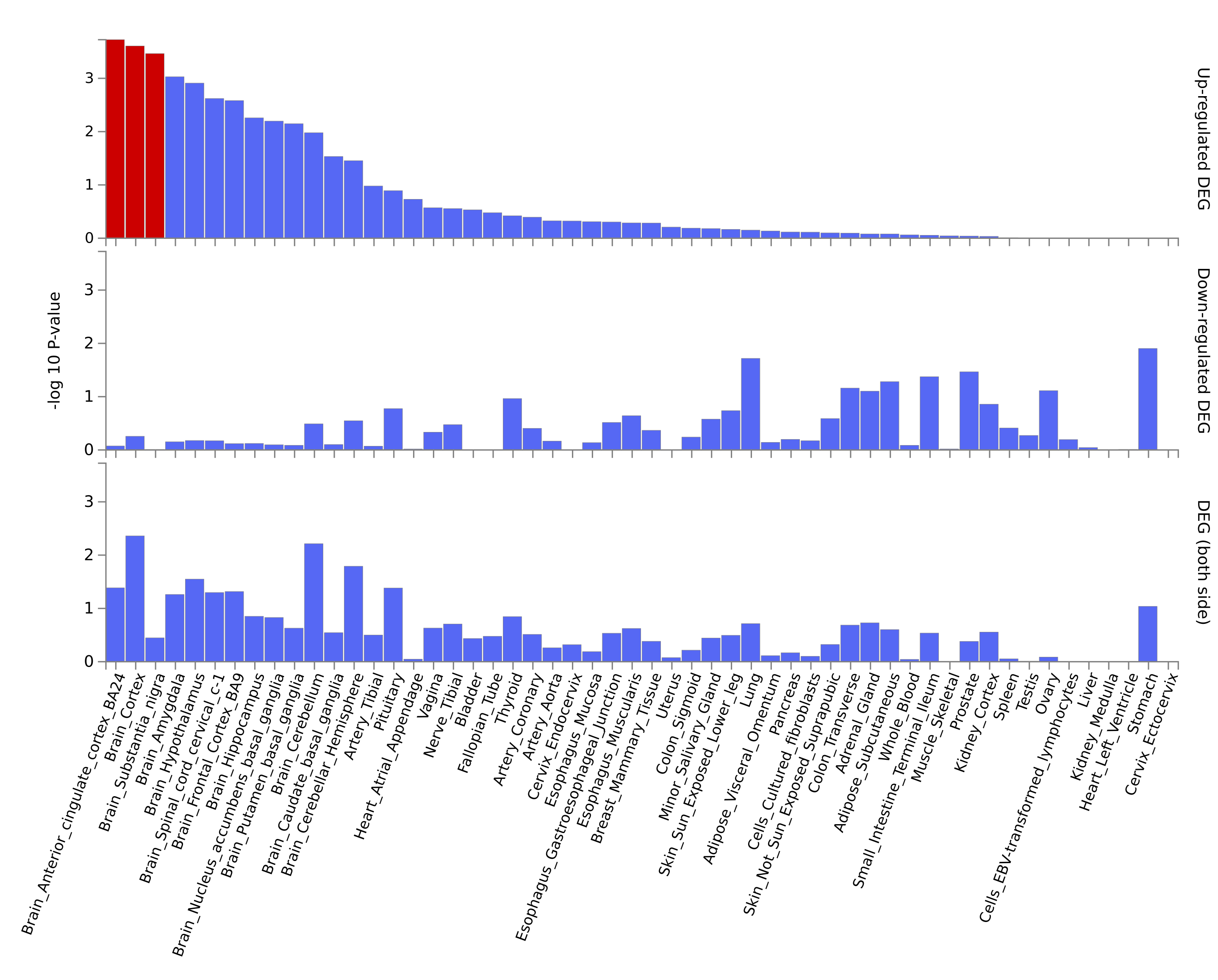}
    \caption{}
    \label{fig:area_deg}
    \end{subfigure}%
  }\cr
  \noalign{\hfill}
  \hbox{%
    \begin{subfigure}[]{.48\textwidth}
    \centering
    \includegraphics[width=\textwidth, height=10cm]{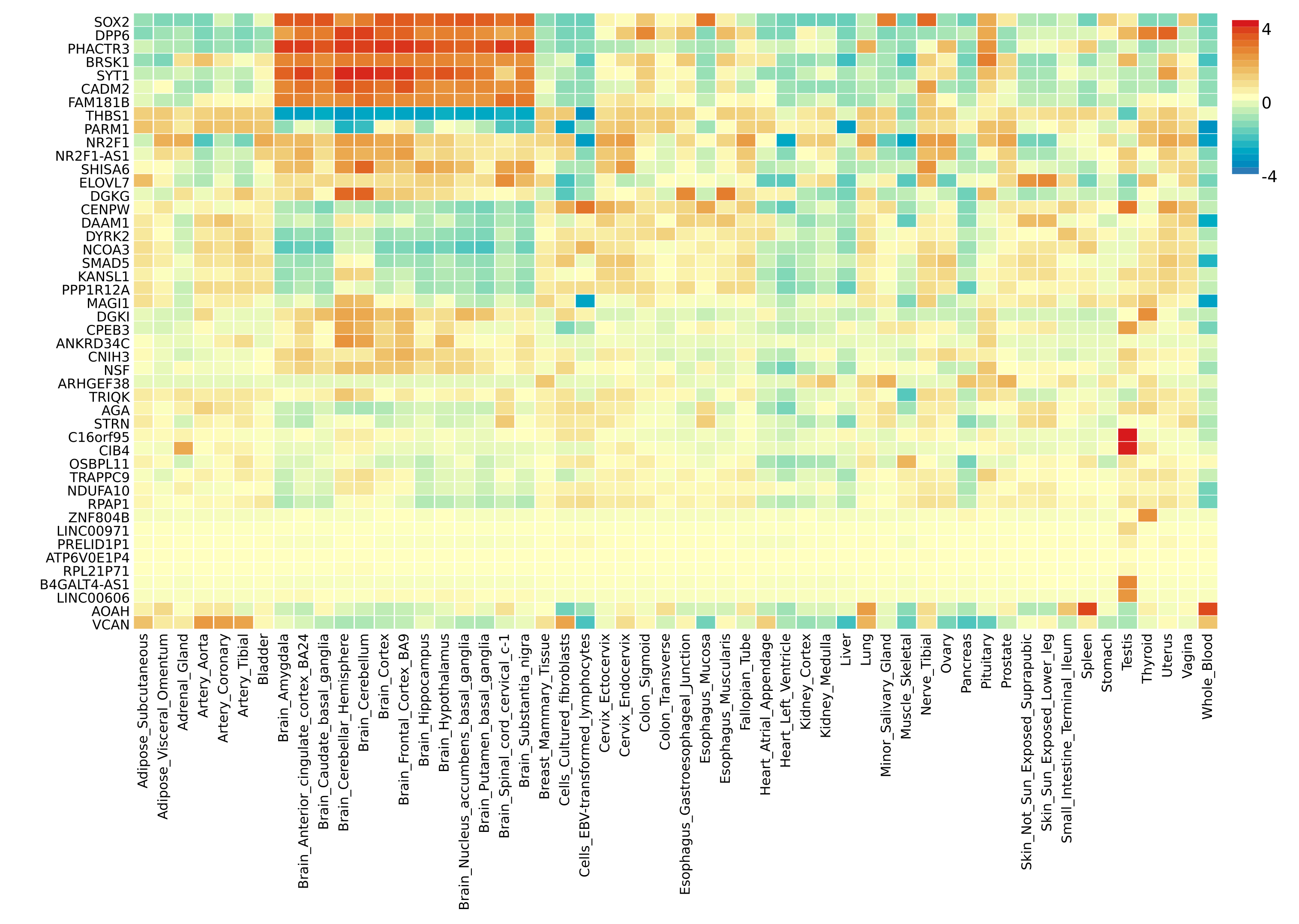}
    \caption{}
    \label{fig:area_heat}
    \end{subfigure}%
  }\cr
}

\caption{Analysis results of the 212 \textit{Area} IDPs. (a) Manhattan plot of the SNPs identified by MTAFS. For (b) and (c), we use the GTEx data over 54 tissue types. (b) Tissue expression analysis for genes uniquely identified by MTAFS for volume. Significant enrichment are in red with p-values less than 0.05 after Bonferroni correction; (c) The expression heatmap of all genes identified by MTAFS for volume. The red clusters have higher relative expression.}
\label{fig:area_realdata}
\end{figure}

This analysis considered a much larger set of traits. In this case, our preliminary analysis found that metaUSAT has numerical issues; thus, it was excluded from consideration. MTAFS identified 55 SNPs with p-values less than $5 \times 10^{-8}$ (Figure \ref{fig:area_man}). On the other hand, metaMANOVA, aMAT, and MTAR had identified smaller sets of similar number of SNPs (Figure \ref{fig:area_venn}). Single-trait analysis only identified 1 SNP (Table \ref{tab:sigsnp}), and that SNP was identified by all multi-trait methods, indicating that multi-trait methods were more powerful than single-trait analysis in real data applications, mostly because they are less conservative compared to Bonferroni correction for the number of SNPs $\times$ traits combinations.

Among the genes corresponding to significant SNPs identified by MTAFS only, we found that two genes, \textit{DPP6} and \textit{LINC02210-CRHR1}, were previously identified in our first analysis with 58 \textit{Volumetric} IDPs. We further investigated \textit{LINC02210-CRHR1} and found that it was reported to be associated with several brain structures and functions. \citep{zhao2019genome} reported that \textit{LINC02210-CRHR1} was significantly associated with brain volume, and \citep{hibar2015common} found that it was associated specifically with subcortical brain region volumes. Two recent studies showed its relevance in the cortical surface area \citep{shin2020global, grasby2020genetic}. Several genes identified by MTAFS, such as \textit{C16orf95} and \textit{DGKI}, also appeared in other studies using the UK Biobank data: \cite{van2020understanding} analyzed the structural brain imaging data and identified genes \textit{C16orf95}, \textit{DGKI}, \textit{SYT1}, and \textit{VCAN}; \cite{hofer2020genetic} conducted association studies of brain cortical thickness, surface area, and volume, and they also identified gene \textit{C16orf95}, \textit{DAAM1}, \textit{NR2F1}, \textit{NSF}, and \textit{VCAN}.

In the expression heatmap (Figure \ref{fig:area_heat}), it shows a cluster of genes that had higher relative expression in the brain-related tissues than other tissues (the cluster locating at the same position as in the \textit{Volume} analysis). In particular, we saw that \textit{PHACTR3} was highly expressed in all brain-related tissues. Many studies showed that this gene is important in intelligence, cognitive function, and schizophrenia \citep{goes2015genome, turley16gene, davies2018study, hill2019combined}. Figure \ref{fig:area_deg} shows that the gene set consisting of genes identified by MTAFS were significantly enriched in brain three tissues substantia nigra, cortex, and anterior cingulate cortex. In contrast, the gene set consisting of the genes identified by only the other methods was not significantly enriched in any brain tissues (Figure \ref{fig:area_other_deg}), although the expression level in a small cluster had relatively higher expression in brain-related tissues (Figure \ref{fig:area_other_heat}).

\section{Discussion}

GWAS have successfully identified a large number of genetic variants associated with traits or diseases. However, for many traits, a large portion of the heritability is still unaccounted for. In contrast to individual-level data, GWAS summary statistics are usually publicly available and have more potentials for achieving greater statistical power through combining a large amount of information. Our method utilizes z scores which are usually available in GWAS summary statistics along with their p-values. In rare cases where only p-values are available, we can transform the p-values to z scores by using the normality assumption. Although methods are available for joint analyses of a large number of traits from deep phenotyping data, inconsistent performance, computational inefficiency, and numerical issues when a large number of traits is considered are issues that are yet to be resolved. Our proposed MTAFS is an attempt in this direction. Our simulation study shows that MTAFS can control type 1 error well and has consistent performance under a variety of settings, underscoring its robustness. In real data applications, we see that, in contrast to single-trait analysis, MTAFS identified many more significant SNPs without omitting any detected by the former. Further, MTAFS identified more significant SNPs than the existing multi-trait analysis methods, and the genes identified by MTAFS are supported by evidence in the literature. Moreover, the expression of the gene set identified by MTAFS are more highly expressed in a biologically relevant manner. In contrast, the expression of gene sets identified by the existing methods do not lead to significant enrichment in brain tissues. Taken together, the two analyses show the power of MTAFS and provide some insights on genes that may be related to brain \textit{Volumetric} and \textit{Area} IDPs.

In general, MTAFS exhibits desirable properties, and have several advantages over existing methods as a whole. First, MTAFS controls type 1 error well, even with small significance levels. Second, MTAFS has robust performance given various correlation matrices, underlying association models, and different levels of signal sparsity. Third, MTAFS is an efficient method in practice, though it is not as computationally efficient as some existing methods (Table \ref{tab:time}). It is much faster than methods using permutation tests like minP and aSPU. For example, it took about 2 hours to analyze 593,416 SNPs for 58 \textit{Volumetric} IDPs with a single core of 4GB memory. Moreover, parallel computing can greatly reduce its computational time making it acceptable in pratice. As a demonstration, we analyzed the \textit{Area} IDP data for 593,416 SNPs and 212 traits, and MTAFS finished the analysis in only 10 minutes by using 60 cores of 4GB memory.

The advantages notwithstanding, there are limitations of the proposed method. First, because we transform raw z score vectors by eigendecomposition, it is difficult to interpret the association between one SNP and one single trait. Second, our choice of the levels of variance explained and the number of levels are both ad hoc. Third, MTAFS currently only considers common variants; thus further development is warranted for including rare ones.

%Existing methods take the assumption of normality of z scores for granted and few literature has discussed the deviation from normality. In our preliminary sensitivity analysis, it appears that SUM and HOM are the ones that are the least dependent on the normality assumption as their type 1 error inflation are slight. Therefore, considering a potential violation of the normality assumption, we would recommend using SUM or HOM to avoid severe type 1 error inflation.

%%%%%%%%%%%%%%%%%%%%%%%%%%%%%%% TABLES & FIGURES %%%%%%%%%%%%%%%%%%%%%%%%%%%%%%%
\newpage

% table: simulation parameters
\begin{table}[htbp]
  \centering
  \begin{threeparttable}[htbp]
  \caption{Combinations of parameter settings in the simulation for power study}
  \label{tab:scenarios}
\begin{tabular}{ccccc}
Correlation & \# Traits   & Association Models & Effect Sizes$^a$    & Settings$^b$                            \\ \hline
UKCOR1      & 58  & M1                 & {[}0.6, 1.6{]}  & J=2                                 \\
            &     &                    & {[}0.6, 1.6{]}  & J=11                                \\
            &     &                    & {[}0.6,  1.6{]} & J=25                                \\ 
            &     & M2                 & {[}3, 6{]}      & $\mu_{56} - \mu_{58} $ \\
            &     &                    & {[}0.07, 1.3{]} & $\mu_{46} - \mu_{58} $   \\
            &     &                    & {[}0.3, 0.7{]}  & $\mu_{29} - \mu_{58} $   \\ \hline
UKCOR2      & 139 & M1                 & {[}0.5, 1.4{]}  & J=2                                 \\
            &     &                    & {[}0.5, 1.1{]}  & J=27                                \\
            &     &                    & {[}0.5,  1.1{]} & J=69                                \\
            &     & M2                 & {[}1.3, 2.5{]}  & $\mu_{133},\mu_{134} $        \\
            &     &                    & {[}1.5, 3{]}    & $\mu_{107} - \mu_{134} $ \\
            &     &                    & {[}1, 2{]}      & $\mu_{70} - \mu_{139} $  \\ \hline
CS(0.3)     & 50  & M2                 & {[}2, 6{]}      & $\mu_{1},\mu_{2} $            \\
            &     &                    & {[}1, 3{]}      & $\mu_{1} - \mu_{10} $    \\ 
            & 100  & M2                 & {[}4, 7{]}      & $\mu_{1},\mu_{2} $            \\
            &     &                    & {[}1.5, 2.5{]}      & $\mu_{1} - \mu_{20} $    \\ \hline
CS(0.7)     & 50  & M2                 & {[}2, 4{]}      & $\mu_{1},\mu_{2} $            \\
            &     &                    & {[}1, 2{]}      & $\mu_{1} - \mu_{10} $    \\ 
            & 100  & M2                 & {[}2.5, 4.5{]}      & $\mu_{1},\mu_{2} $            \\
            &     &                    & {[}1, 1.5{]}      & $\mu_{1} - \mu_{20} $    \\ \hline
AR(0.3)     & 50  & M2                 & {[}4, 7{]}      & $\mu_{25},\mu_{26} $          \\
            &     &                    & {[}1, 4{]}      & $\mu_{20} - \mu_{30} $   \\ 
            & 100  & M2                 & {[}5, 8{]}      & $\mu_{50},\mu_{51} $          \\
            &     &                    & {[}1, 2.5{]}      & $\mu_{40} - \mu_{60} $   \\ \hline
AR(0.7)     & 50  & M2                 & {[}2, 6{]}      & $\mu_{25},\mu_{26} $          \\
            &     &                    & {[}2, 5{]}      & $\mu_{20} - \mu_{30} $   \\ 
            & 100  & M2                 & {[}4, 6.7{]}      & $\mu_{50},\mu_{51} $          \\
            &     &                    & {[}2, 3.5{]}      & $\mu_{40} - \mu_{60} $   \\ \hline
\end{tabular}
\begin{tablenotes}
    \item[a] For M1, the effect size refers to $c$; for M2, it is the nonzero value of the $\bm \mu$ components. For all, 10 different effect sizes are considered, which are evenly distributed in the range specified, inclusive.
    \item[b] For M1, the number specified is for $J$, the number of eigenvectors having a nonzero effect; for M2, we list the range of the $\bm \mu$ components that have nonzero values, inclusive.
  \end{tablenotes}
 \end{threeparttable}
\end{table}

% table: type 1 error

\begin{table}[htbp]
  \centering
  \begin{threeparttable}[htbp]
  \caption{Type 1 error$^a$ with correlation matrix UKCOR1}
  \label{tab:type1}
  \begin{tabular}{cccccc}
  \hline
               & \multicolumn{5}{c}{Significance Levels}    \\            
Methods & $5 \times 10^{-2}$ & $1 \times 10^{-2}$ & $1 \times 10^{-3}$ & $1 \times 10^{-4}$ & $1 \times 10^{-5}$ \\ \hline
metaMANOVA     & 1.01     & 1.01     & 1.02          & 0.963         & 0.95          \\
metaUSAT       & 1        & 1.08     & 1.17          & \textbf{1.72} & \textbf{1.87} \\
SUM            & 1        & 1        & 1             & 1.08          & 1.05          \\
SSU            & 0.92     & 1.01     & \textbf{1.34} & \textbf{2}    & \textbf{3.16} \\
HOM            & 1.01     & 1        & 1.01          & 1.03          & 1.03          \\
Cauchy         & 1.14     & 1.13     & 1.07          & 1.03          & 1             \\
aMAT           & 0.94     & 0.94     & 0.97          & 1             & 1.22 \\
MTAR           & 1        & 0.99     & 0.981         & 0.95          & 1.04          \\
MTAFS          & 1.16     & 1.14     & 1.1           & 1.13          & 1.03         \\ \hline
\end{tabular}
  \begin{tablenotes}
    \item[a] The values in the table are ratios of empirical Type I errors divided by the corresponding significance levels. Values larger than 1.3 are bold (let $\alpha= 1 \times 10^{-5}, \alpha + 3 \sqrt{\frac{\alpha(1-\alpha)}{10^7}} \approx 1.3 \times 10^{-5}$).

  \end{tablenotes}
 \end{threeparttable}
\end{table}

% figure: power, volume, M1

% figure: power, volume, M2

% volume: real data

% area: real data

\clearpage
\newpage

\bibliographystyle{natbib}
\bibliography{ref2}

\newpage

\newcommand{\beginsupplement}{%
        \setcounter{table}{0}
        \renewcommand{\thetable}{S\arabic{table}}%
        \setcounter{figure}{0}
        \renewcommand{\thefigure}{S\arabic{figure}}%
     }

%\section*{Supplementary Materials}
%\label{ch:supp}

\begin{center}
  \textbf{\large An Adaptive and Robust Method for Multi-trait Analysis of Genome-wide Association Studies Using Summary Statistics\\Supplementary Material}\\[.2cm]
  Qiaolan Deng,$^{1,3}$ Chi Song,$^{2}$ and Shili Lin$^{1,3}$\\[.1cm]
  {\itshape ${}^1$Interdisciplinary Ph.D. Program in Biostatistics\\
  ${}^2$Division of Biostatistics, College of Public Health\\
  ${}^3$Department of Statistics, The Ohio State University, Columbus, Ohio\\}
\end{center}

\beginsupplement

\newpage

% work flow
\begin{figure}[hbpt]
    \centering
    \includegraphics[width=\textwidth]{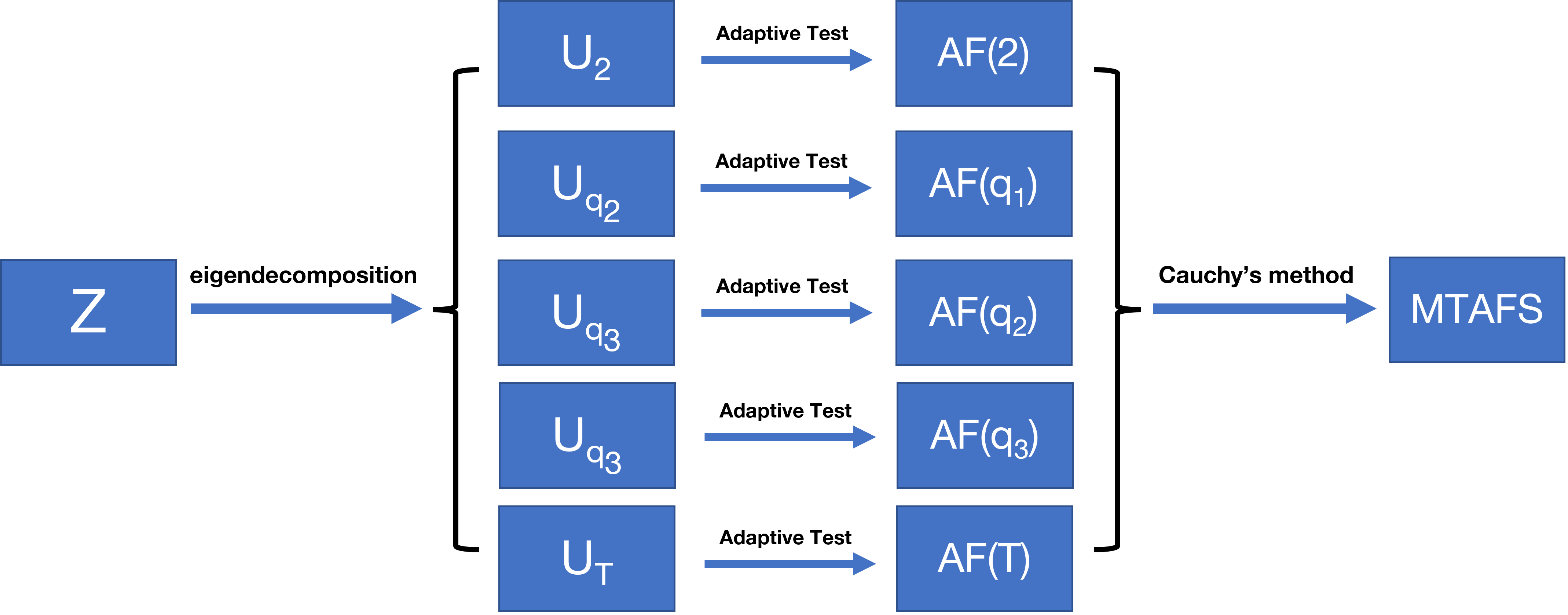}
    \caption{workflow of MTAFS}
    \label{fig:v0}
\end{figure}

% Volume
\begin{figure}[htbp]
    \centering
    \includegraphics[width=\linewidth]{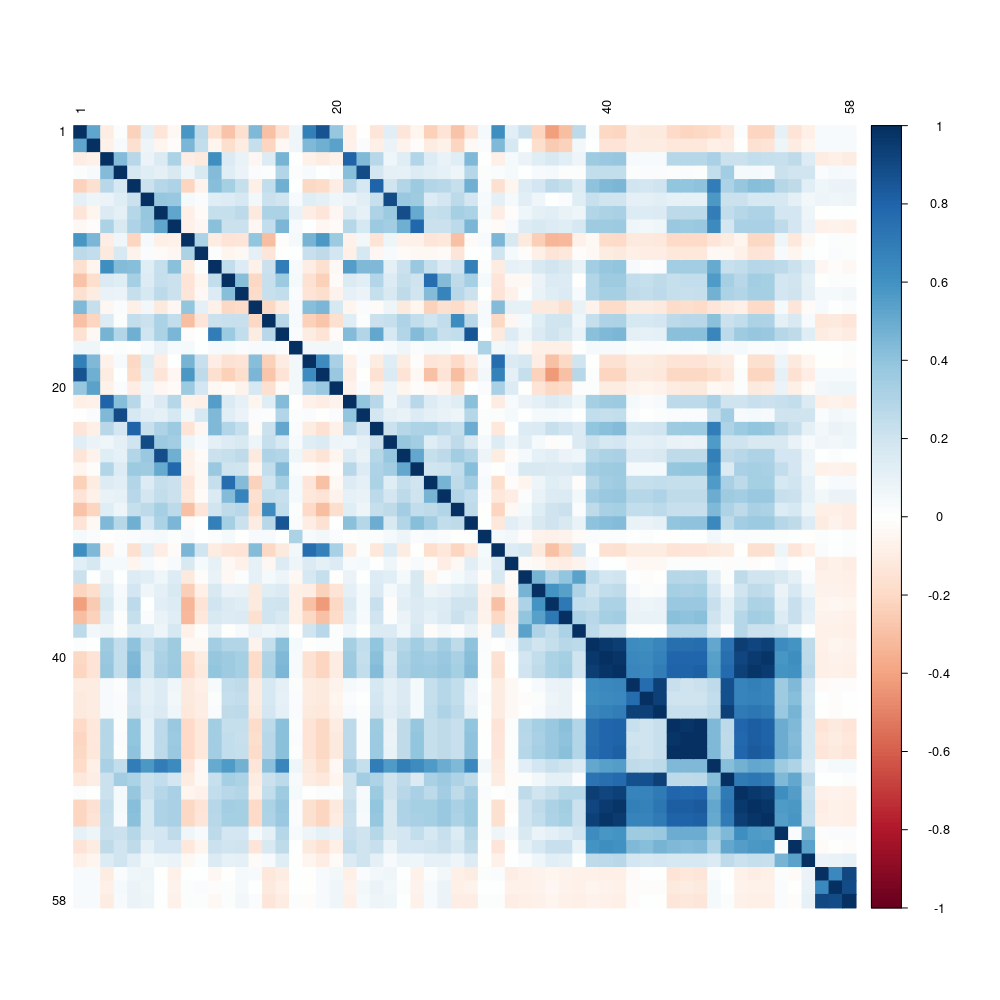}
    \caption{The LDSC estimated trait correlation matrix of Volume. Volume consists of 58 IDPs}
    \label{fig:cor_volume}
\end{figure}

% T1FAST
\begin{figure}[htbp]
    \centering
    \includegraphics[width=\linewidth]{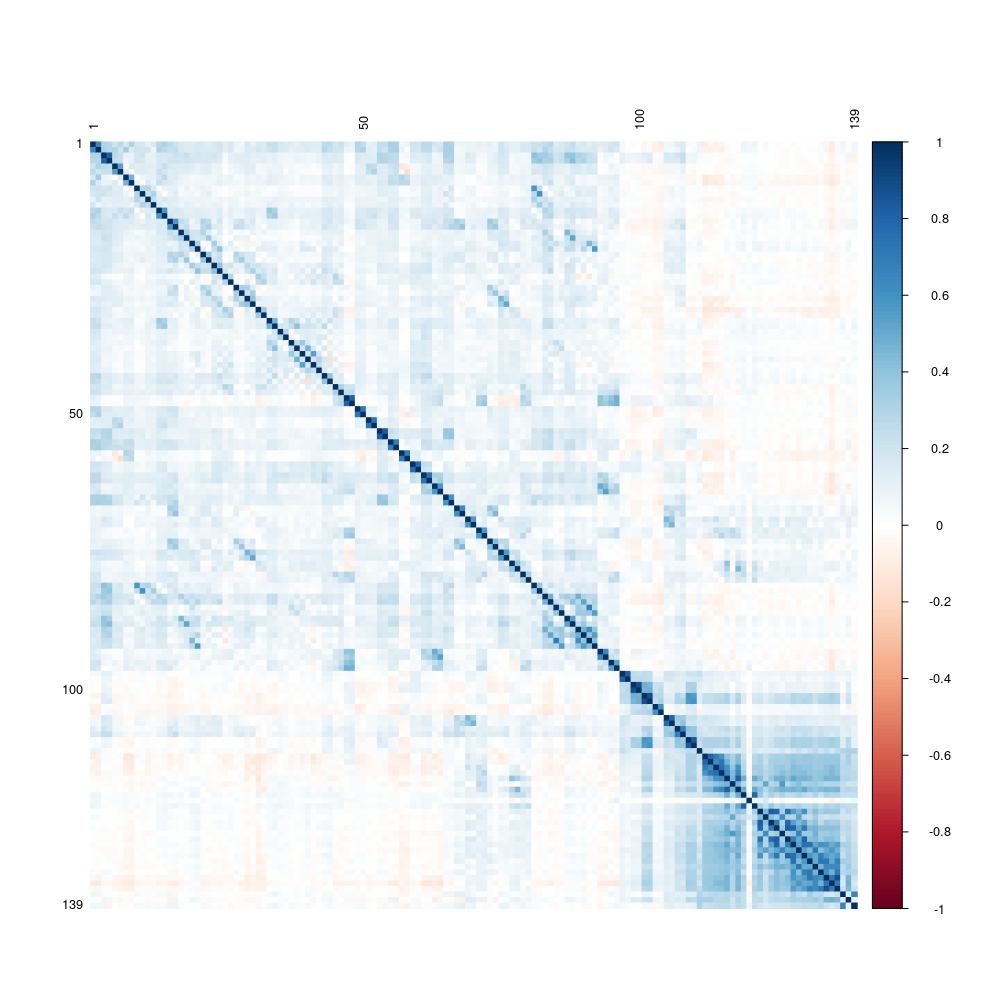}
    \caption{The LDSC estimated trait correlation matrix of T1FAST. T1FAST consists of 138 IDPs}
    \label{fig:cor_t1fast}
\end{figure}

% optimal ratio

%% Volume
\begin{figure}[htbp]
  \begin{subfigure}[t]{.4\textwidth}
    \centering
    \includegraphics[width=\linewidth]{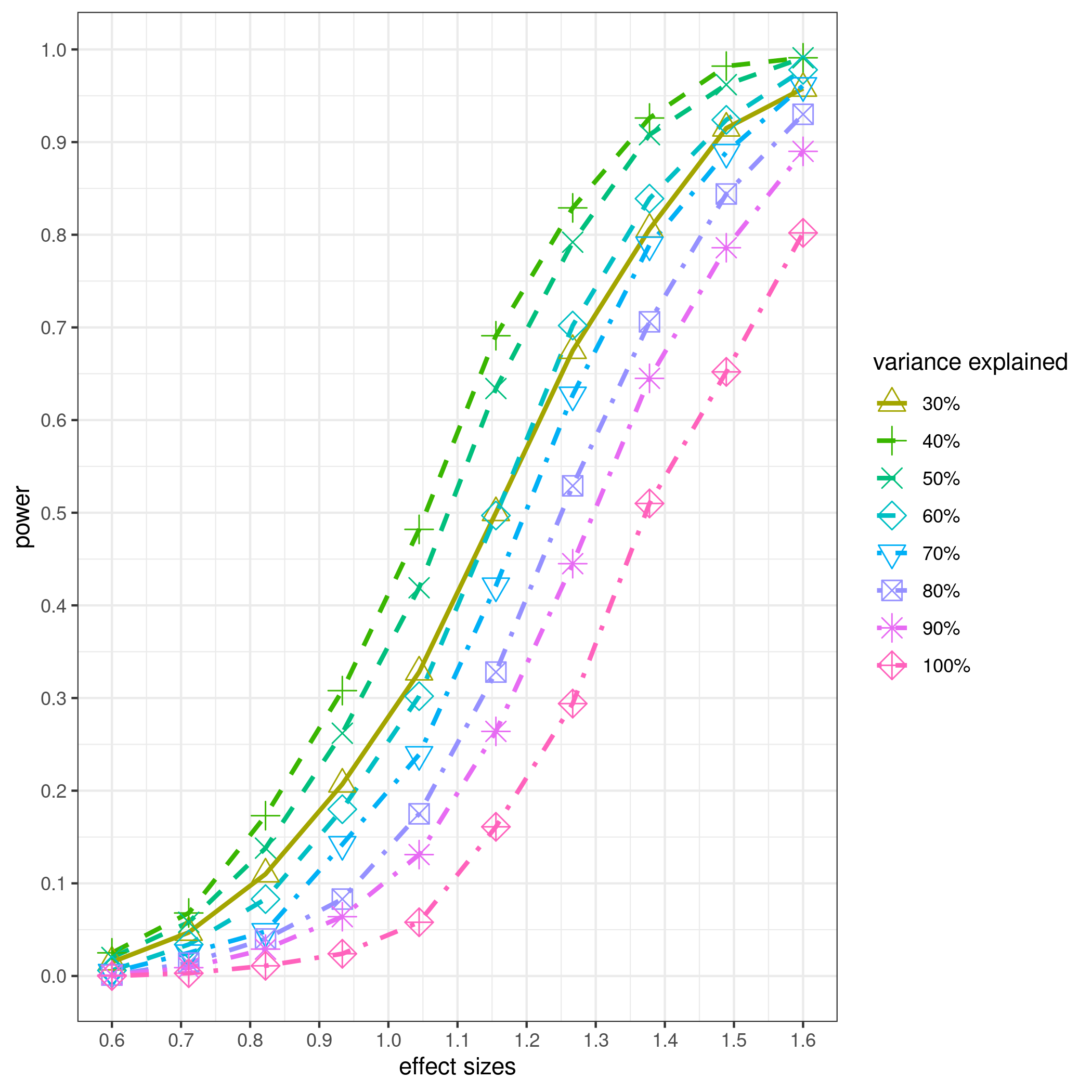}
    \caption{}
    \label{fig:s2a}
  \end{subfigure}
  \hfill
  \begin{subfigure}[t]{.4\textwidth}
    \centering
    \includegraphics[width=\linewidth]{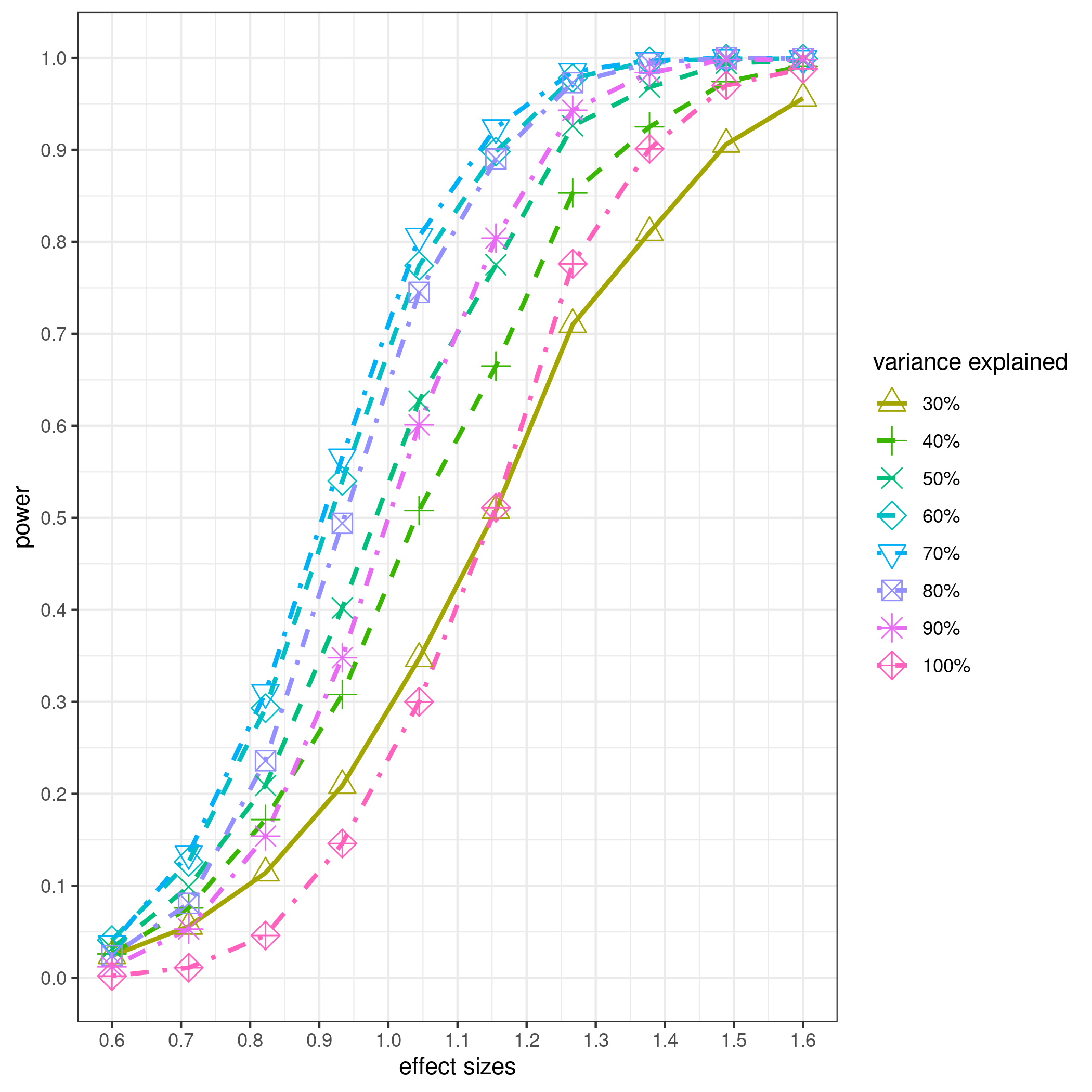}
    \caption{}
    \label{fig:s2b}
  \end{subfigure}

  \medskip

  \begin{subfigure}[t]{.4\textwidth}
    \centering
    \includegraphics[width=\linewidth]{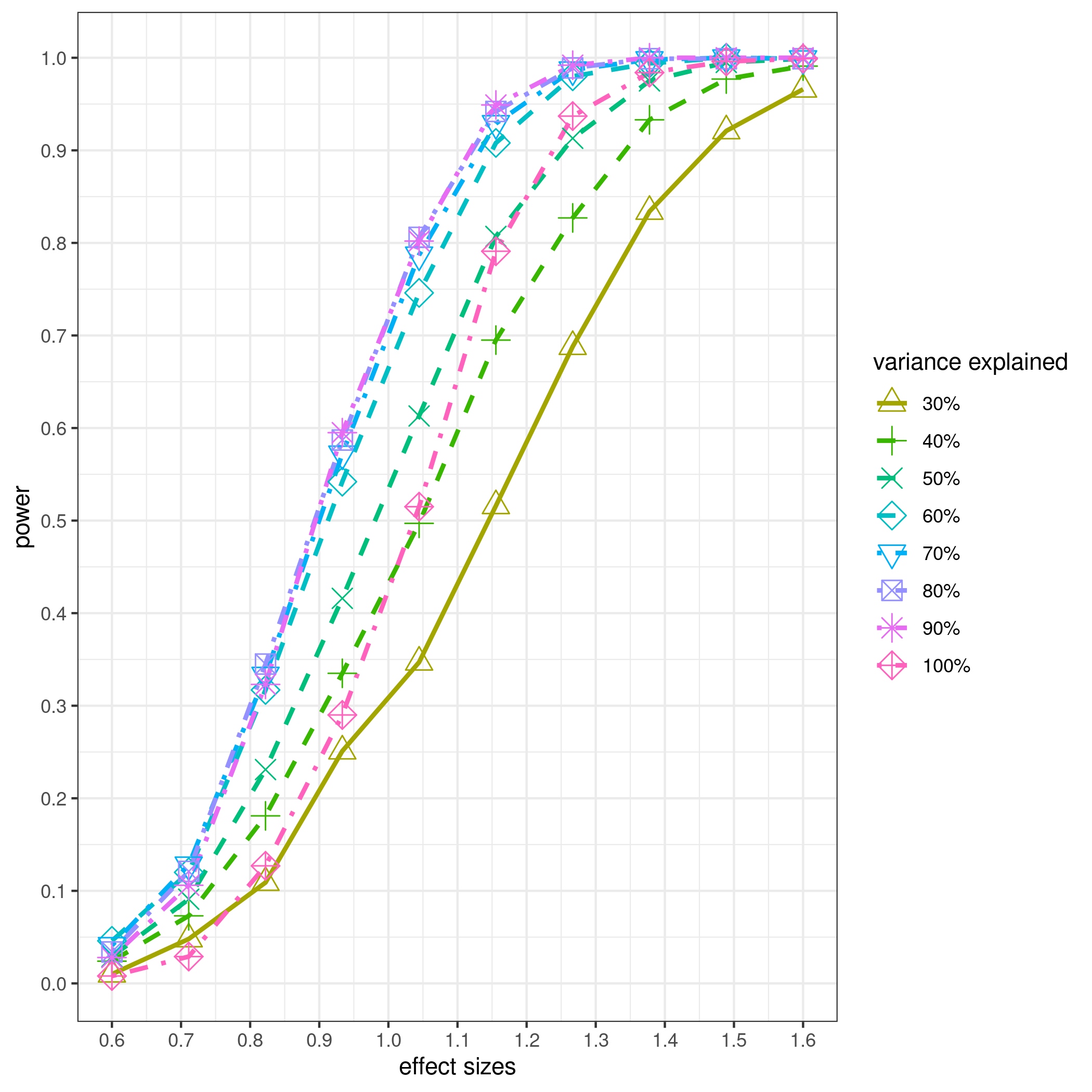}
    \caption{}
    \label{fig:s2c}
  \end{subfigure}
  
  \caption{Power comparisons with UKCOR1 and M1. The first eigenvector explained at least 30\%, thus the lines for 10\% and 20\% were excluded in the plots. (a) In the sparse scenario, the model using eigenvectors which explained 40\% of variance gave the maximum power. (b) In the intermediate scenario, including eigenvectors which explained 70\% of variance gave the maximum power. (c) In the dense scenario, including eigenvectors corresponding to 80\% or 90\% of variance gave comparable power.}
  \label{fig:s2}
\end{figure}

%% T1FAST

\begin{figure}
  \begin{subfigure}[t]{.4\textwidth}
    \centering
    \includegraphics[width=\linewidth]{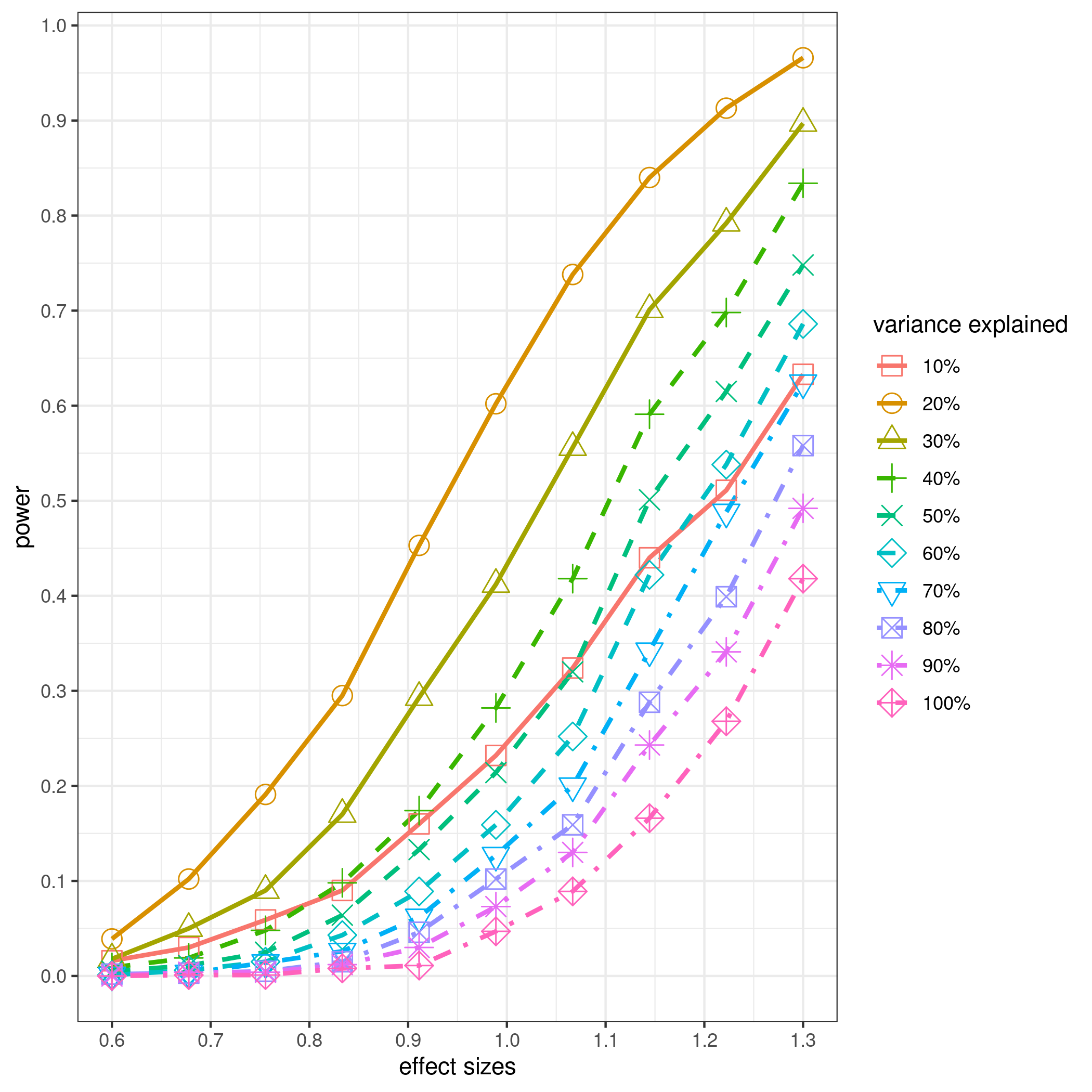}
    \caption{}
    \label{fig:s3a}
  \end{subfigure}
  \hfill
  \begin{subfigure}[t]{.4\textwidth}
    \centering
    \includegraphics[width=\linewidth]{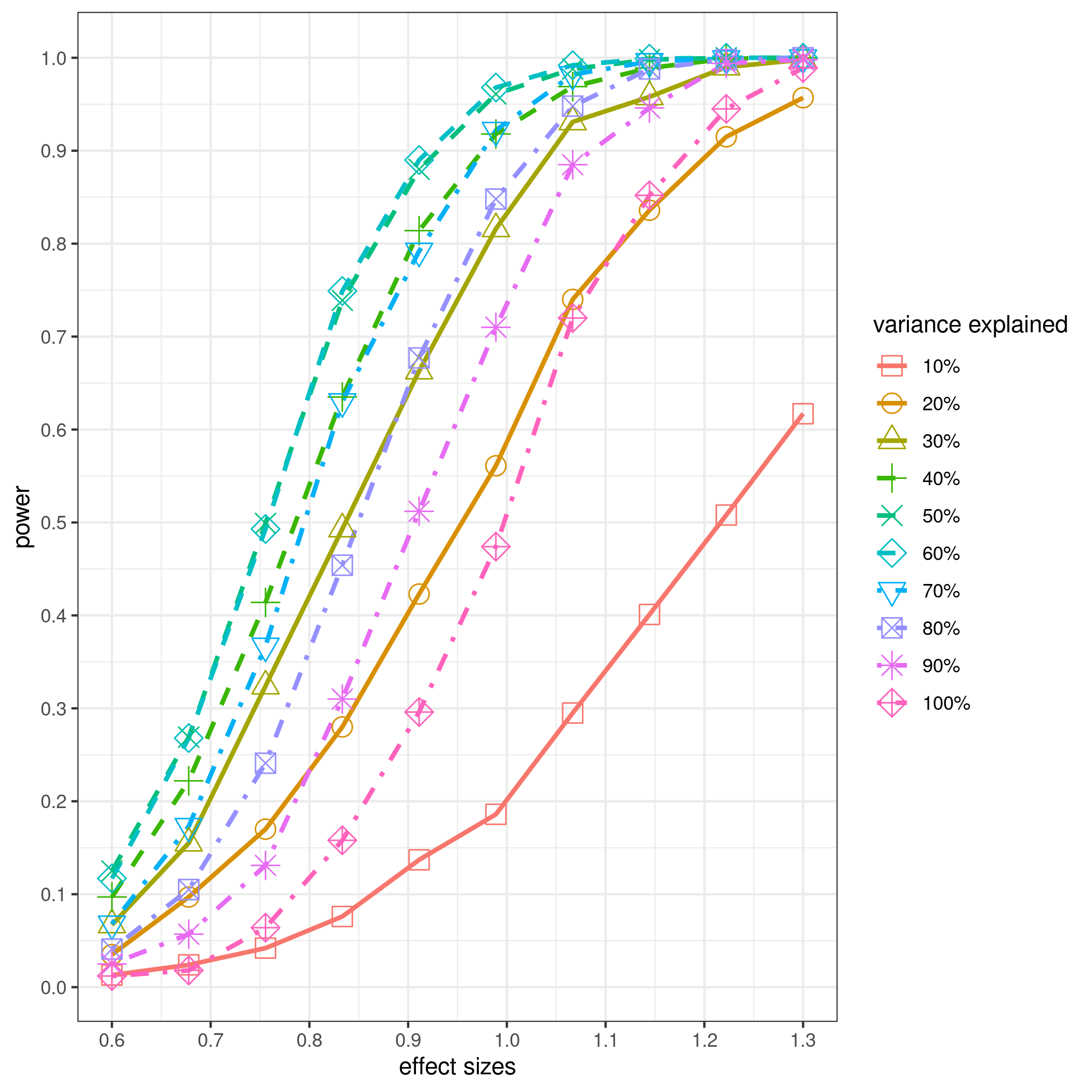}
    \caption{}
    \label{fig:s3b}
  \end{subfigure}

  \medskip

  \begin{subfigure}[t]{.4\textwidth}
    \centering
    \includegraphics[width=\linewidth]{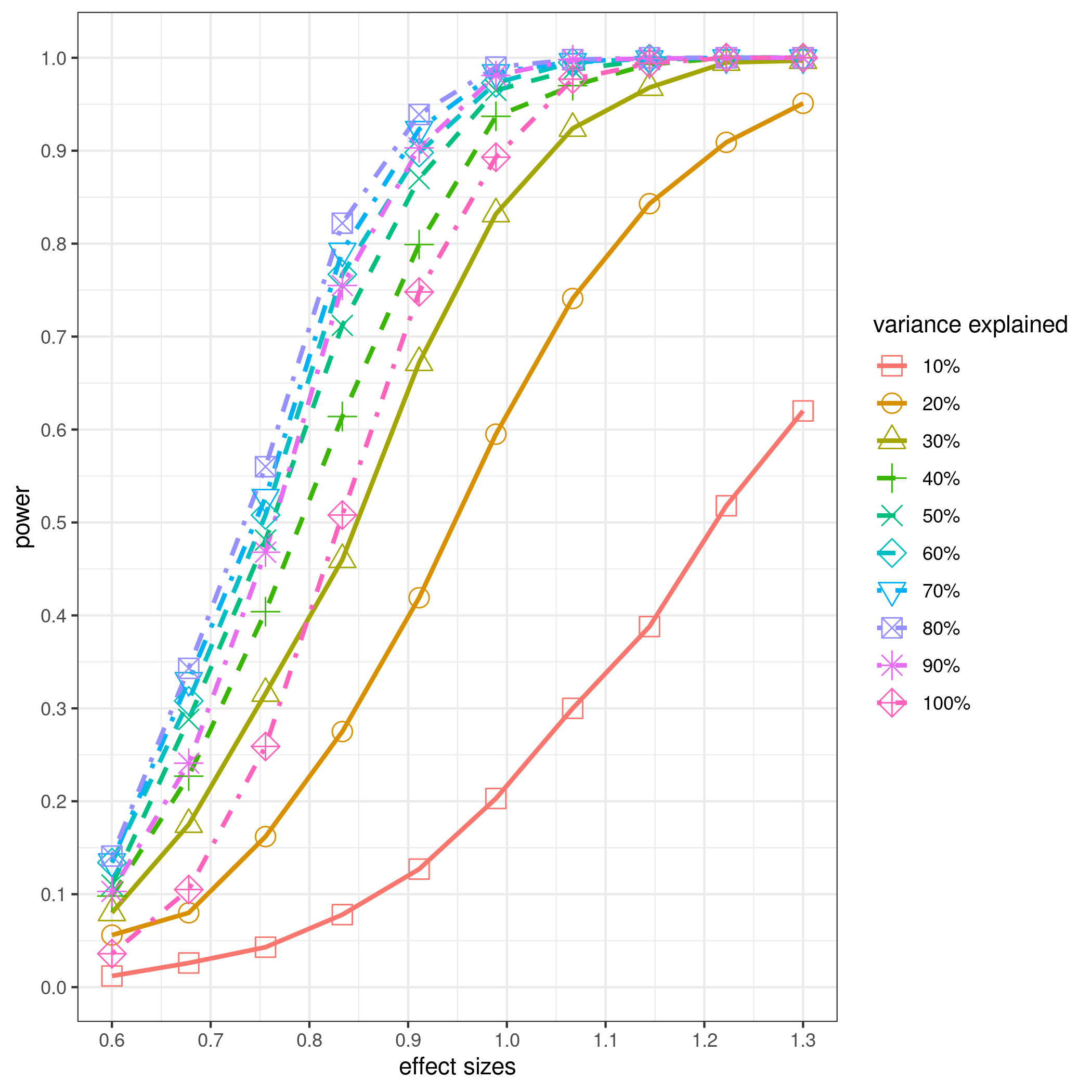}
    \caption{}
    \label{fig:s3c}
  \end{subfigure}
  
  \caption{Power comparisons with UKCOR2 and M1. (a) In the sparse scenario, the model using only the first eigenvector (10\% line) was not the most powerful. (b) In the intermediate scenario, using eigenvectors which explained 50\% or 60\% of variance gave the maximum power. (c) In the dense scenario, including eigenvectors corresponding to 80\% of variance gave the maximal power.}
  \label{fig:s3}
\end{figure}

%% CS
\begin{figure}
  \begin{subfigure}[t]{.4\textwidth}
    \centering
    \includegraphics[width=\linewidth]{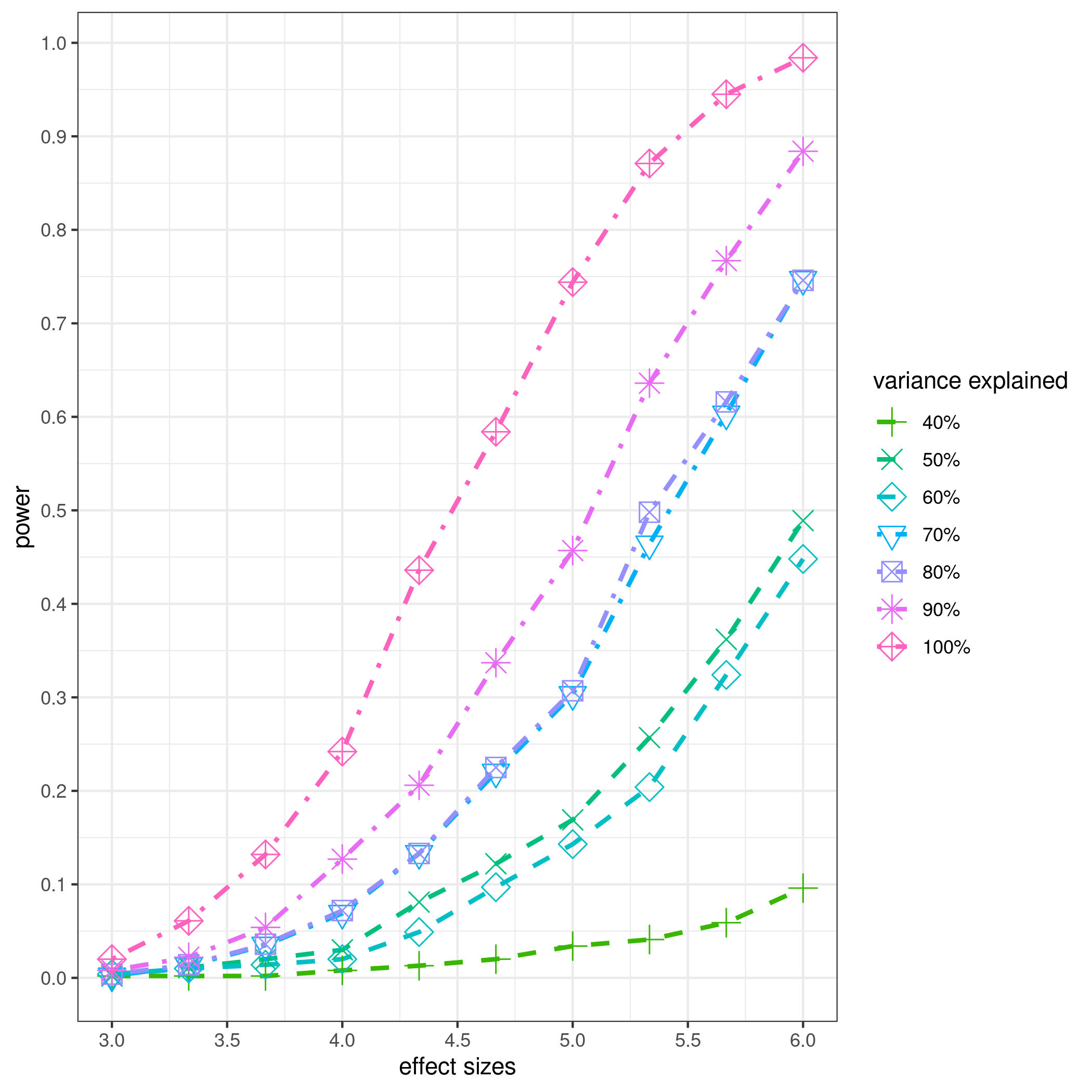}
    \caption{}
    \label{fig:s4a}
  \end{subfigure}
  \hfill
  \begin{subfigure}[t]{.4\textwidth}
    \centering
    \includegraphics[width=\linewidth]{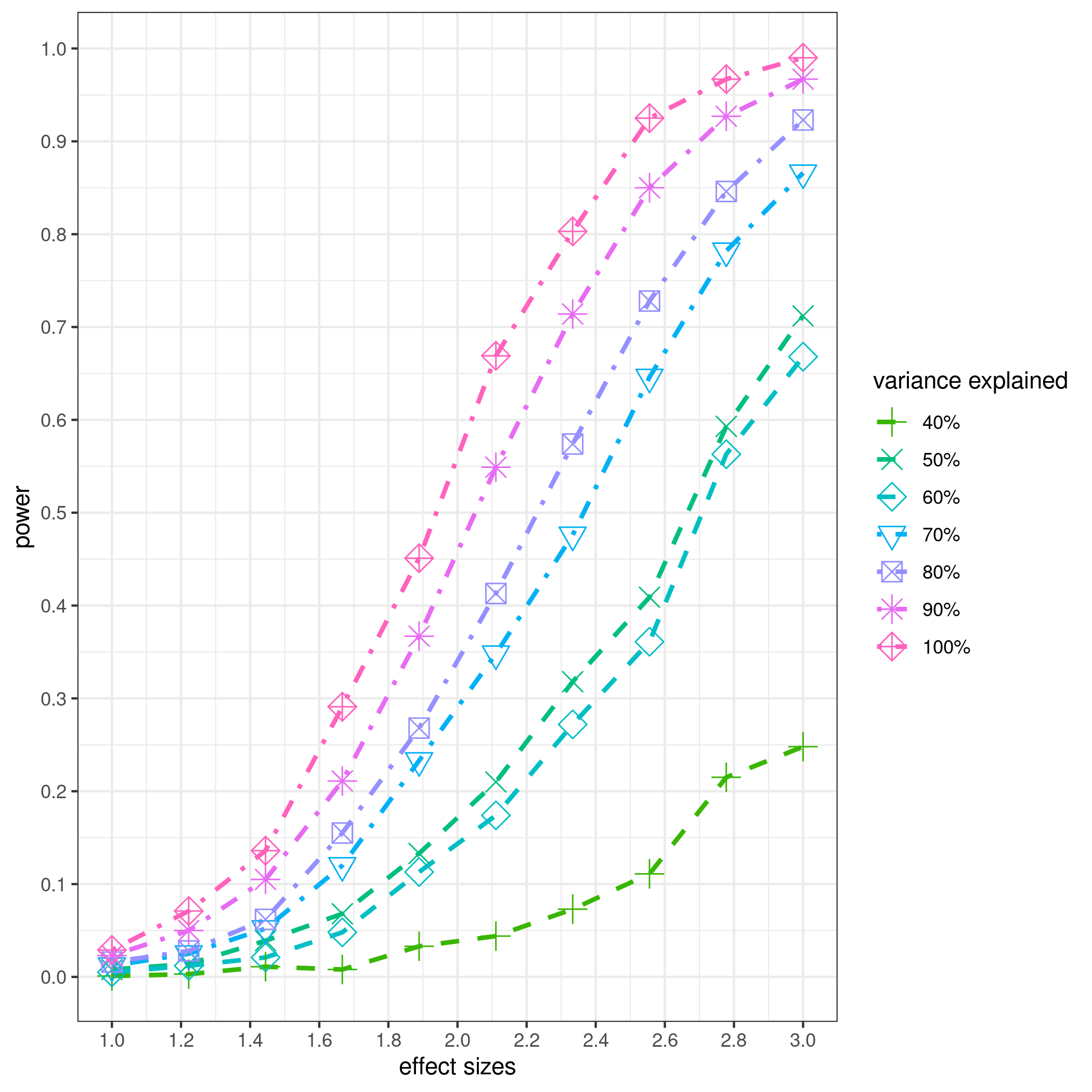}
    \caption{}
    \label{fig:s4b}
  \end{subfigure}

  \medskip

  \begin{subfigure}[t]{.4\textwidth}
    \centering
    \includegraphics[width=\linewidth]{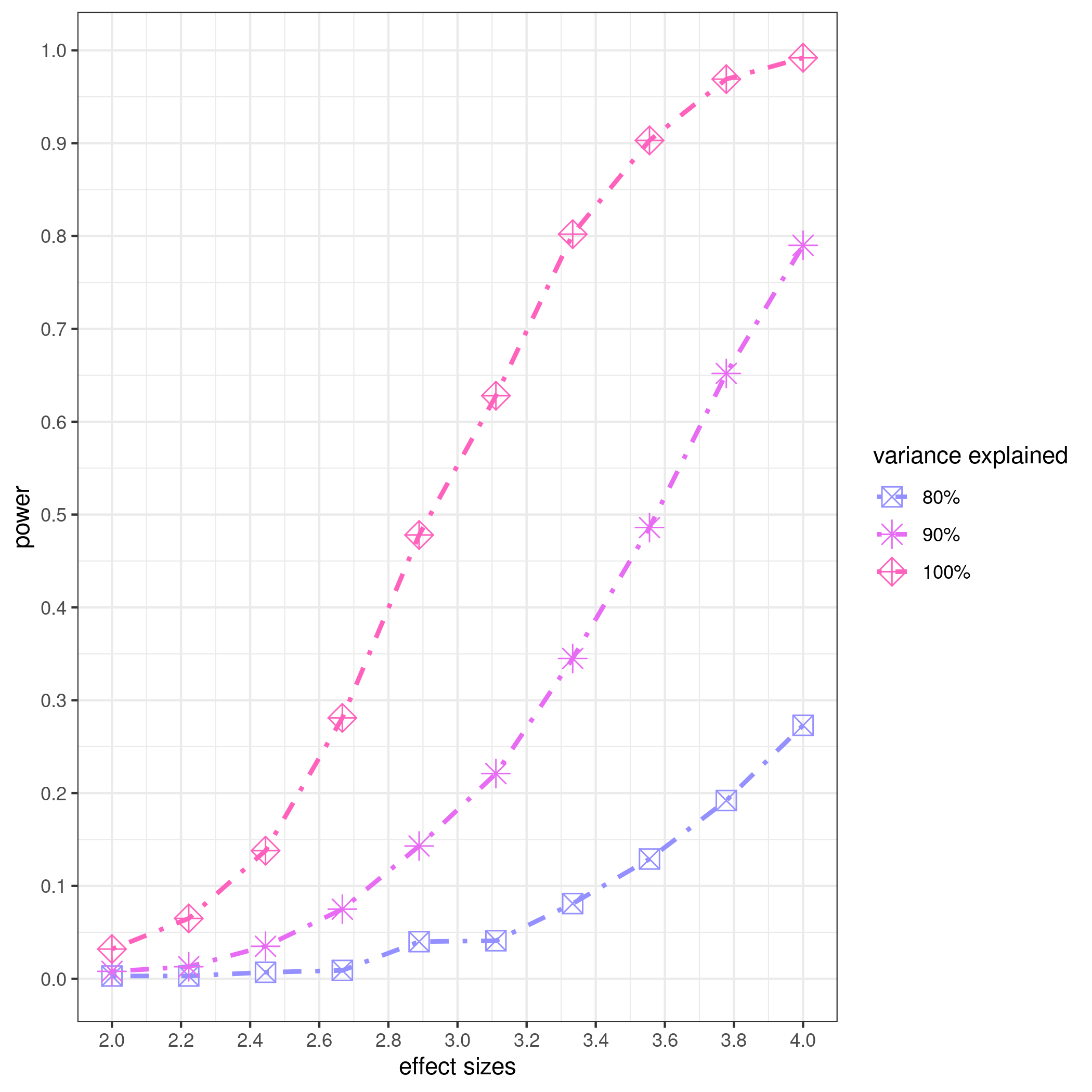}
    \caption{}
    \label{fig:s4c}
  \end{subfigure}
  \hfill
  \begin{subfigure}[t]{.4\textwidth}
    \centering
    \includegraphics[width=\linewidth]{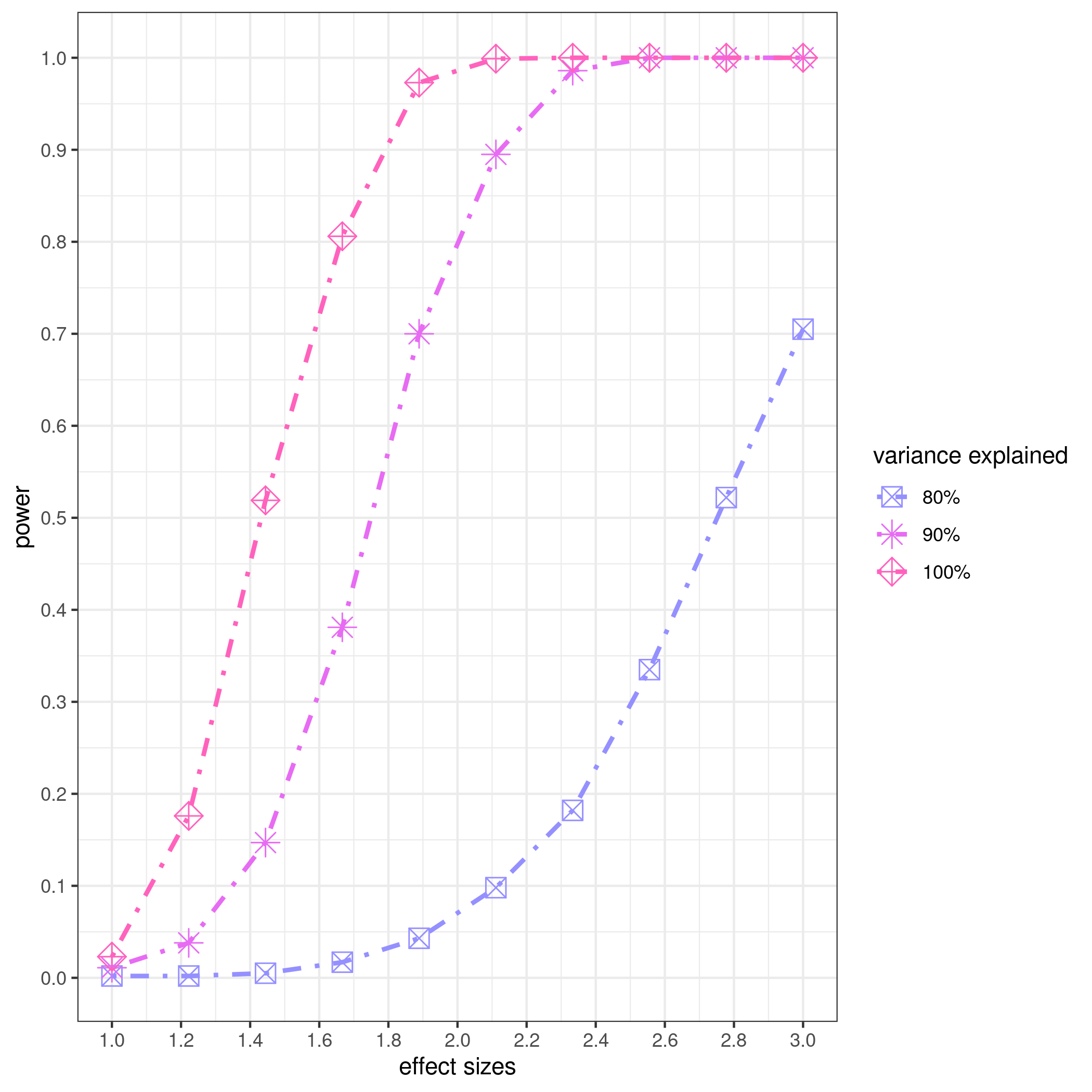}
    \caption{}
    \label{fig:s4D}
  \end{subfigure}
  \caption{Power comparisons with UKCOR2 and M1. (a) and (b) has CS(0.3), and (c) and (d) has CS(0.7). Regardless of sparse ((a) and (c)) or dense scenarios ((b) and (d)), including all eigenvectors had the maximal power.}
  \label{fig:s4}
\end{figure}

% type 1: t1fast

\begin{table}[H]
  \centering
  \begin{threeparttable}[htbp]
  \caption{Type 1 error$^a$ with UKCOR2}
  \label{tab:type1_t1fast}
\begin{tabular}{cccccc}
\hline
\multicolumn{1}{l}{} & \multicolumn{5}{c}{Significance Levels}              \\ 
Methods               & $5 \times 10^{-2}$ & $1 \times 10^{-2}$ & $1 \times 10^{-3}$ & $1 \times 10^{-4}$ & $1 \times 10^{-5}$ \\ \hline
metaMANOVA           & 1.03     & 1.03     & 1.05     & 1.014    & 0.93     \\
metaUSAT             & 1        & 1.1      & \textbf{1.22}     & \textbf{1.79}     & \textbf{1.95}     \\
SUM                  & 1        & 1        & 1.01     & 1.06     & 1.01     \\
SSU                  & 0.94     & 1.04     & \textbf{1.41}     & \textbf{1.98}     & \textbf{3.25}     \\
HOM                  & 1        & 1.01     & 1.02     & 1.04     & 1.06     \\
Cauchy               & 1.14     & 1.1      & 1.05     & 1.06     & 1.15     \\
aMAT                 & 0.97     & 0.97     & 0.99     & 1        & 1.17     \\
MTAR                 & 1.02     & 1.03     & 0.99     & 1.062    & 0.99     \\
MTAFS                & 1.19     & 1.16     & 1.1      & 1.04     & 0.95    \\ \hline
\end{tabular}
  \begin{tablenotes}
    \item[a] Values are ratios of empirical Type I errors divided by the corresponding significance levels. Inflated values are bold.

  \end{tablenotes}
 \end{threeparttable}
\end{table}

% type 1: cs0.3, 50

\begin{table}[htbp]
  \centering
  \begin{threeparttable}[htbp]
  \caption{Type 1 error$^a$ with correlation matrix CS(0.3) and 50 traits}
  \label{tab:type1_cs03}
\begin{tabular}{cccccc}
\hline
               & \multicolumn{5}{c}{Significance Levels}              \\ 
Methods & $5 \times 10^{-2}$ & $1 \times 10^{-2}$ & $1 \times 10^{-3}$ & $1 \times 10^{-4}$ & $1 \times 10^{-5}$ \\ \hline
metaMANOVA     & 1.01     & 1.01     & 1.01     & 1.02     & 1.17     \\
metaUSAT       & 1        & 1.06     & 1.09     & \textbf{1.48}    & \textbf{1.66}     \\
SUM            & 1        & 1        & 1        & 1        & 1.16     \\
SSU            & 0.96     & 1.01     & 1.13     & \textbf{1.3}      & \textbf{1.76}     \\
HOM            & 1        & 1        & 1        & 1        & 1.16     \\
Cauchy         & 1.27     & 1.23     & 1.1      & 1.02     & 1.02     \\
aMAT           & 0.99     & 1        & 0.99     & 1.05     & 1.2      \\
MTAR           & 1        & 0.99     & 0.97     & 1.01     & 0.91     \\
MTAFS          & 1.05     & 0.99     & 0.91     & 0.87     & 1.03    \\ \hline
\end{tabular}
  \begin{tablenotes}
    \item[a] The values in the table are ratios of empirical Type I errors divided by the corresponding significance levels. Values larger than 1.3 are bold.

  \end{tablenotes}
 \end{threeparttable}
\end{table}

% type 1: cs0.3, 100

\begin{table}[htbp]
  \centering
  \begin{threeparttable}[htbp]
  \caption{Type 1 error$^a$ with correlation matrix CS(0.3) and 100 traits}
  \label{tab:type1_cs03_100}
\begin{tabular}{cccccc}
\hline
               & \multicolumn{5}{c}{Significance Levels}              \\ 
Methods & $5 \times 10^{-2}$ & $1 \times 10^{-2}$ & $1 \times 10^{-3}$ & $1 \times 10^{-4}$ & $1 \times 10^{-5}$ \\ \hline
metaMANOVA & 1.01          & 1.02     & 1.04     & 1.09          & 1.23          \\
metaUSAT   & 1.02          & 1.06     & 1.02     & \textbf{1.38} & \textbf{1.52} \\
SUM        & 1             & 1        & 1.01     & 1.02          & 0.97          \\
SSU        & 0.97          & 1        & 1.08     & 1.2           & \textbf{1.33} \\
HOM        & 1             & 1        & 1.02     & 1.03          & 0.96          \\
Cauchy     & \textbf{1.33} & 1.3      & 1.11     & 1.04          & 1.12          \\
aMAT       & 1             & 1.01     & 103      & 1.06          & 1.1           \\
MTAR       & 1.01          & 0.99     & 1.01     & 1.07          & 0.88          \\
MTAFS      & 0.97          & 0.89     & 0.81     & 0.75          & 0.73     \\ \hline
\end{tabular}
  \begin{tablenotes}
    \item[a] The values in the table are ratios of empirical Type I errors divided by the corresponding significance levels. Values larger than 1.3 are bold.

  \end{tablenotes}
 \end{threeparttable}
\end{table}

% type 1: cs0.7, 50

\begin{table}[htbp]
  \centering
  \begin{threeparttable}[htbp]
  \caption{Type 1 error$^a$ with correlation matrix CS(0.7) and 50 traits}
  \label{tab:type1_cs07}
\begin{tabular}{cccccc}
\hline
               & \multicolumn{5}{c}{Significance Levels}              \\ 
Methods & $5 \times 10^{-2}$ & $1 \times 10^{-2}$ & $1 \times 10^{-3}$ & $1 \times 10^{-4}$ & $1 \times 10^{-5}$ \\ \hline 
metaMANOVA     & 1.01     & 1.01     & 1.03     & 1.08     & 1.14     \\
metaUSAT       & 1.02     & 1.05     & 1.03     & 1.19     & \textbf{1.41}     \\
SUM            & 1        & 1        & 1        & 1.01     & 1.06     \\
SSU            & 1        & 1        & 1        & 1.03     & 1.06     \\
HOM            & 1        & 1        & 1        & 1.01     & 1.05     \\
Cauchy         & 1.26     & 1.28     & 1.23     & 1.18     & 1.15     \\
aMAT           & 1        & 1        & 1        & 1.01     & 1.06     \\
MTAR           & 1        & 0.98     & 1.03     & 1        & 0.86     \\
MTAFS          & 1.05     & 0.99     & 0.93     & 0.86     & 0.9     \\ \hline
\end{tabular}
  \begin{tablenotes}
    \item[a] The values in the table are ratios of empirical Type I errors divided by the corresponding significance levels. Values larger than 1.3 are bold.

  \end{tablenotes}
 \end{threeparttable}
\end{table}

% type 1: cs0.7, 100

\begin{table}[htbp]
  \centering
  \begin{threeparttable}[htbp]
  \caption{Type 1 error$^a$ with correlation matrix CS(0.7) and 100 traits}
  \label{tab:type1_cs07_100}
\begin{tabular}{cccccc}
\hline
               & \multicolumn{5}{c}{Significance Levels}              \\ 
Methods & $5 \times 10^{-2}$ & $1 \times 10^{-2}$ & $1 \times 10^{-3}$ & $1 \times 10^{-4}$ & $1 \times 10^{-5}$ \\ \hline 
metaMANOVA & 1.03     & 1.04          & 1.06     & 1.1      & 1.09          \\
metaUSAT   & 1.04     & 1.04          & 0.93     & 1.16     & \textbf{1.35} \\
SUM        & 1        & 1             & 1.01     & 0.97     & 1.05          \\
SSU        & 1        & 1             & 1.02     & 0.97     & 1.06          \\
HOM        & 1        & 1             & 1.02     & 0.97     & 1.02          \\
Cauchy     & 1.28     & \textbf{1.31} & 1.28     & 1.22     & \textbf{1.31} \\
aMAT       & 1        & 1             & 1.01     & 0.97     & 1.05          \\
MTAR       & 1        & 1             & 0.99     & 1.03     & 1.02          \\
MTAFS      & 0.97     & 0.89          & 0.81     & 0.77     & 0.81         \\ \hline
\end{tabular}
  \begin{tablenotes}
    \item[a] The values in the table are ratios of empirical Type I errors divided by the corresponding significance levels. Values larger than 1.3 are bold.

  \end{tablenotes}
 \end{threeparttable}
\end{table}

% type 1: ar0.3, 50 

\begin{table}[htbp]
  \centering
  \begin{threeparttable}[htbp]
  \caption{Type 1 error$^a$ with correlation matrix AR(0.3) and 50 traits}
  \label{tab:type1_ar03}
\begin{tabular}{cccccc}
\hline
               & \multicolumn{5}{c}{Significance Levels}              \\ 
Methods & $5 \times 10^{-2}$ & $1 \times 10^{-2}$ & $1 \times 10^{-3}$ & $1 \times 10^{-4}$ & $1 \times 10^{-5}$ \\ \hline
metaMANOVA     & 1.01     & 1.01     & 1.01     & 1.02     & 1.04     \\
metaUSAT       & 0.719    & 0.77     & 0.83     & 0.96     & 1.06     \\
SUM            & 1        & 1        & 0.99     & 0.96     & 1.18     \\
SSU            & 0.99     & 1.01     & 1.06     & 1.15     & \textbf{1.36}     \\
HOM            & 1        & 1        & 0.99     & 0.99     & 1.22    \\
Cauchy         & 1.02     & 1.01     & 1.01     & 1.06     & 0.93     \\
aMAT           & 0.98     & 0.98     & 0.97     & 0.96     & 1.01     \\
MTAR           & 1        & 1        & 1.04     & 0.99     & 1        \\
MTAFS          & 1.16     & 1.14     & 1.09     & 1.06     & 1.02    \\ \hline
\end{tabular}
  \begin{tablenotes}
    \item[a] The values in the table are ratios of empirical Type I errors divided by the corresponding significance levels. Values larger than 1.3 are bold.

  \end{tablenotes}
 \end{threeparttable}
\end{table}

% type 1: ar0.3, 100

\begin{table}[htbp]
  \centering
  \begin{threeparttable}[htbp]
  \caption{Type 1 error$^a$ with correlation matrix AR(0.3) and 100 traits}
  \label{tab:type1_ar03_100}
\begin{tabular}{cccccc}
\hline
               & \multicolumn{5}{c}{Significance Levels}              \\ 
Methods & $5 \times 10^{-2}$ & $1 \times 10^{-2}$ & $1 \times 10^{-3}$ & $1 \times 10^{-4}$ & $1 \times 10^{-5}$ \\ \hline
metaMANOVA     & 1.02     & 1.02     & 1.04     & 1        & 0.94     \\
metaUSAT       & 0.72     & 0.77     & 0.83     & 0.88    & 0.94     \\
SUM            & 1        & 1        & 1        & 0.99    & 1.06     \\
SSU            & 1        & 1        & 1.02     & 1.02     & 1.06     \\
HOM            & 1.01     & 1        & 1.02     & 0.99     & 1.08     \\
Cauchy         & 1.02     & 1.01     & 1        & 0.96     & 1.1      \\
aMAT           & 0.98     & 0.97     & 0.97     & 0.94     & 0.84     \\
MTAR           & 1        & 0.95     & 1.15     & 1.07     & 1.1      \\
MTAFS          & 1.17     & 1.13     & 1.07     & 1.03     & 1.02    \\ \hline
\end{tabular}
  \begin{tablenotes}
    \item[a] The values in the table are ratios of empirical Type I errors divided by the corresponding significance levels.

  \end{tablenotes}
 \end{threeparttable}
\end{table}

% type 1: ar0.7, 50

\begin{table}[htbp]
  \centering
  \begin{threeparttable}[htbp]
  \caption{Type 1 error$^a$ with correlation matrix AR(0.7) and 50 traits}
  \label{tab:type1_ar07}
\begin{tabular}{cccccc}
\hline
               & \multicolumn{5}{c}{Significance Levels}              \\ 
Methods & $5 \times 10^{-2}$ & $1 \times 10^{-2}$ & $1 \times 10^{-3}$ & $1 \times 10^{-4}$ & $1 \times 10^{-5}$ \\ \hline
metaMANOVA     & 1.01     & 1.01     & 1.01     & 1        & 1.08     \\
metaUSAT       & 0.93     & 1.03     & 1.14     & \textbf{1.43}     & \textbf{1.76}     \\
SUM            & 1        & 1        & 1        & 1        & 0.92     \\
SSU            & 0.98     & 1.02     & 1.15     & \textbf{1.37}     & \textbf{1.52}     \\
HOM            & 1        & 1        & 1        & 0.98     & 1.04     \\
Cauchy         & 1.1      & 1.09     & 1.05     & 1.04     & 0.98     \\
aMAT           & 0.96     & 0.96     & 0.98     & 1.03     & \textbf{1.29}     \\
MTAR           & 1        & 1.01     & 1.02     & 1.11     & 0.95     \\
MTAFS          & 1.16     & 1.14     & 1.1      & 1.06     & 1.13    \\ \hline
\end{tabular}
  \begin{tablenotes}
    \item[a] The values in the table are ratios of empirical Type I errors divided by the corresponding significance levels. Values larger than 1.3 are bold.

  \end{tablenotes}
 \end{threeparttable}
\end{table}

% type 1: ar0.7, 100

\begin{table}[htbp]
  \centering
  \begin{threeparttable}[htbp]
  \caption{Type 1 error$^a$ with correlation matrix AR(0.7) and 100 traits}
  \label{tab:type1_ar07_100}
\begin{tabular}{cccccc}
\hline
               & \multicolumn{5}{c}{Significance Levels}              \\ 
Methods & $5 \times 10^{-2}$ & $1 \times 10^{-2}$ & $1 \times 10^{-3}$ & $1 \times 10^{-4}$ & $1 \times 10^{-5}$ \\ \hline
metaMANOVA     & 1.02     & 1.03     & 1.05     & 1.08          & 1.17          \\
metaUSAT       & 0.95     & 1.05     & 1.17     & \textbf{1.46} & \textbf{1.73} \\
SUM            & 1        & 1        & 0.98     & 0.95          & 1.02          \\
SSU            & 0.99     & 1.02     & 1.08     & 1.29          & \textbf{1.6}  \\
HOM            & 1        & 1        & 0.99     & 0.99          & 1.03          \\
Cauchy         & 1.09     & 1.07     & 1.04     & 1.03          & 1.18          \\
aMAT           & 0.98     & 0.98     & 0.99     & 1.05          & 1.19          \\
MTAR           & 1        & 0.98     & 0.97     & 1.02          & 0.94          \\
MTAFS          & 1.18     & 1.15     & 1.1      & 1.1           & 0.94         \\ \hline
\end{tabular}
  \begin{tablenotes}
    \item[a] The values in the table are ratios of empirical Type I errors divided by the corresponding significance levels. Values larger than 1.3 are bold.

  \end{tablenotes}
 \end{threeparttable}
\end{table}

% type 1: t dist
\iffalse
\begin{table}
  \centering
  \begin{threeparttable}[b]
  \caption{Type 1 error with the estimated Volume trait correlation matrix and the t distribution}
  \label{tab:s2}
  \begin{tabular}{cccccc}
  \hline
  method & $5 \times 10^{-2}$ & $1 \times 10^{-2}$ & $1 \times 10^{-3}$ & $1 \times 10^{-4}$ & $1 \times 10^{-5}$ \\
  \hline
  metaMANOVA & 5 & 16.7 & 101.9 & 660.3 & 4455.9 \\
  metaUSAT & 4.6 & 15.1 & 92 & 633.7 & 4328.2 \\
  SUM & 1.3 & 1.8 & 3.7 & 9.1 & 26.8 \\
  SSU & 2 & 3.6 & 9.8 & 32.1 & 118.8 \\
  HOM & 1.3 & 1.8 & 3.6 & 8.9 & 26.3 \\
  Cauchy & 2.7 & 5.2 & 14.5 & 47 & 169.4 \\
  aMAT & 3.4 & 9.4 & 47 & 262.1 & 1576.9 \\
  MTAR & 6.2 & 23.6 & 174.2 & 1313.3 & 10378.4 \\
  MTAFS & 4.4 & 13.6 & 79.1 & 506.9 & 3424.5 \\
  
  \hline
  \end{tabular}
  \begin{tablenotes}
    \item Notes:  The values in the table are ratios of empirical Type I errors divided by the significance levels. Values larger than 1.3 are bold.

  \end{tablenotes}
 \end{threeparttable}
\end{table}
\fi

%% power: t1fast ep1
\begin{figure}
  \begin{subfigure}[t]{.3\textwidth}
    \centering
    \includegraphics[width=\textwidth]{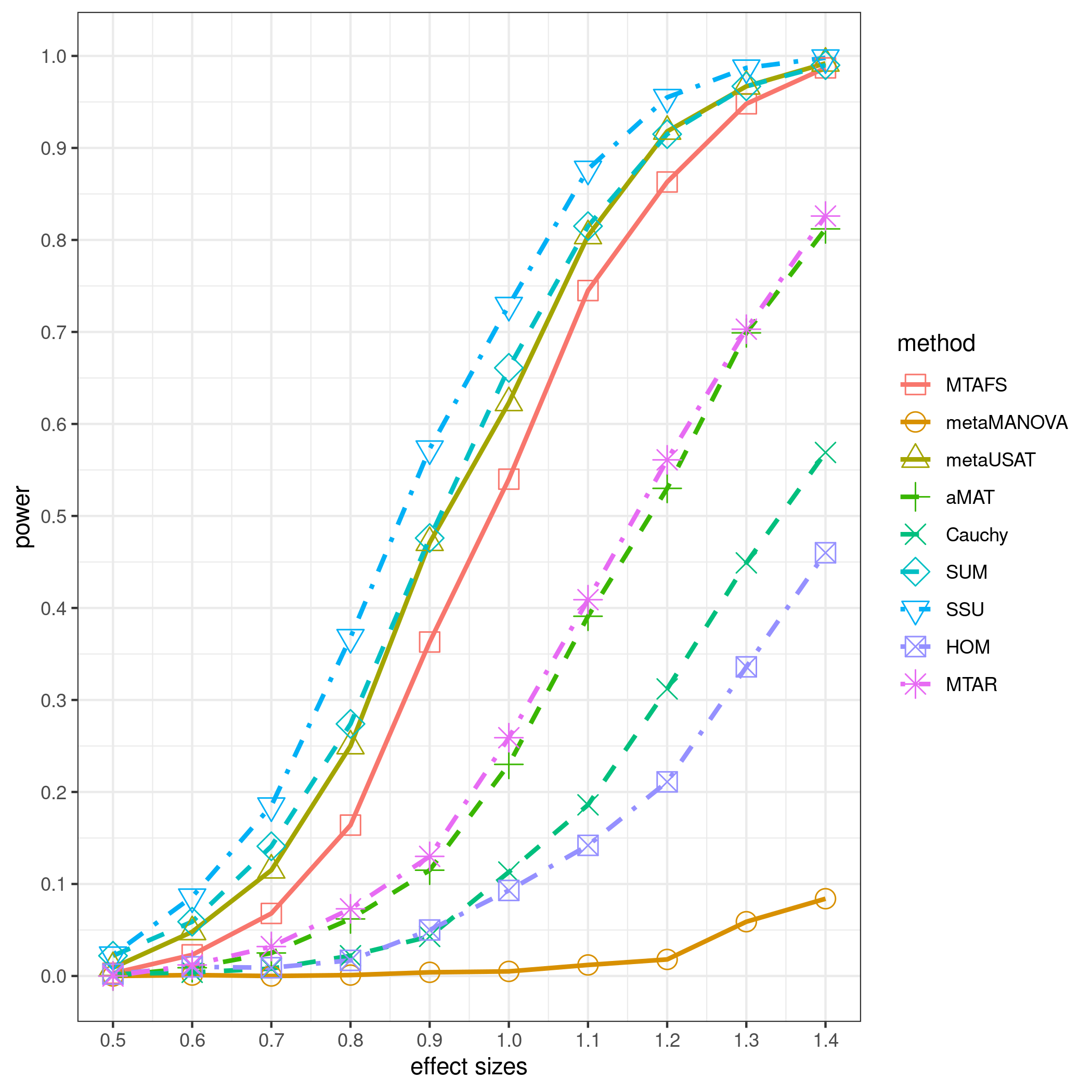}
    \caption{}
    \label{fig:s5a}
  \end{subfigure}
  \hfill
  \begin{subfigure}[t]{.3\textwidth}
    \centering
    \includegraphics[width=\textwidth]{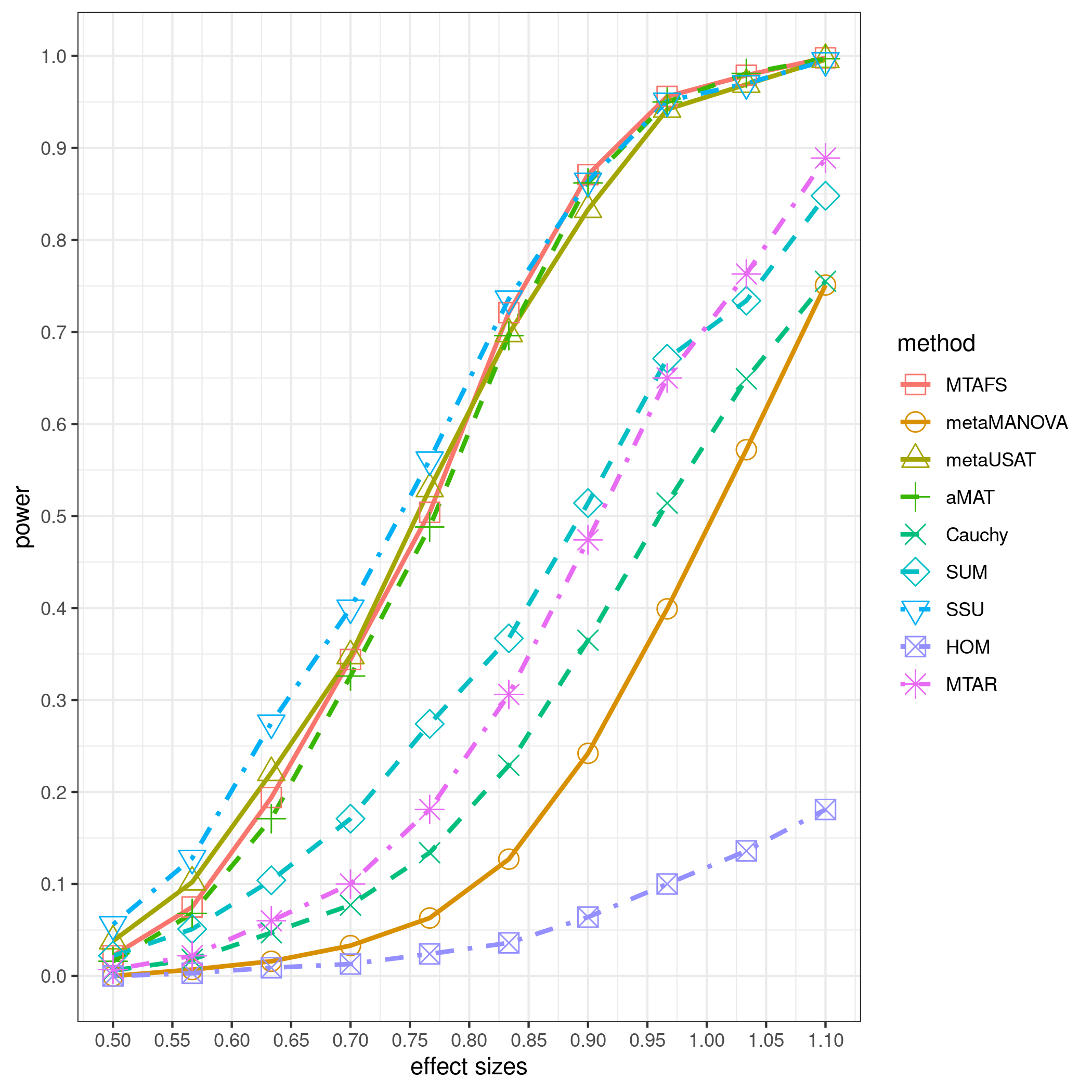}
    \caption{}
    \label{fig:s5b}
  \end{subfigure}
    \hfill
  \begin{subfigure}[t]{.3\textwidth}
    \centering
    \includegraphics[width=\textwidth]{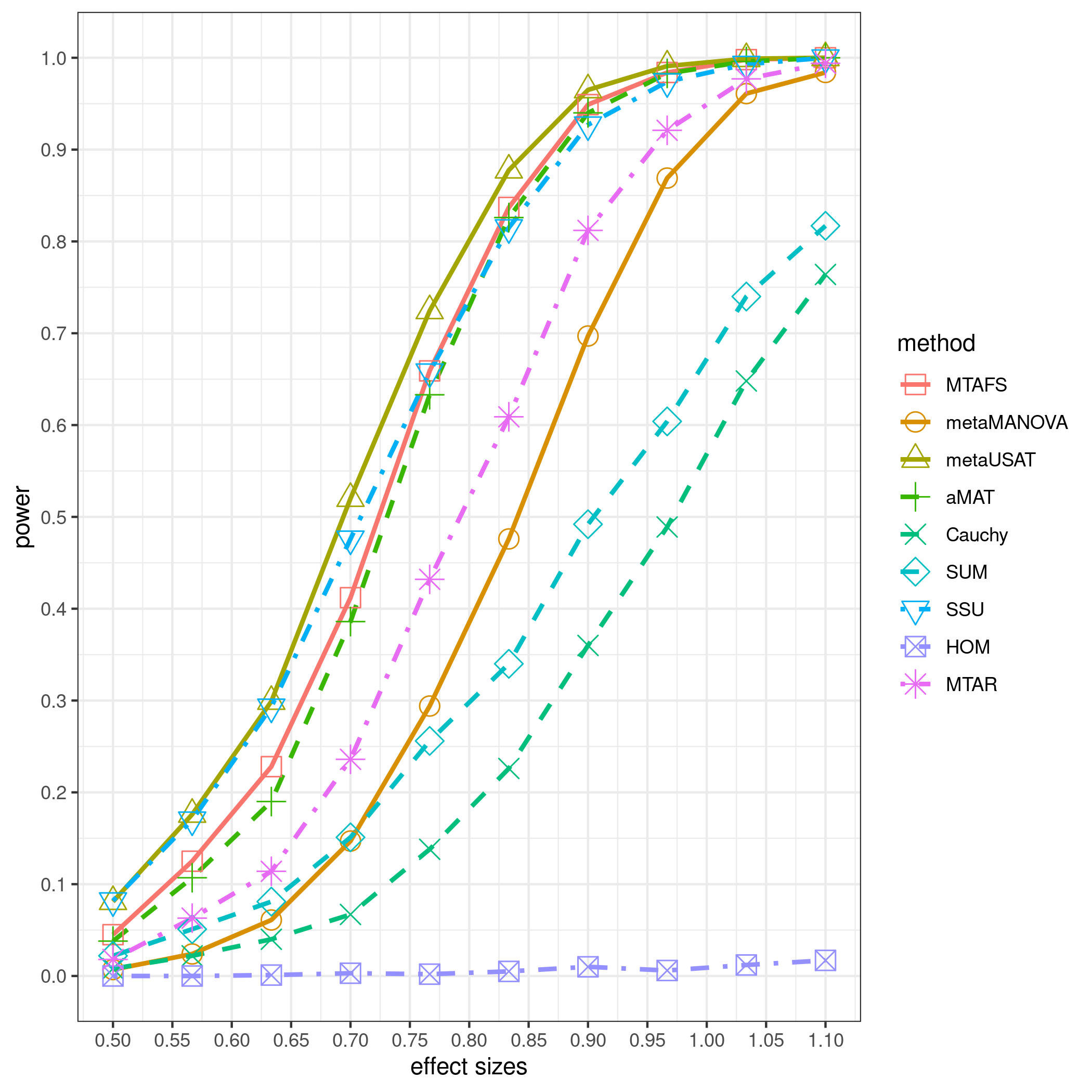}
    \caption{}
    \label{fig:s5c}
  \end{subfigure}

  \medskip

  \begin{subfigure}[t]{.3\textwidth}
    \centering
    \includegraphics[width=\textwidth]{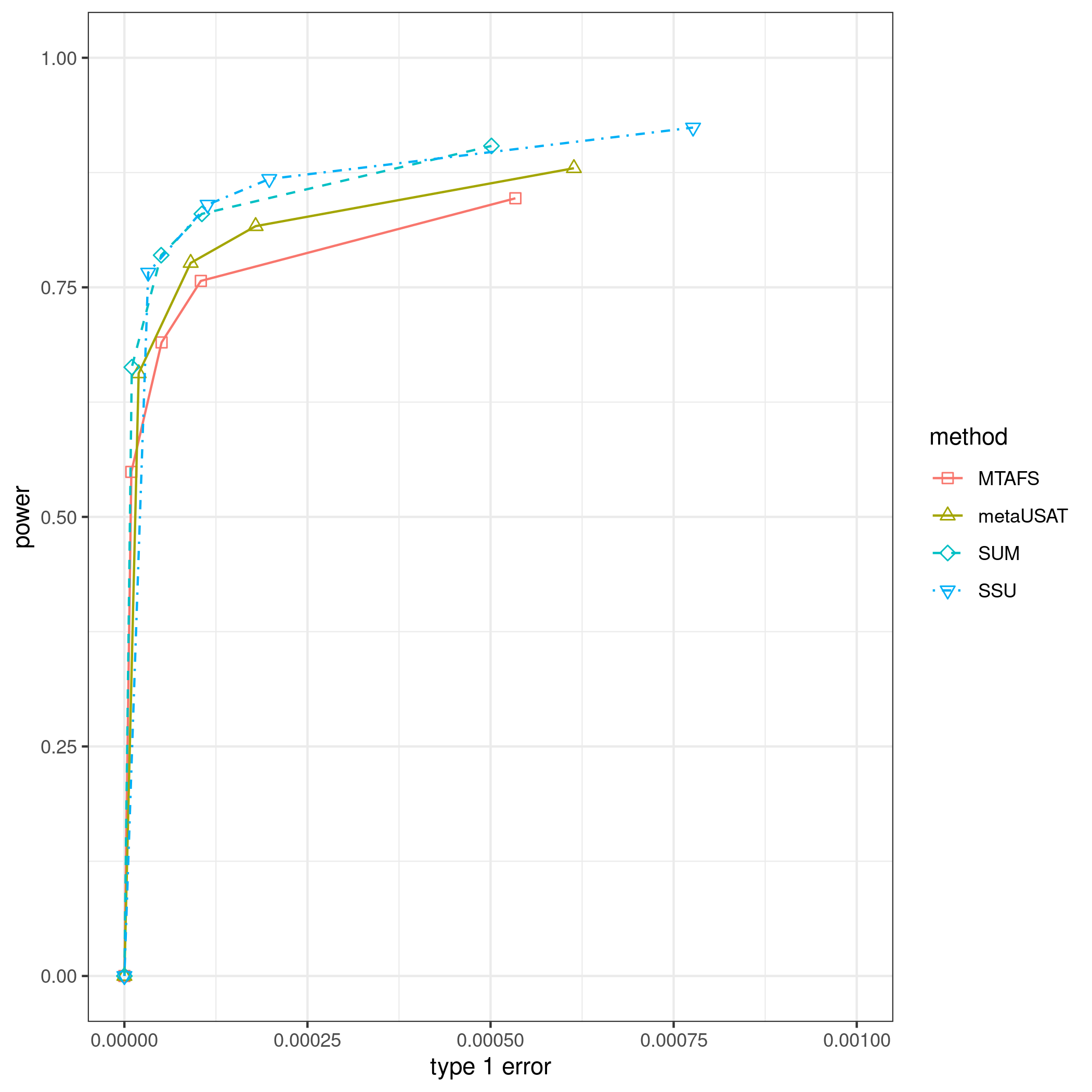}
    \caption{}
    \label{fig:s5d}
  \end{subfigure}
  \hfill
  \begin{subfigure}[t]{.3\textwidth}
    \centering
    \includegraphics[width=\textwidth]{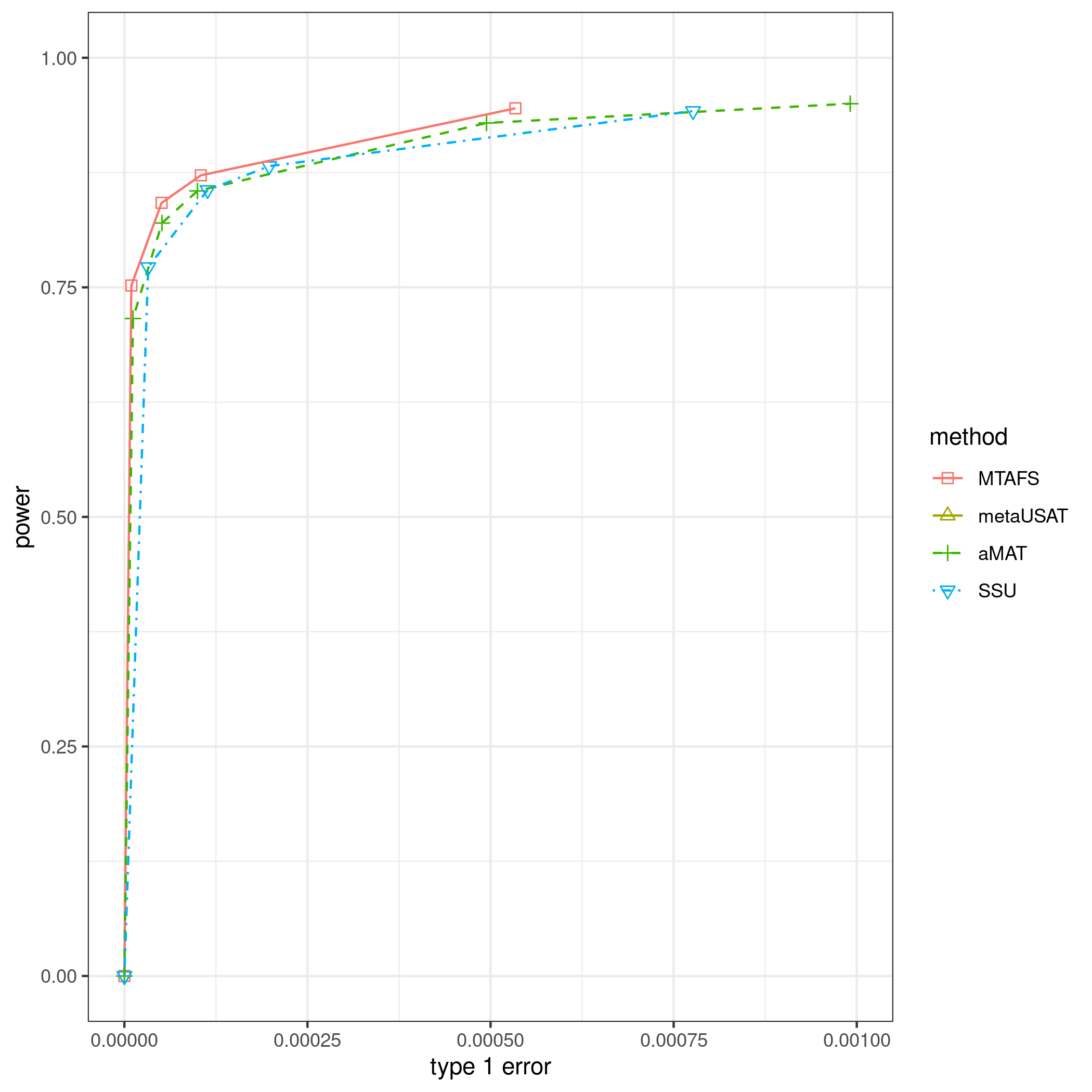}
    \caption{}
    \label{fig:s5e}
  \end{subfigure}
    \hfill
  \begin{subfigure}[t]{.3\textwidth}
    \centering
    \includegraphics[width=\textwidth]{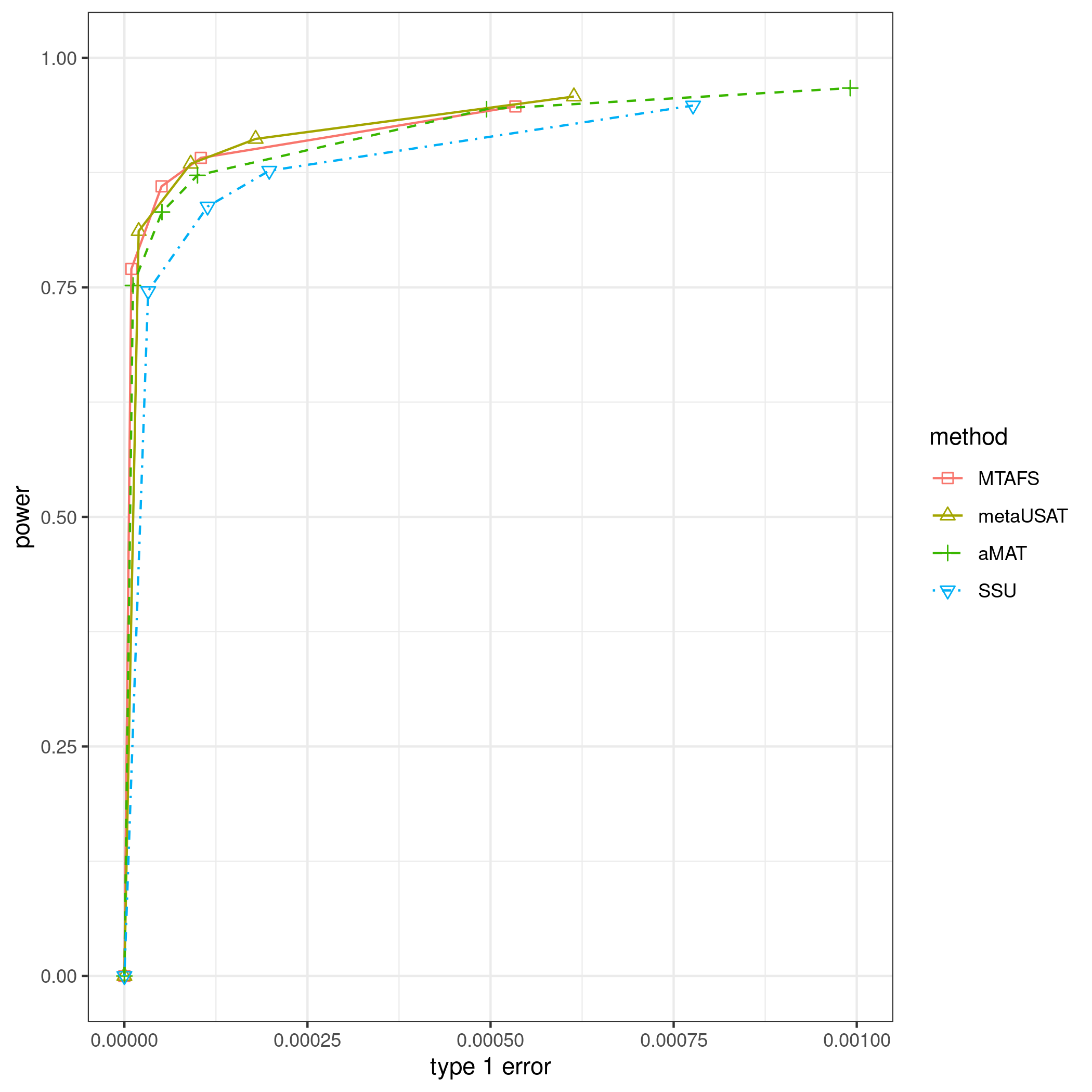}
    \caption{}
    \label{fig:s5f}
  \end{subfigure}

  \caption{Comparison of methods for model M1 using the UKCOR2 correlation matrix. (a) high sparsity, with only top 2 eigenvectors informative; (b) intermediate sparsity, with top 27 eigenvectors informative; (c) low sparsity, with top 69 eigenvectors informative; (d) partial ROC curves for the four best methods with comparable power in (a); (e) partial ROC curves for the four best methods with comparable power in (b); (f) partial ROC curves for the four best methods with comparable power in (c).}
  \label{fig:power_t1fast_m1}
\end{figure}

%% power: t1fast ep2
\begin{figure}
  \begin{subfigure}[t]{.3\textwidth}
    \centering
    \includegraphics[width=\textwidth]{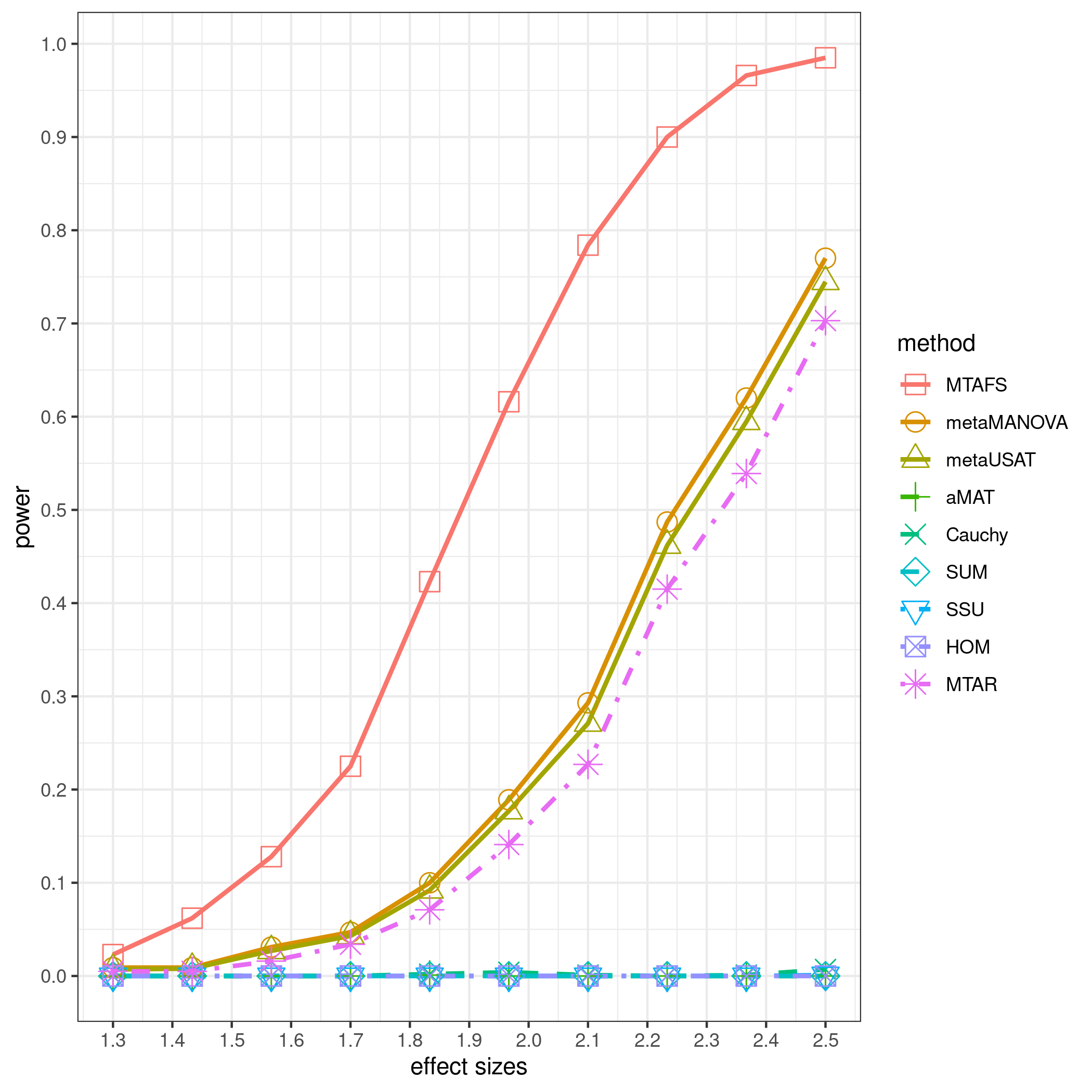}
    \caption{}
    \label{fig:s6a}
  \end{subfigure}
  \hfill
  \begin{subfigure}[t]{.3\textwidth}
    \centering
    \includegraphics[width=\textwidth]{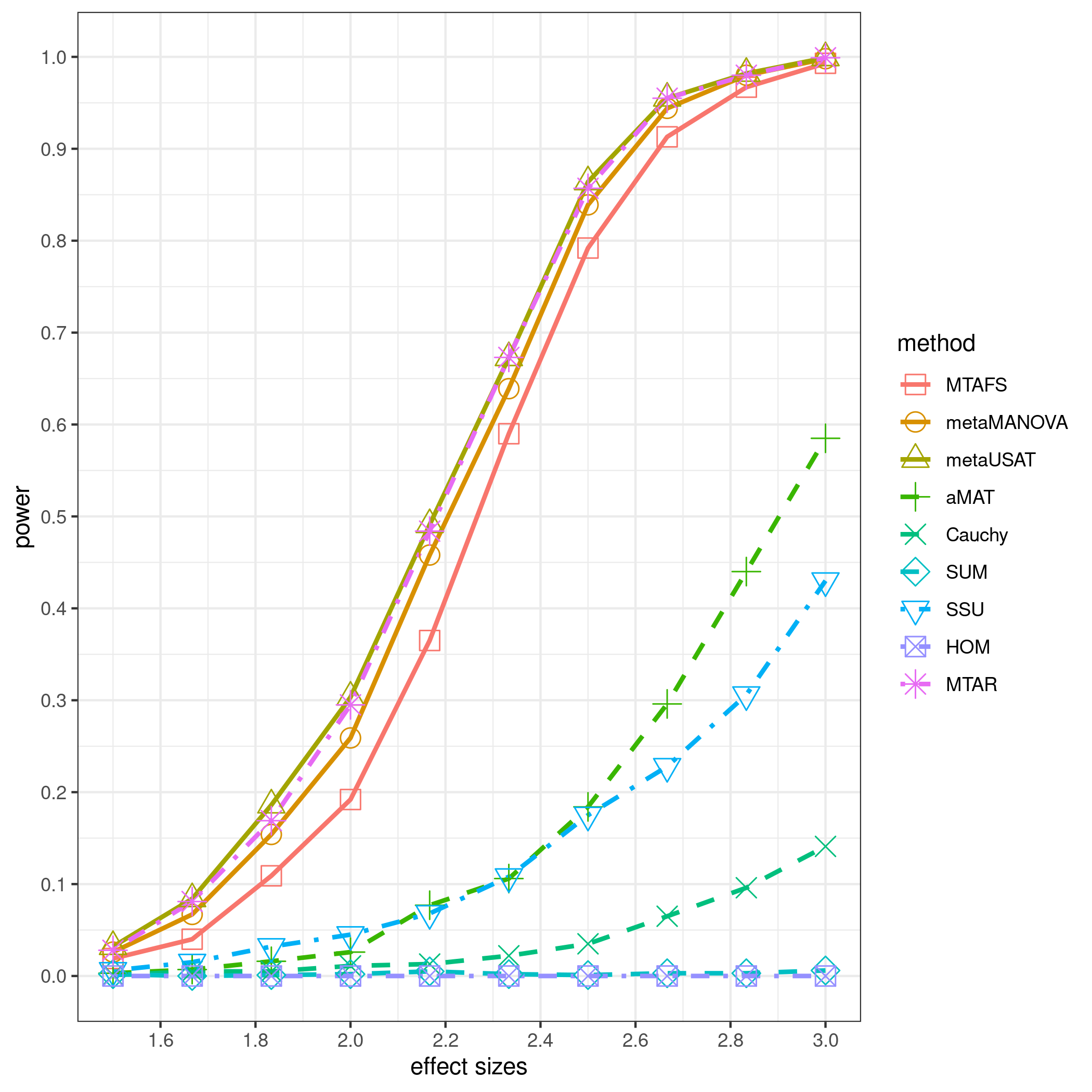}
    \caption{}
    \label{fig:s6b}
  \end{subfigure}
    \hfill
  \begin{subfigure}[t]{.3\textwidth}
    \centering
    \includegraphics[width=\textwidth]{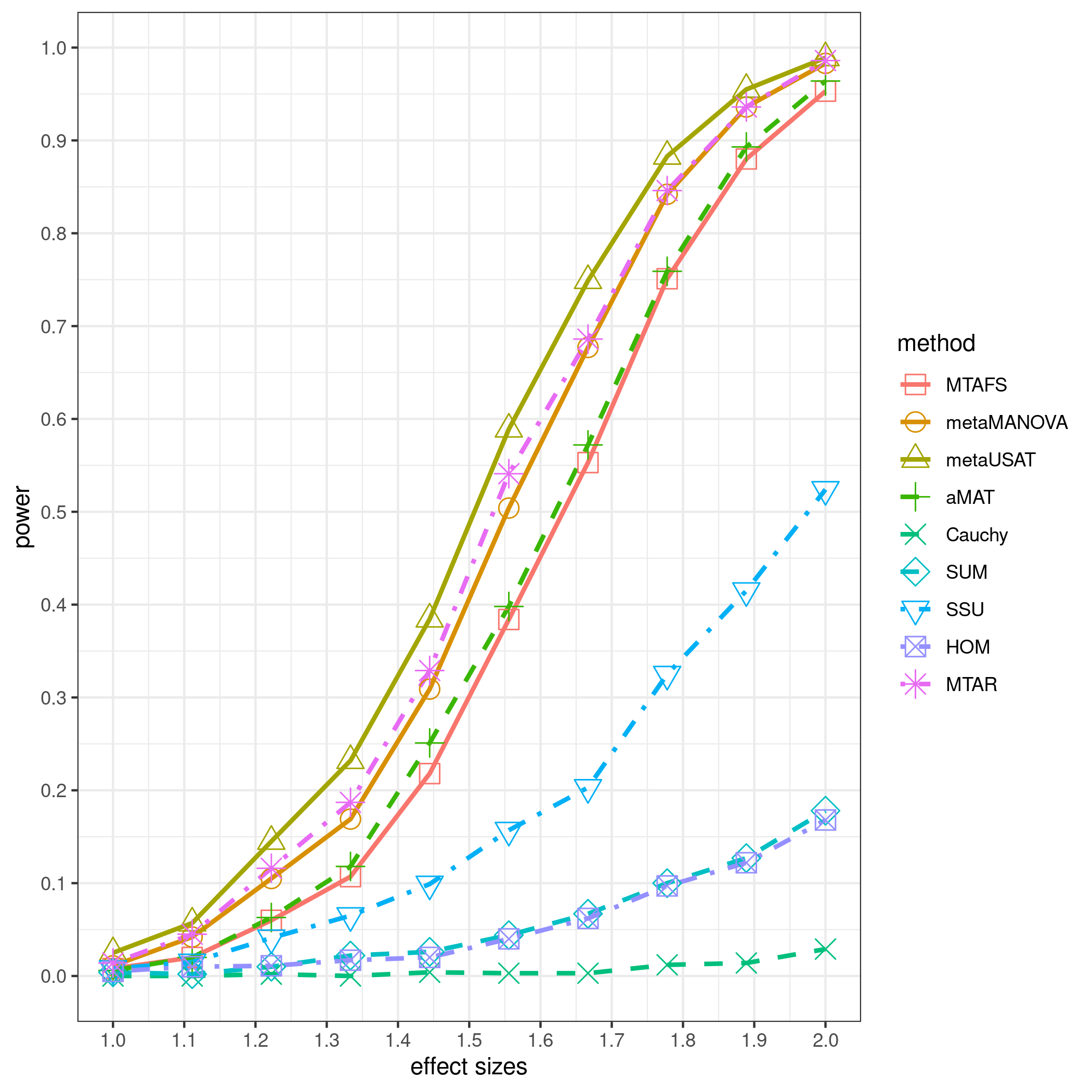}
    \caption{}
    \label{fig:s6c}
  \end{subfigure}

  \medskip

  \begin{subfigure}[t]{.3\textwidth}
    \centering
    \includegraphics[width=\textwidth]{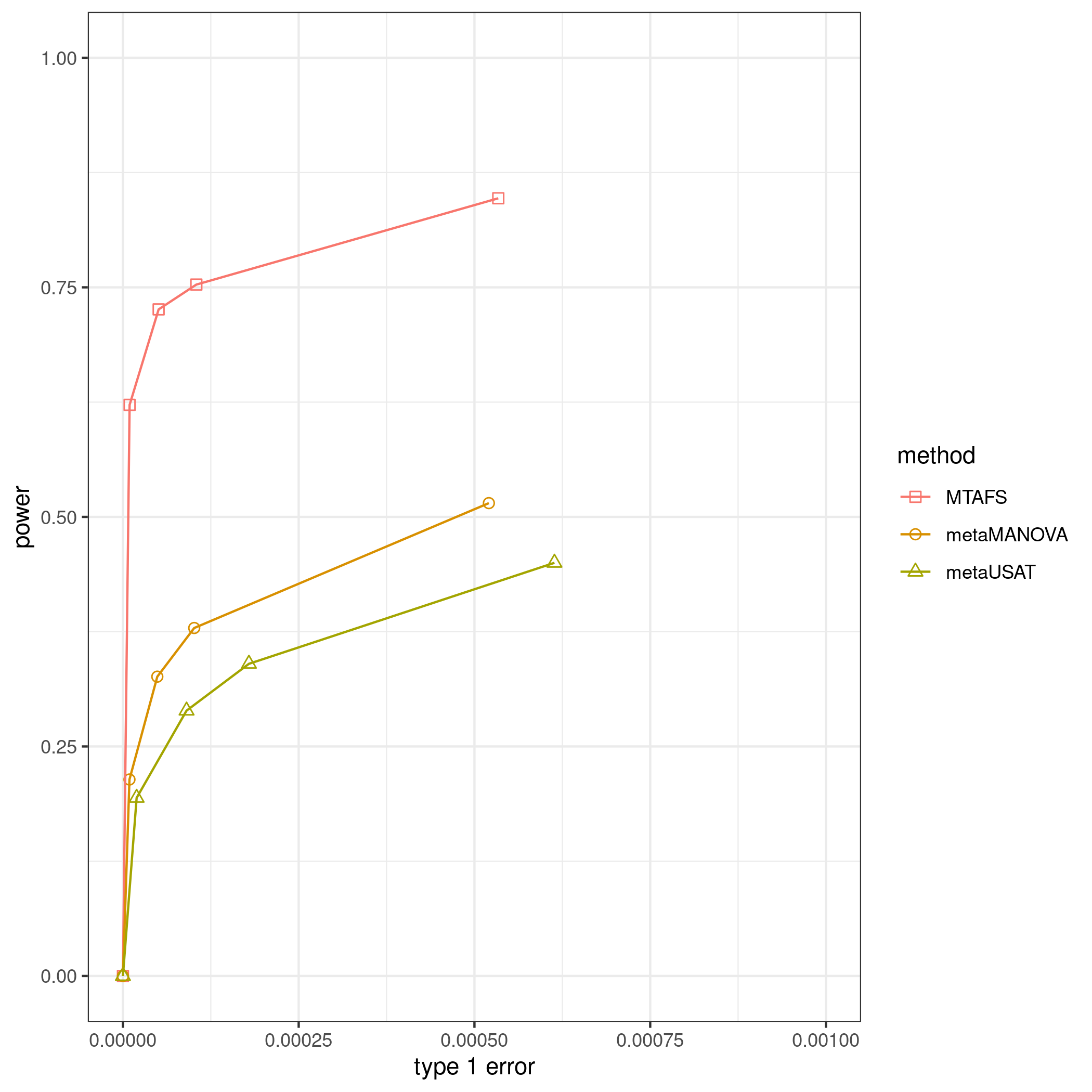}
    \caption{}
    \label{fig:s6d}
  \end{subfigure}
  \hfill
  \begin{subfigure}[t]{.3\textwidth}
    \centering
    \includegraphics[width=\textwidth]{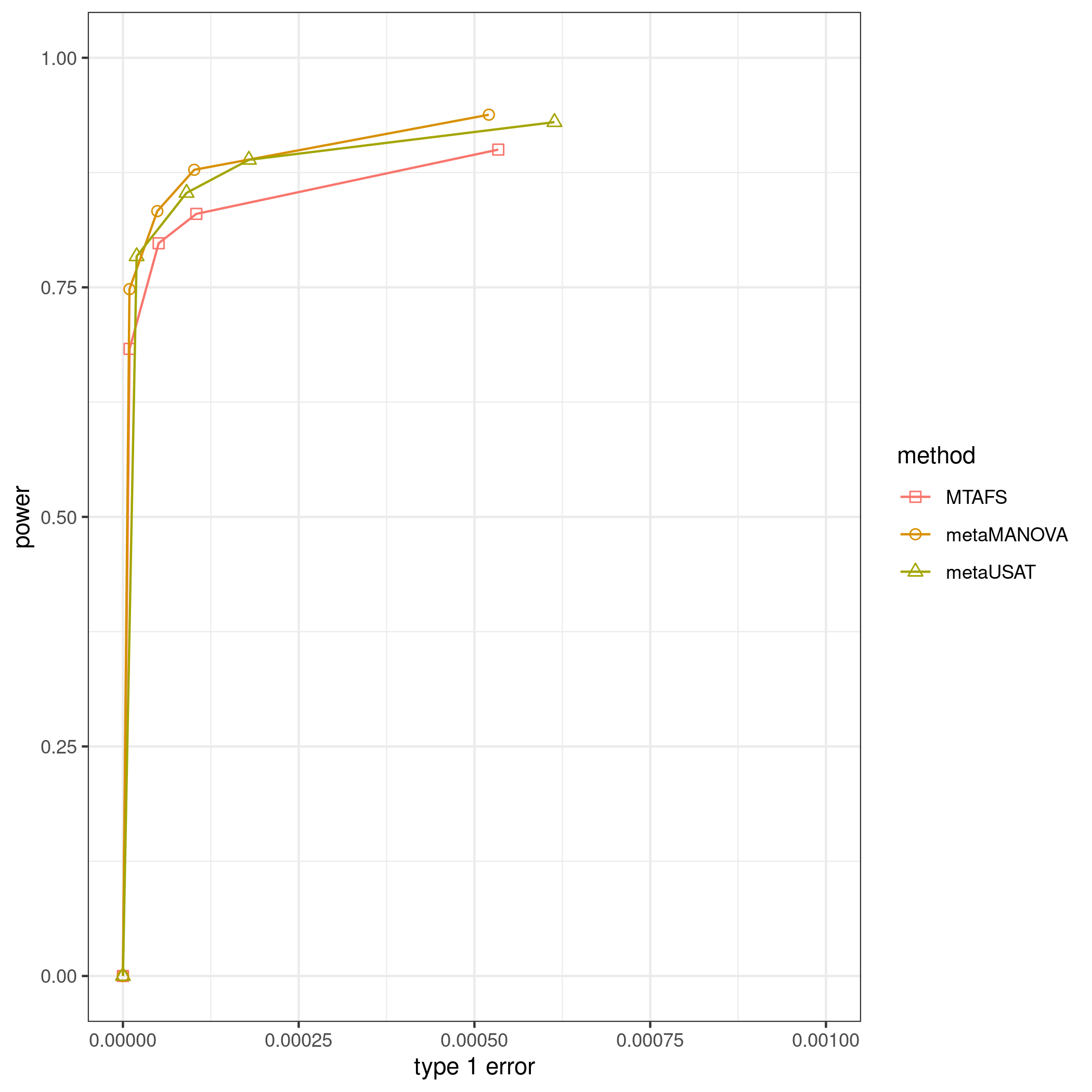}
    \caption{}
    \label{fig:s6e}
  \end{subfigure}
    \hfill
  \begin{subfigure}[t]{.3\textwidth}
    \centering
    \includegraphics[width=\textwidth]{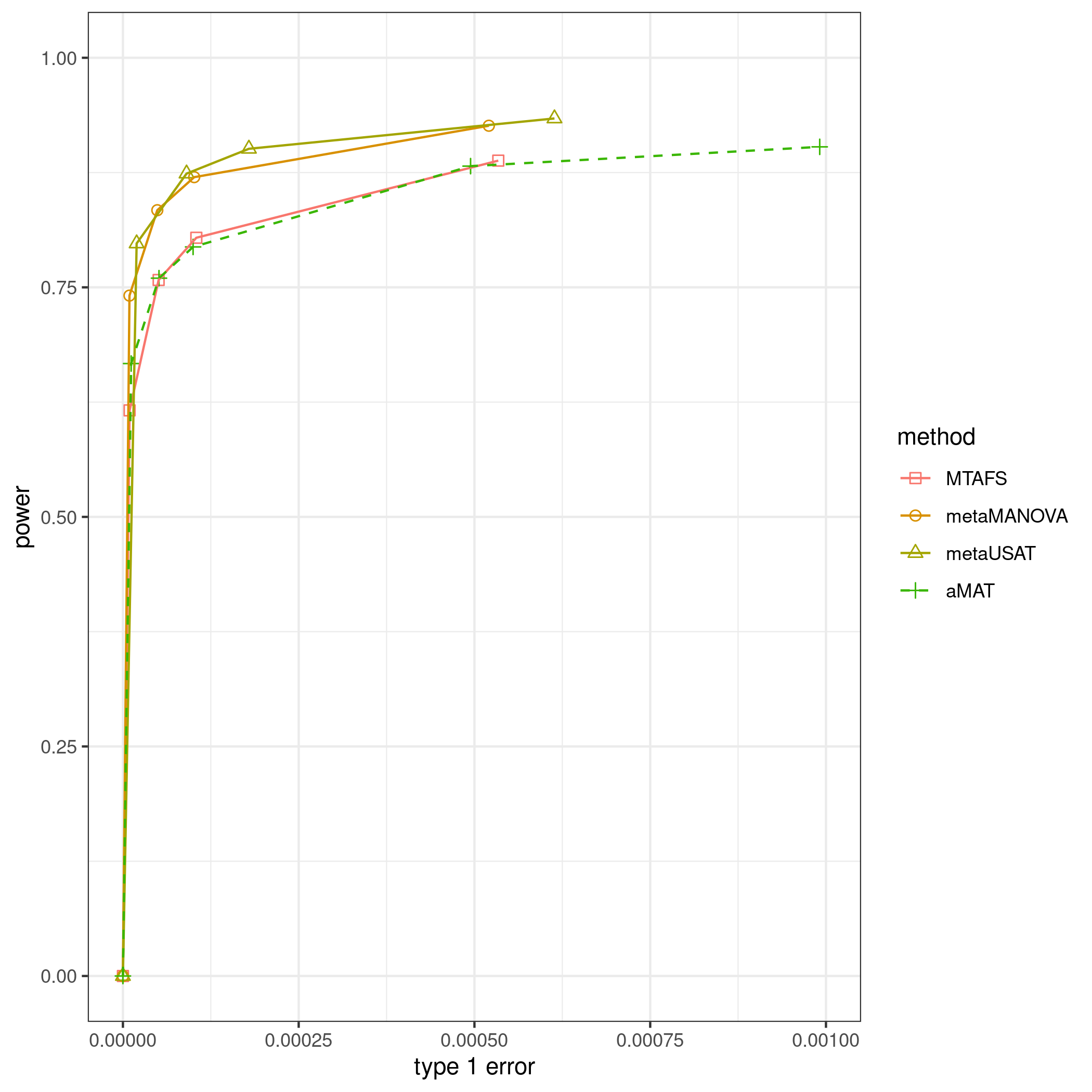}
    \caption{}
    \label{fig:s6f}
  \end{subfigure}

  \caption{Comparison of methods for model M2 using the UKCOR2 correlation matrix. (a) high sparsity, with only 2 nonzero components of $\bm \mu$; (b) intermediate sparsity, with 28 nonzero components of $\bm \mu$; (c) low sparsity, with 70 nonzero components of $\bm \mu$ out of a total of 139; (d) partial ROC curves for the three best methods with comparable power in (a); (e) partial ROC curves for the four best methods with comparable power in (b); (f) partial ROC curves for the four best methods with comparable power in (c).}
  \label{fig:power_t1fast_m2}
\end{figure}

%% Power: zeros, CS 50 traits
\begin{figure}
  \begin{subfigure}[t]{.4\textwidth}
    \centering
    \includegraphics[width=\linewidth]{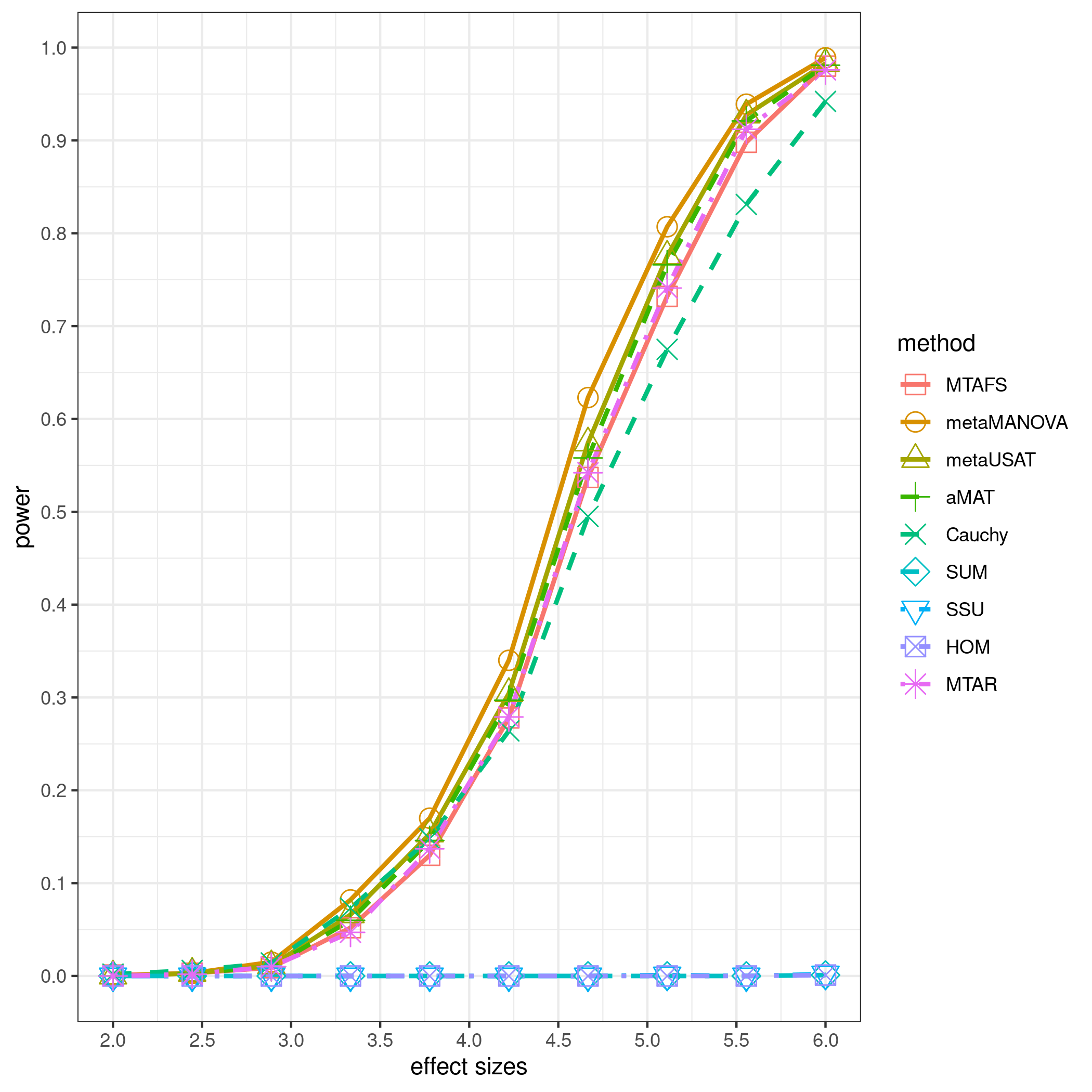}
    \caption{}
    \label{fig:s7a}
  \end{subfigure}
  \hfill
  \begin{subfigure}[t]{.4\textwidth}
    \centering
    \includegraphics[width=\linewidth]{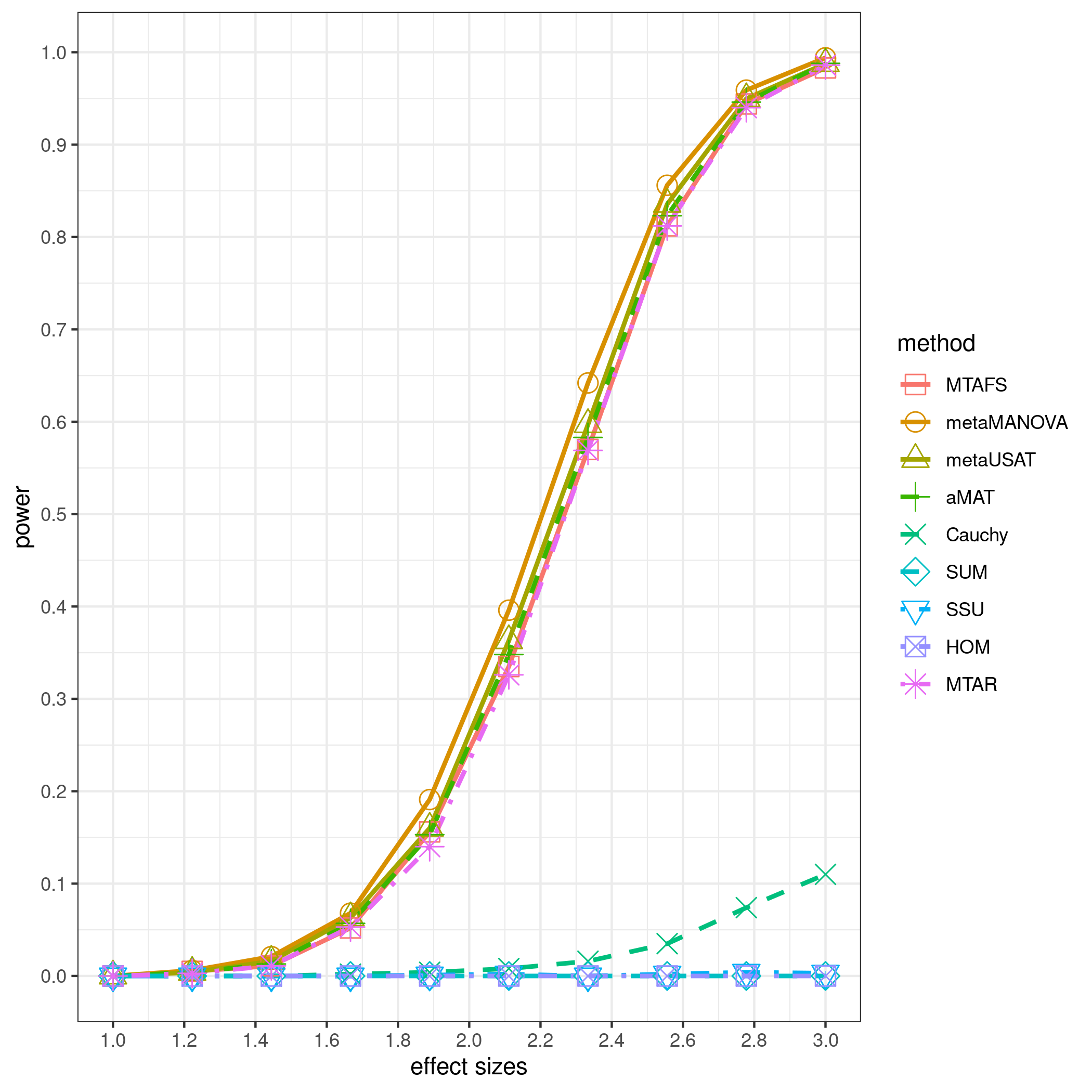}
    \caption{}
    \label{fig:s7b}
  \end{subfigure}

  \medskip

  \begin{subfigure}[t]{.4\textwidth}
    \centering
    \includegraphics[width=\linewidth]{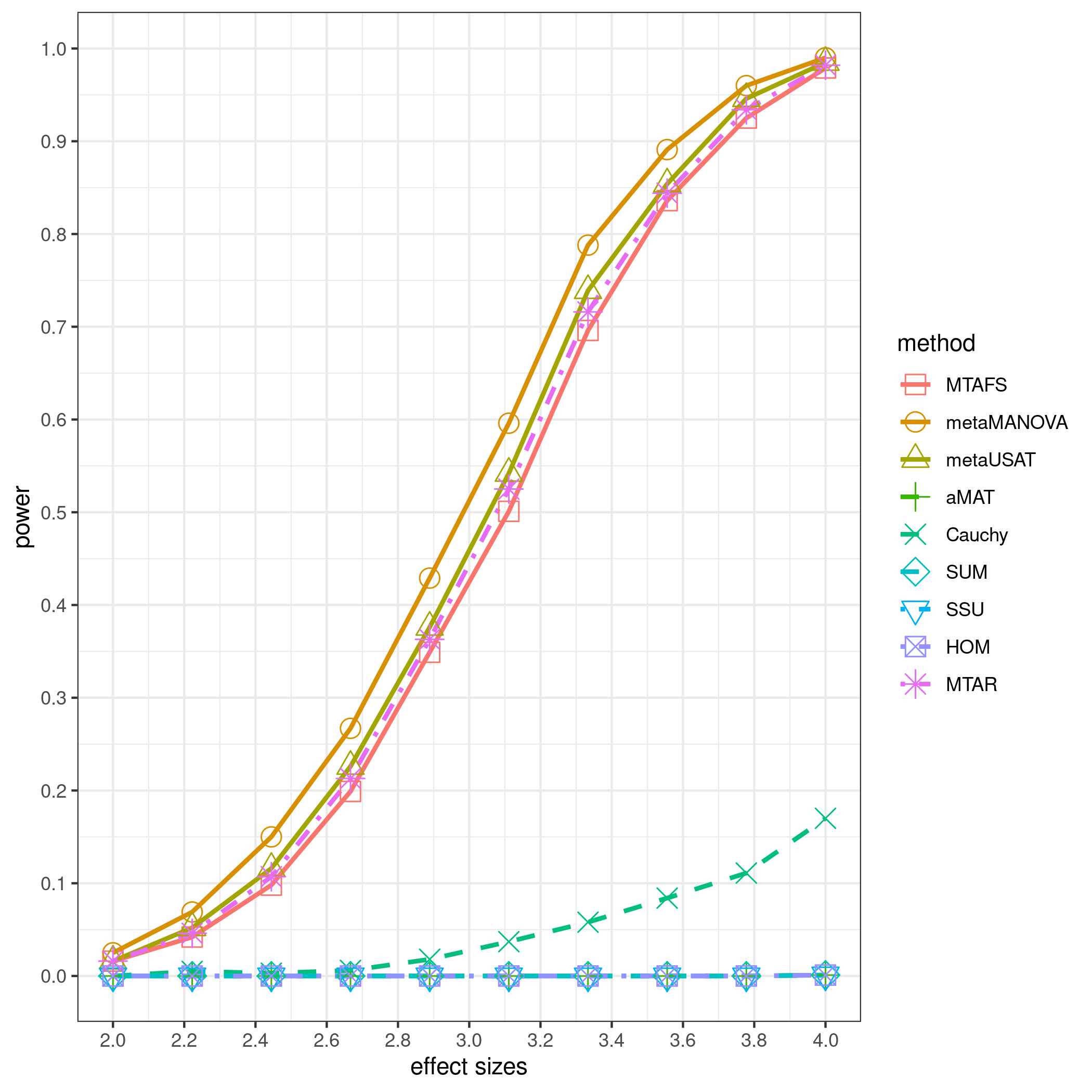}
    \caption{}
  \end{subfigure}
  \hfill
  \begin{subfigure}[t]{.4\textwidth}
    \centering
    \includegraphics[width=\linewidth]{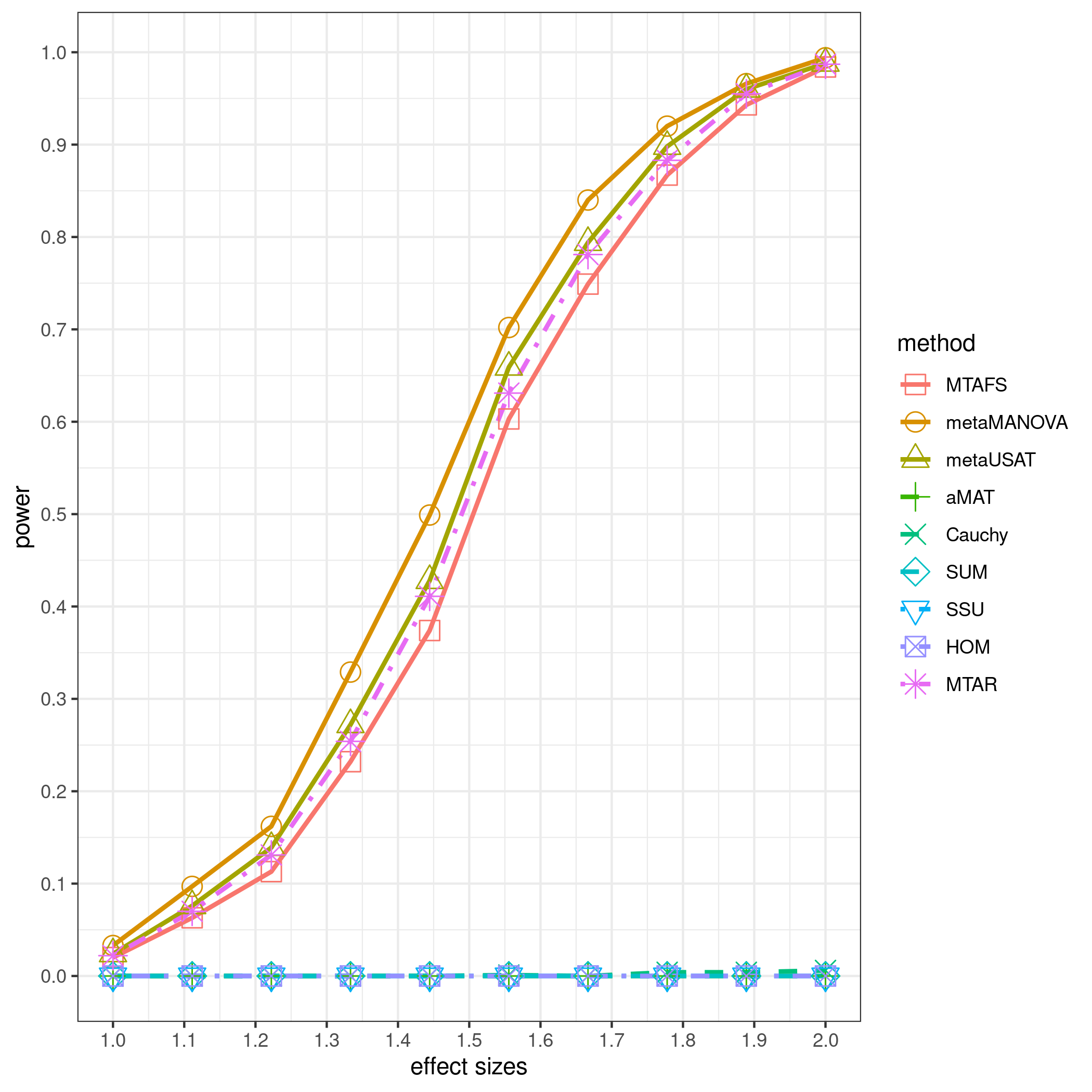}
    \caption{}
  \end{subfigure}
  \caption{Comparison of methods for model M2 using the CS correlation matrix and 50 traits. (a) and (b) have CS(0.3), and (c) and (d) have CS(0.7). (a) and (c) have high sparsity, with only 2 nonzero components of $\bm \mu$; (b) and (d) have intermediate sparsity, with 10 nonzero components of $\bm \mu$ out of a total of 50 traits.}
  \label{fig:power_cs50}
\end{figure}

%% Power: zeros, CS 100 traits
\begin{figure}
  \begin{subfigure}[t]{.4\textwidth}
    \centering
    \includegraphics[width=\linewidth]{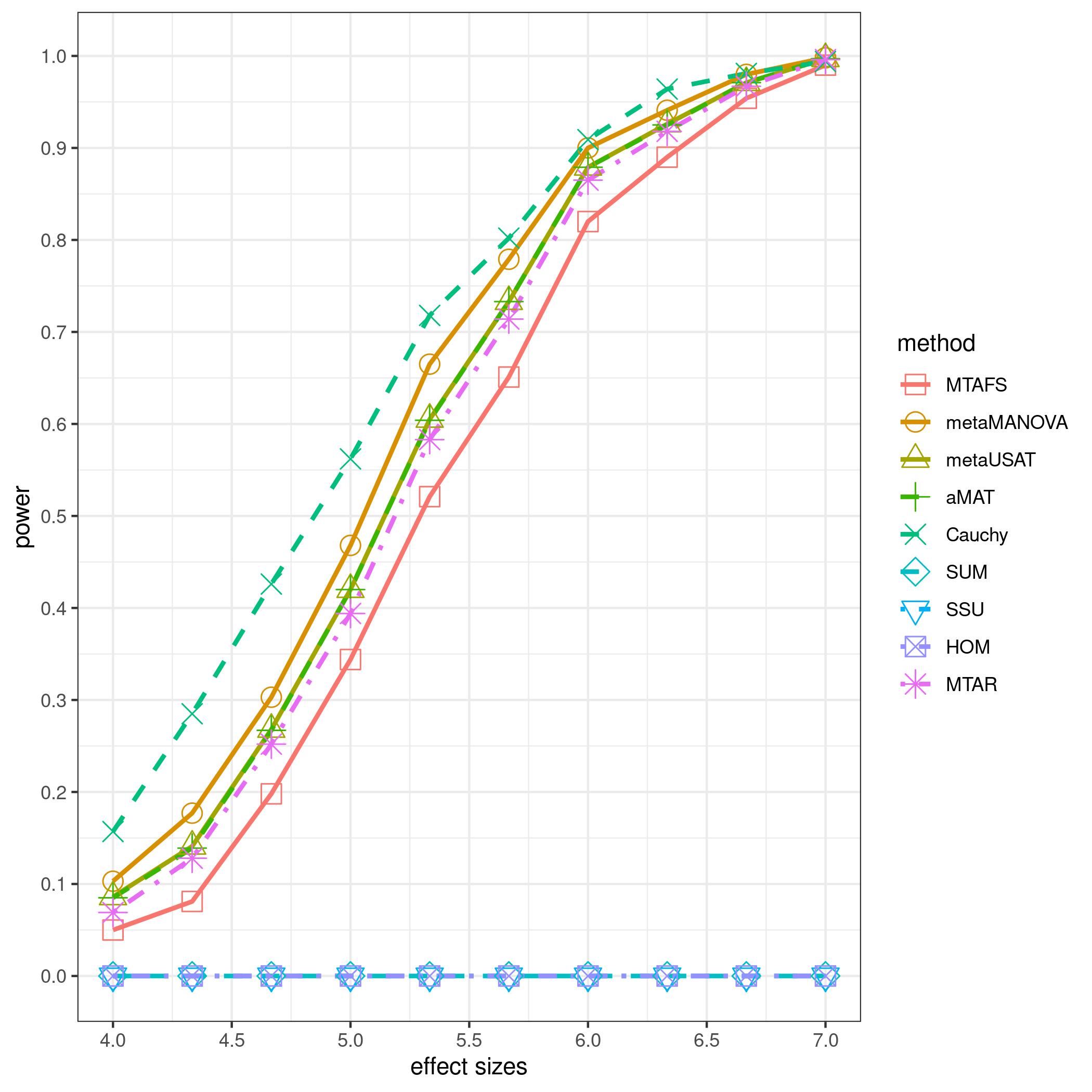}
    \caption{}
  \end{subfigure}
  \hfill
  \begin{subfigure}[t]{.4\textwidth}
    \centering
    \includegraphics[width=\linewidth]{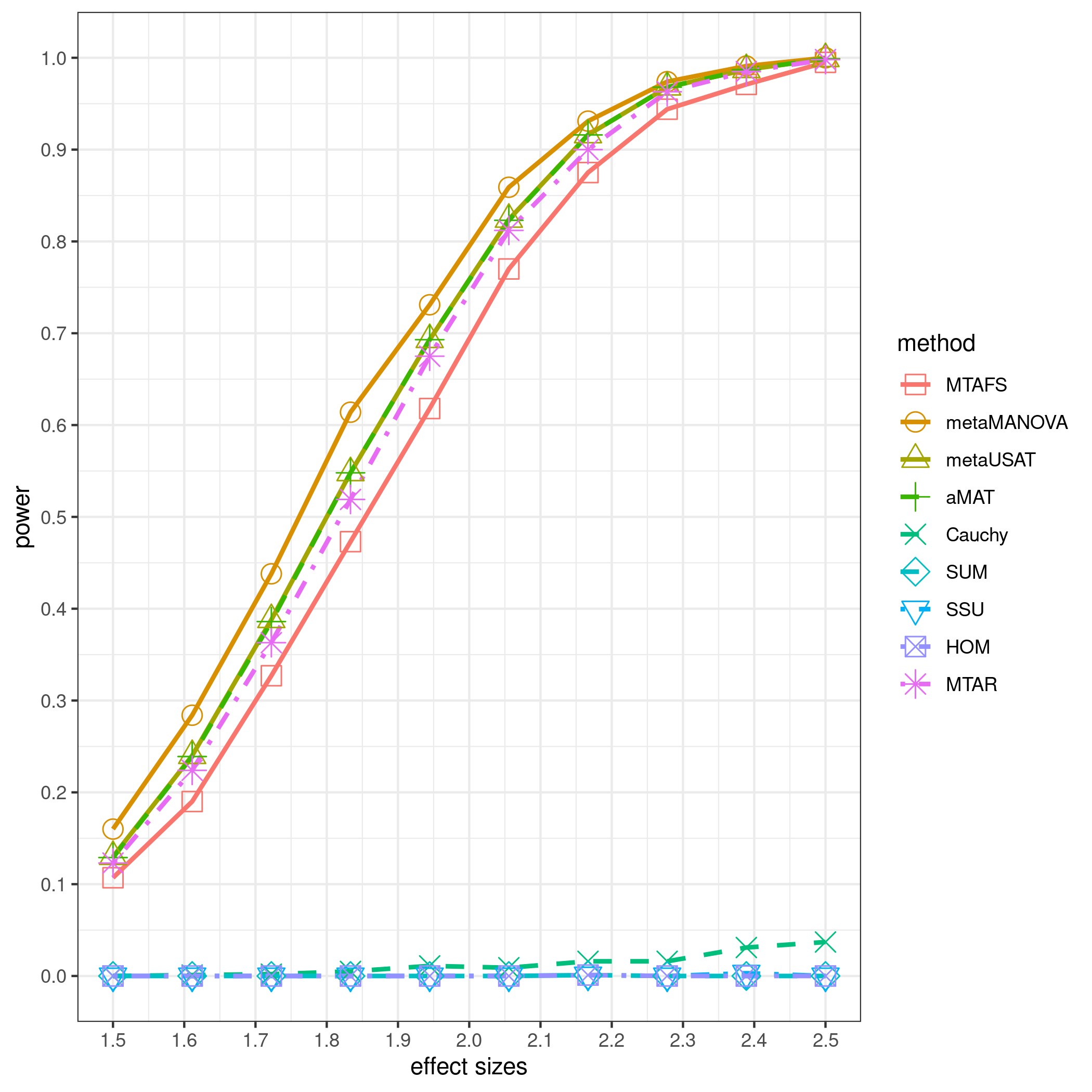}
    \caption{}
  \end{subfigure}

  \medskip

  \begin{subfigure}[t]{.4\textwidth}
    \centering
    \includegraphics[width=\linewidth]{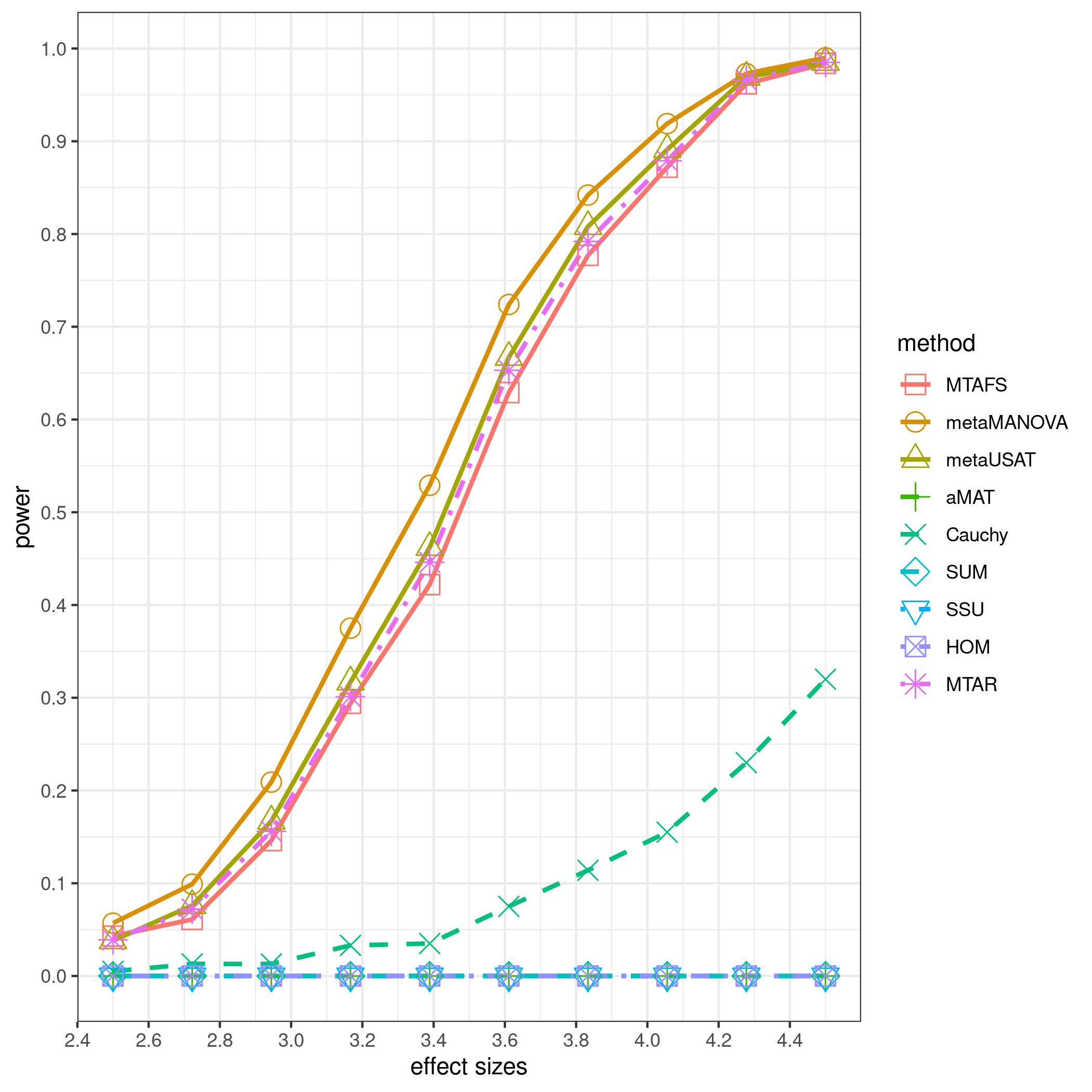}
    \caption{}
  \end{subfigure}
  \hfill
  \begin{subfigure}[t]{.4\textwidth}
    \centering
    \includegraphics[width=\linewidth]{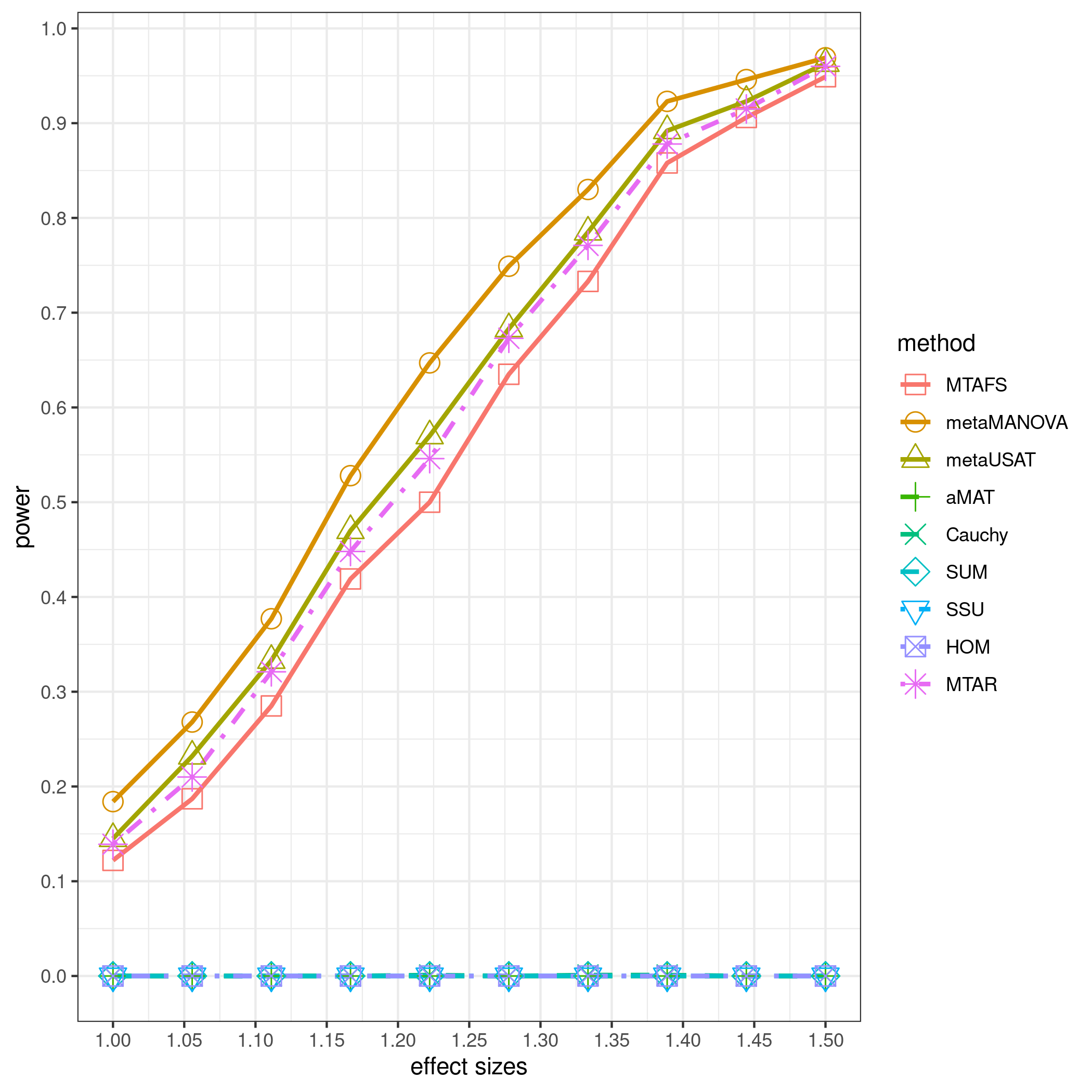}
    \caption{}
  \end{subfigure}
  \caption{Comparison of methods for model M2 using the CS correlation matrix and 100 traits. (a) and (b) have CS(0.3), and (c) and (d) have CS(0.7). (a) and (c) have high sparsity, with only 2 nonzero components of $\bm \mu$; (b) and (d) have intermediate sparsity, with 20 nonzero components of $\bm \mu$ out of a total of 100 traits.}
  \label{fig:power_cs100}
\end{figure}

%% Power: zeros, AR 50 traits
\begin{figure}
  \begin{subfigure}[t]{.4\textwidth}
    \centering
    \includegraphics[width=\linewidth]{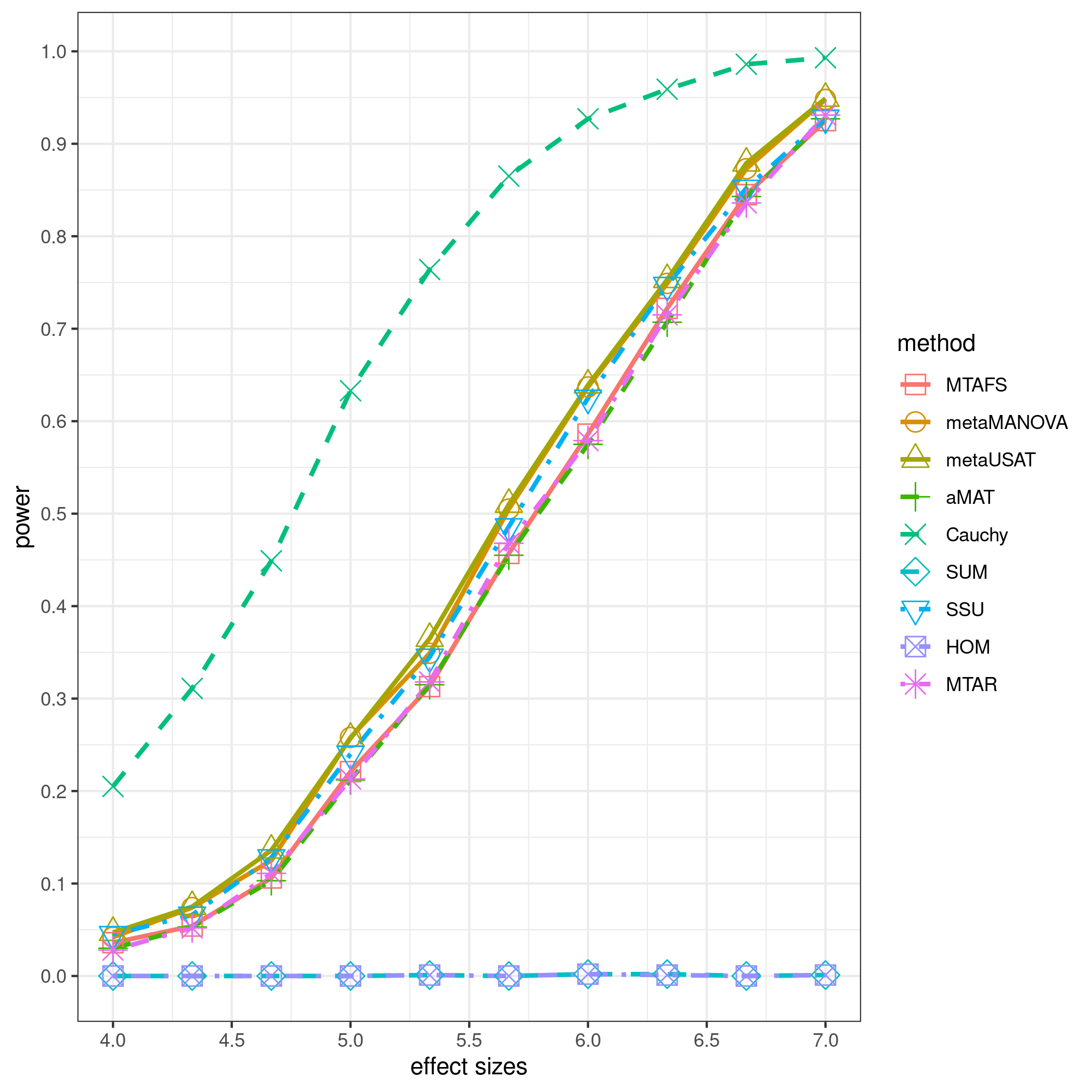}
    \caption{}
  \end{subfigure}
  \hfill
  \begin{subfigure}[t]{.4\textwidth}
    \centering
    \includegraphics[width=\linewidth]{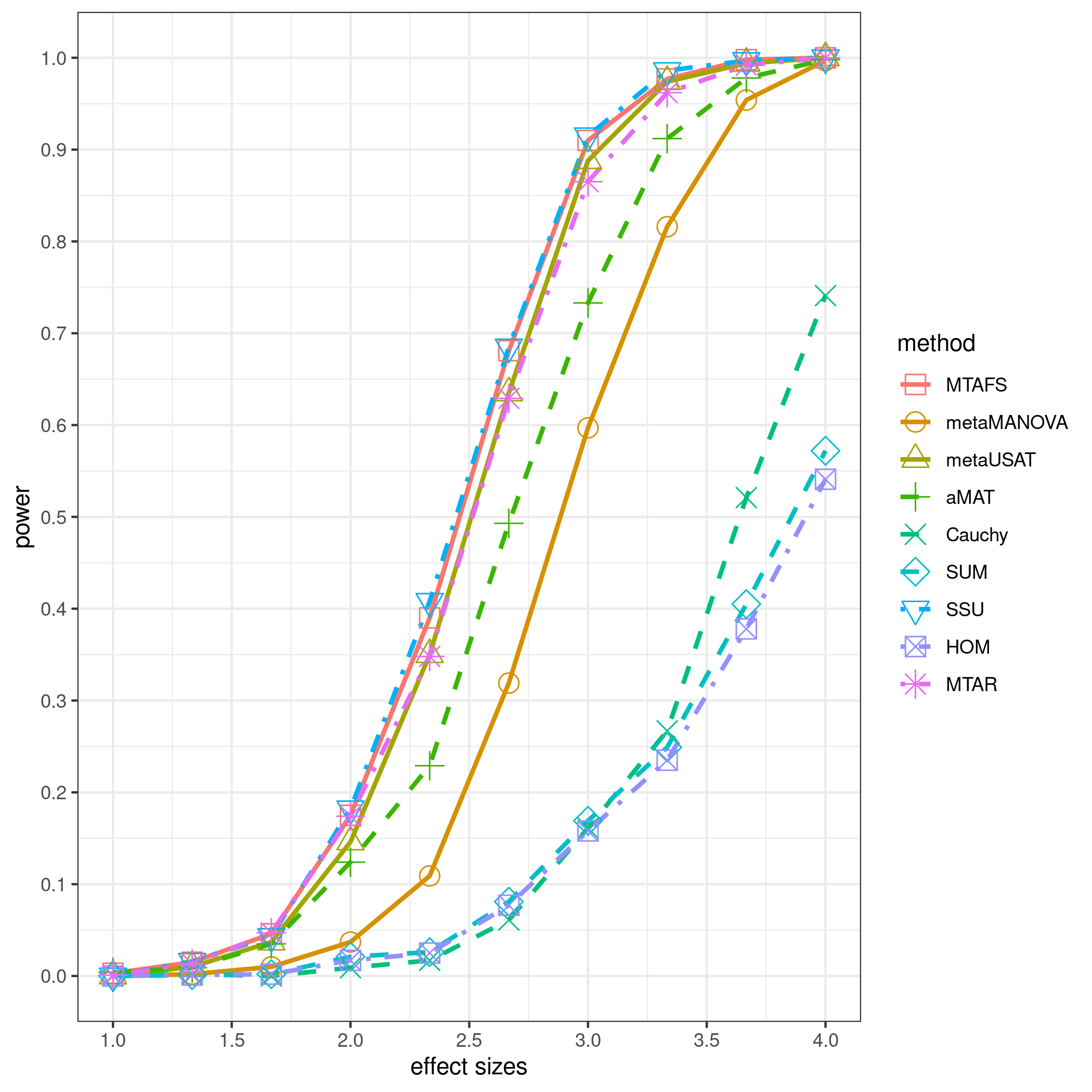}
    \caption{}
  \end{subfigure}

  \medskip

  \begin{subfigure}[t]{.4\textwidth}
    \centering
    \includegraphics[width=\linewidth]{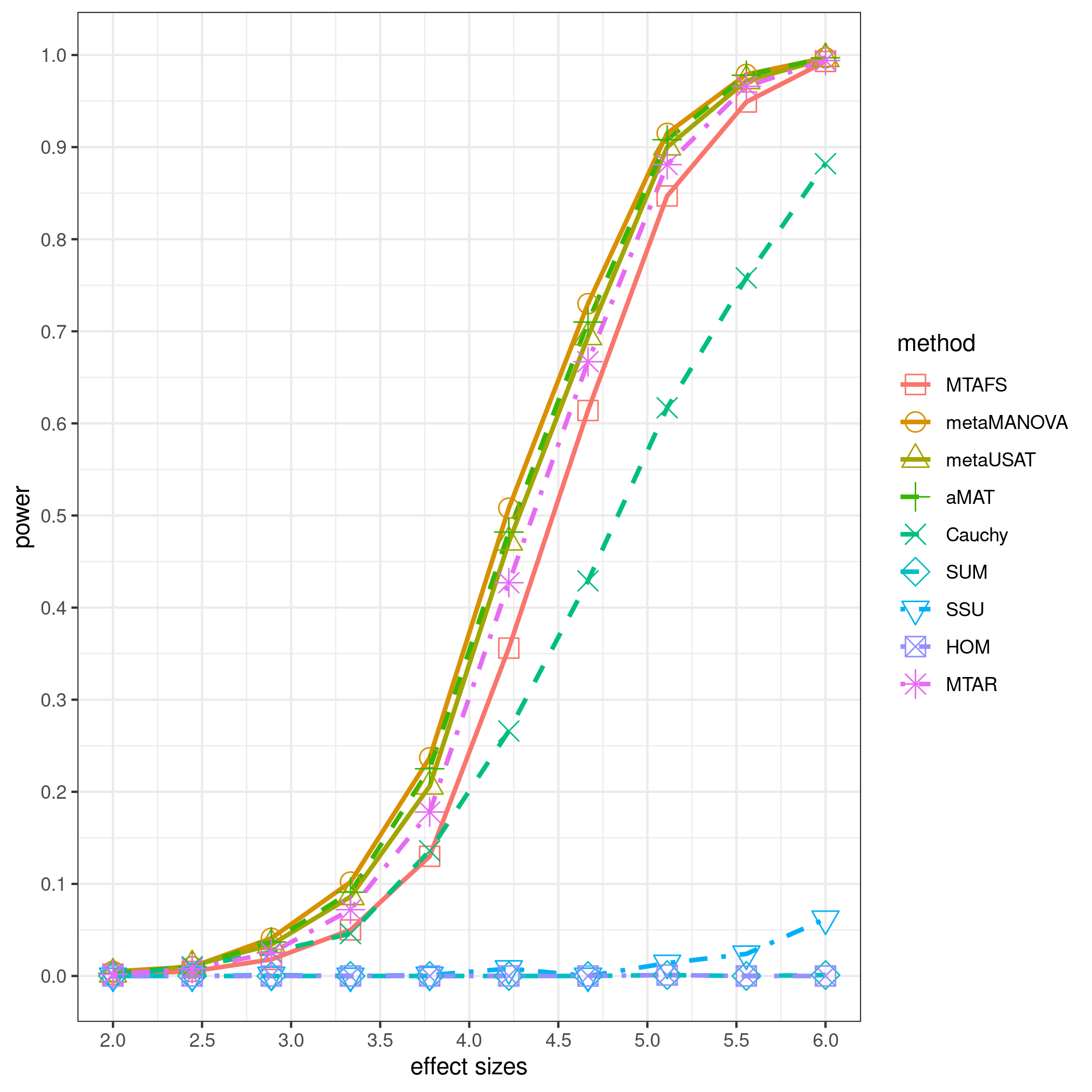}
    \caption{}
  \end{subfigure}
  \hfill
  \begin{subfigure}[t]{.4\textwidth}
    \centering
    \includegraphics[width=\linewidth]{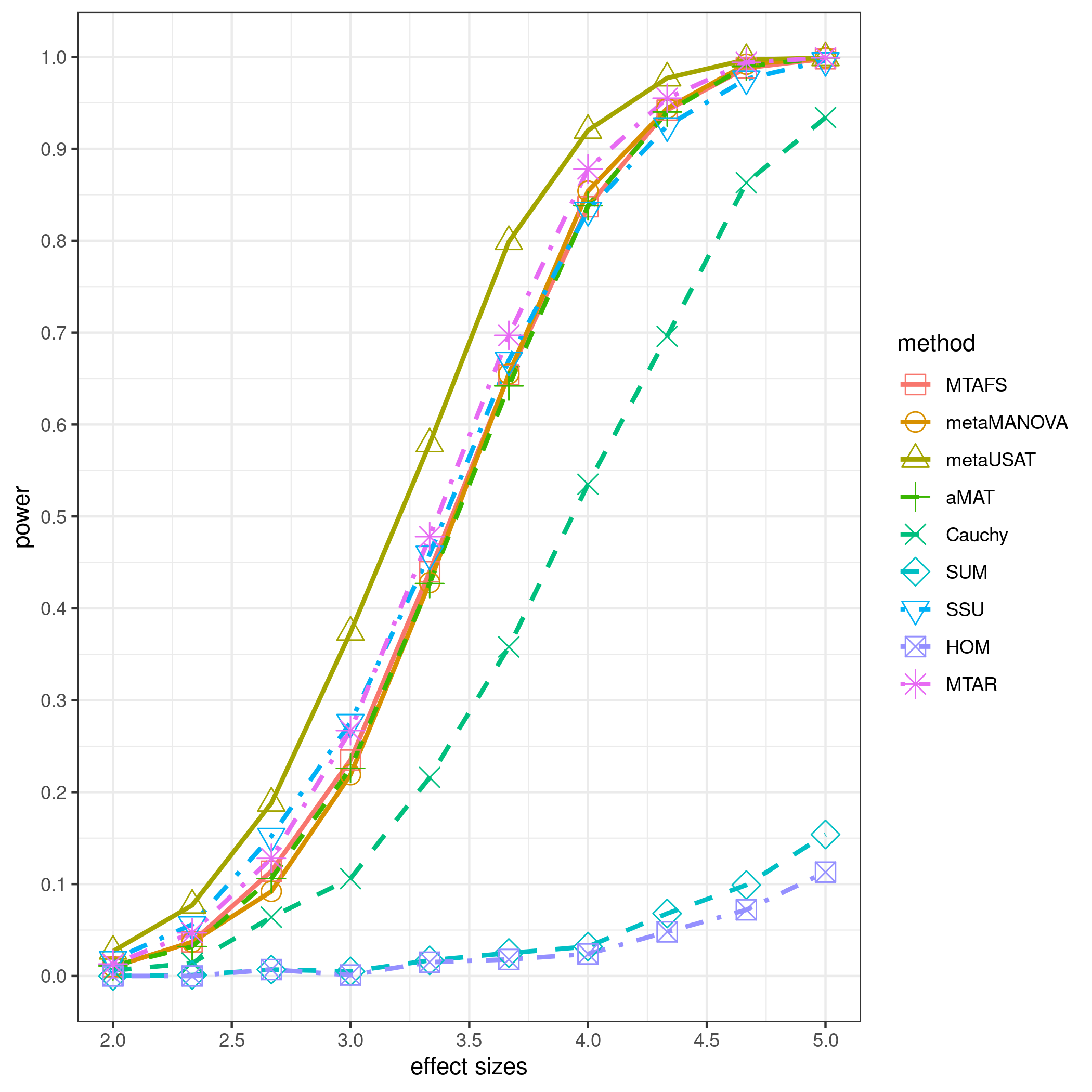}
    \caption{}
  \end{subfigure}
  \caption{Comparison of methods for model M2 using the AR correlation matrix and 50 traits. (a) and (b) have AR(0.3), and (c) and (d) have AR(0.7). (a) and (c) have high sparsity, with only 2 nonzero components of $\bm \mu$; (b) and (d) have intermediate sparsity, with 10 nonzero components of $\bm \mu$ out of a total of 50 traits.}
  \label{fig:power_ar50}
\end{figure}

%% Power: zeros, AR 100 traits
\begin{figure}
  \begin{subfigure}[t]{.4\textwidth}
    \centering
    \includegraphics[width=\linewidth]{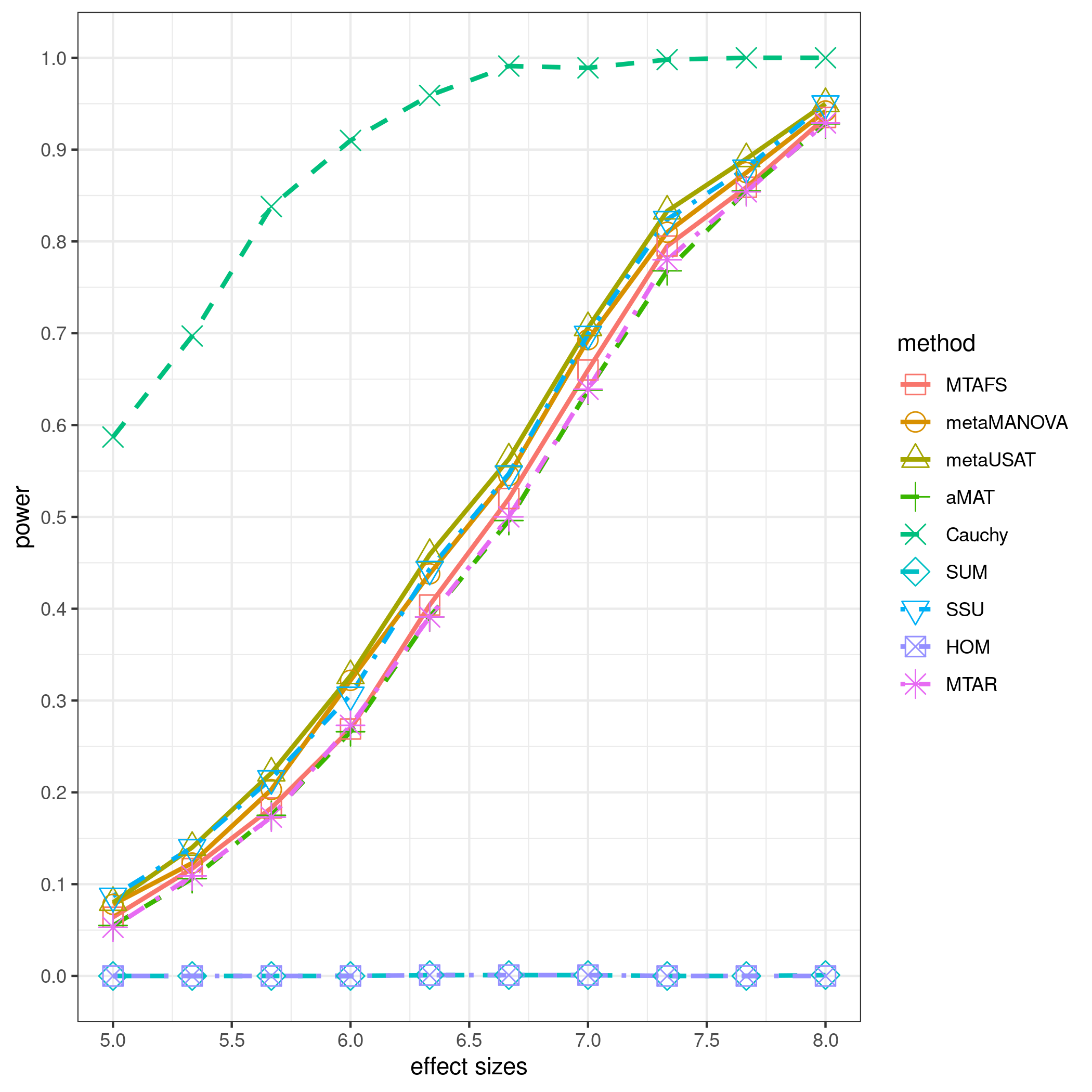}
    \caption{}
  \end{subfigure}
  \hfill
  \begin{subfigure}[t]{.4\textwidth}
    \centering
    \includegraphics[width=\linewidth]{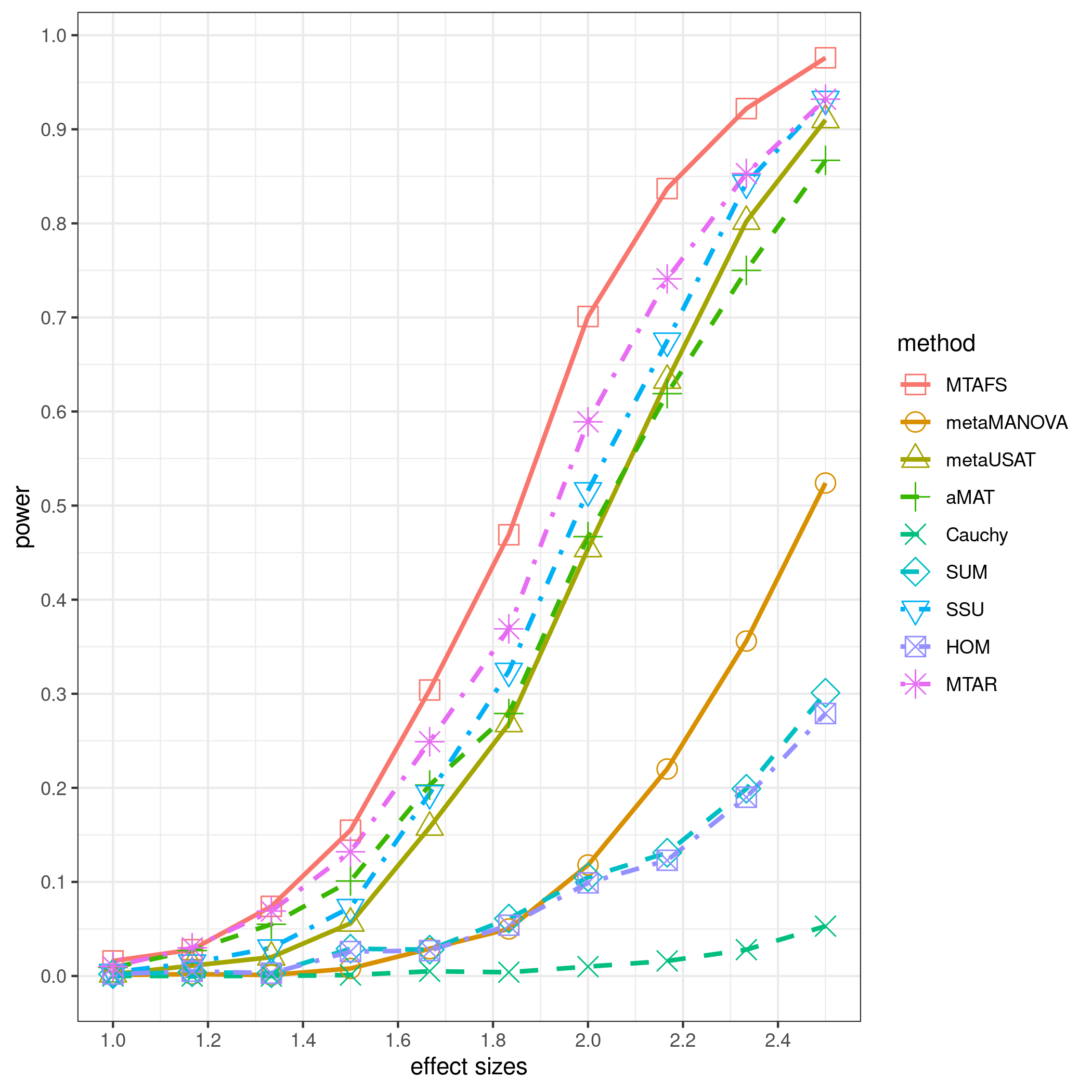}
    \caption{}
    \label{fig:s8b}
  \end{subfigure}

  \medskip

  \begin{subfigure}[t]{.4\textwidth}
    \centering
    \includegraphics[width=\linewidth]{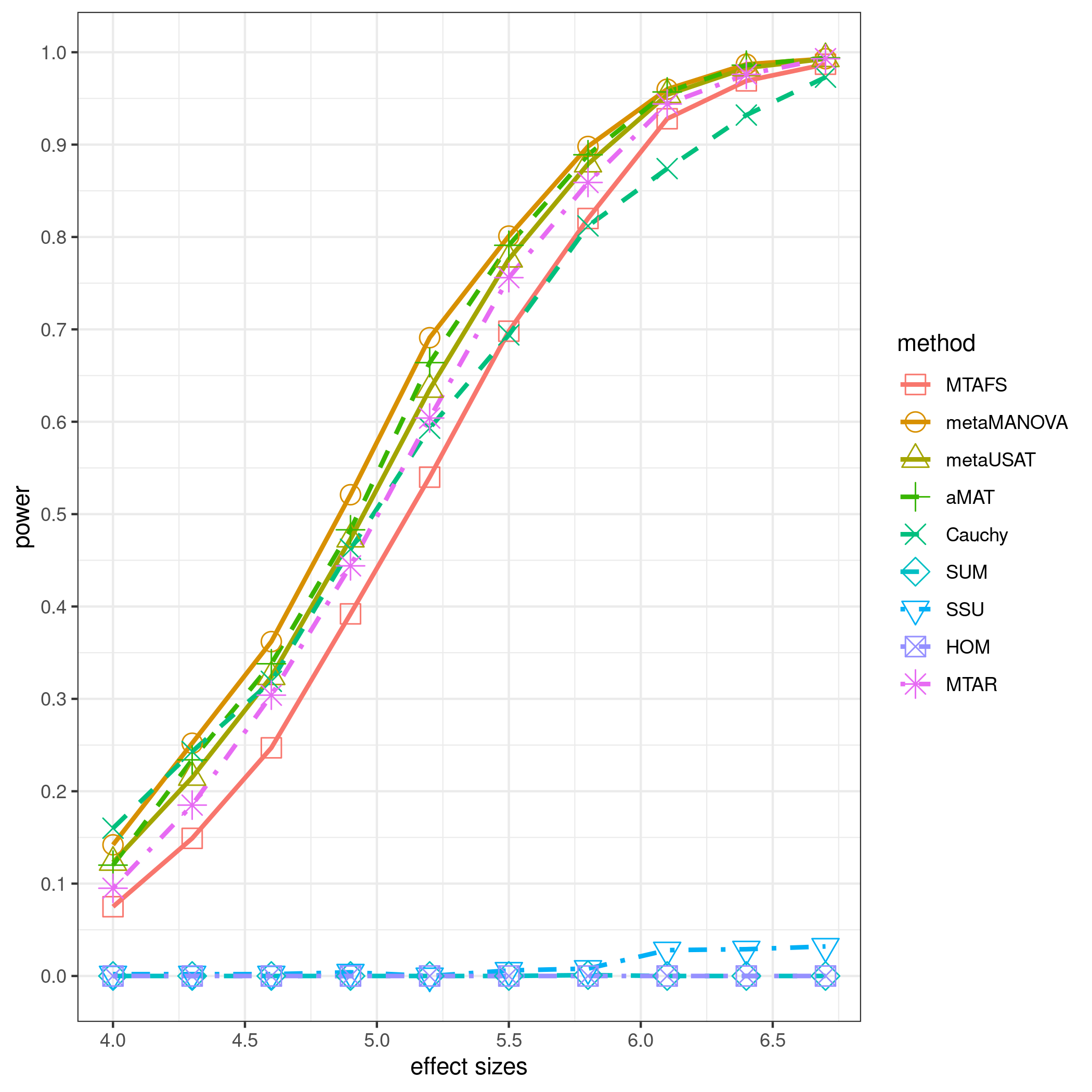}
    \caption{}
  \end{subfigure}
  \hfill
  \begin{subfigure}[t]{.4\textwidth}
    \centering
    \includegraphics[width=\linewidth]{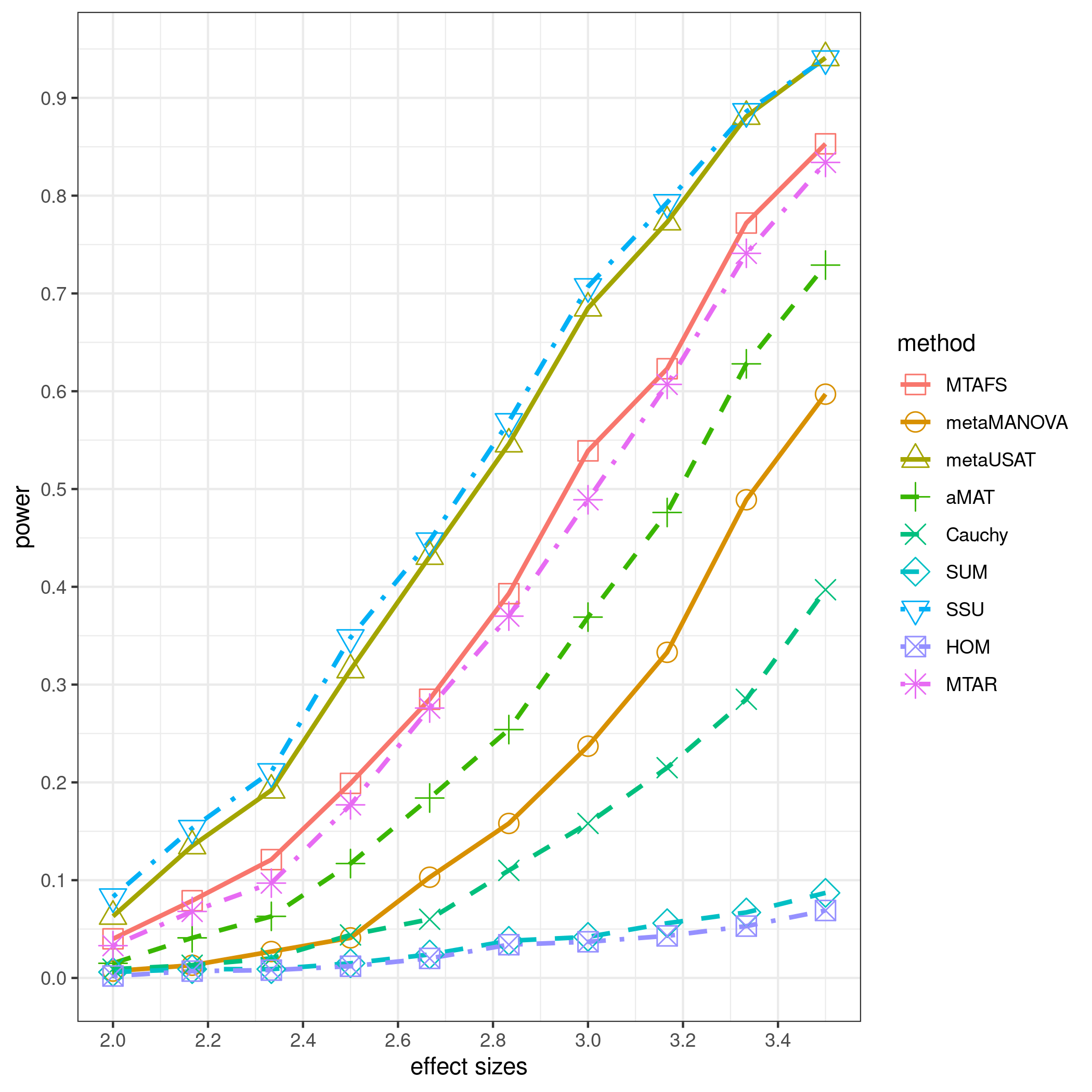}
    \caption{}
  \end{subfigure}
  \caption{Comparison of methods for model M2 using the AR correlation matrix and 100 traits. (a) and (b) have AR(0.3), and (c) and (d) have AR(0.7). (a) and (c) have high sparsity, with only 2 nonzero components of $\bm \mu$; (b) and (d) have intermediate sparsity, with 20 nonzero components of $\bm \mu$ out of a total of 100 traits.}
  \label{fig:power_ar100}
\end{figure}

% cor area
\begin{figure}[htbp]
    \centering
    \includegraphics[width=\linewidth]{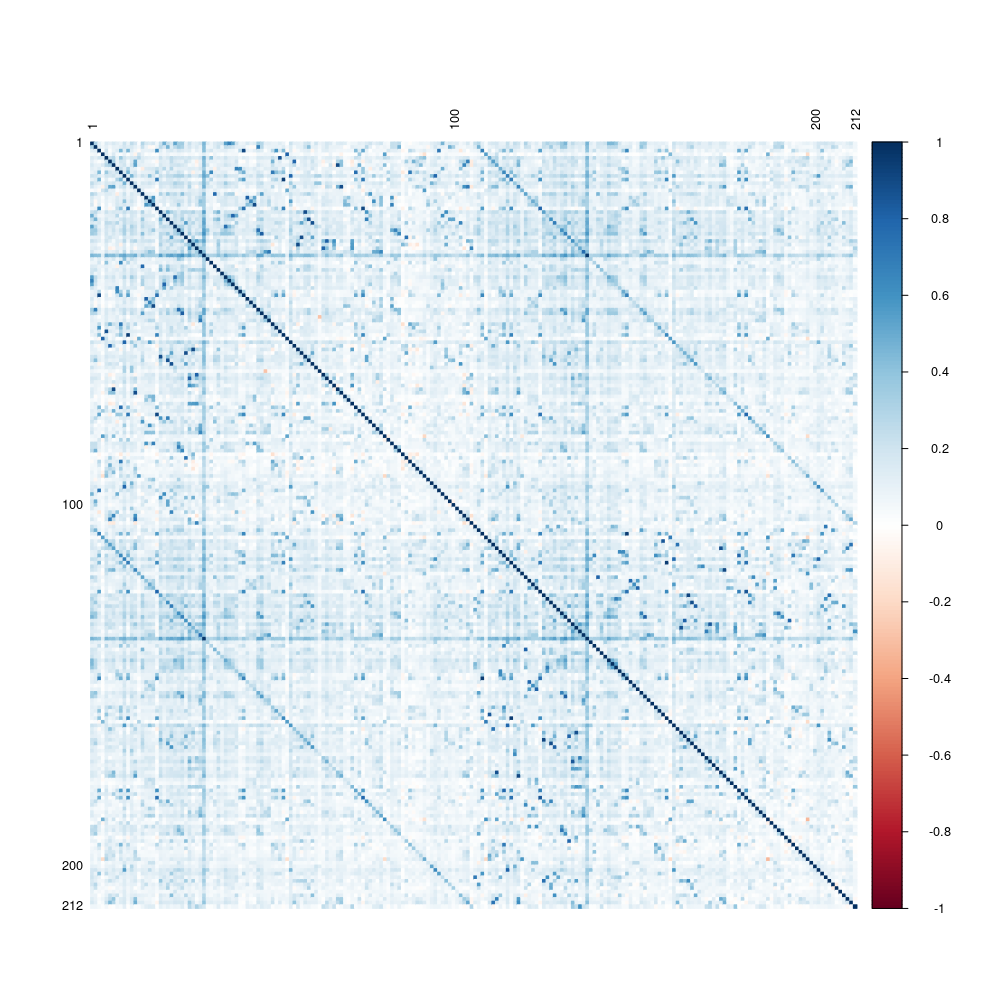}
    \caption{The LDSC estimated trait correlation matrix of Area. Area consists of 212 IDPs}
    \label{fig:cor_area}
\end{figure}

% table: significant SNPs
\begin{table}[ht]
  \centering
  \scalebox{0.9}{
  \begin{threeparttable}[b]
  \caption{The number of significant SNPs identified by methods}
  \label{tab:sigsnp}
  \begin{tabular}{ccccccccc}
  \hline
method & MTAFS & metaMANOVA   & metaUSAT & aMAT & MTAR & SUM & SSU & Single Trait \\
  \hline
Volume & 264 & 90 & 89 & 15  & 78 & 0 & 1 & 6 \\
Area & 55 & 16 & - & 11 & 14 & 0 & 3 & 1 \\ \hline

  \end{tabular}
 \end{threeparttable}
 }
\end{table}

% Venn
\begin{figure}
    \centering
    \includegraphics[width=0.7\linewidth]{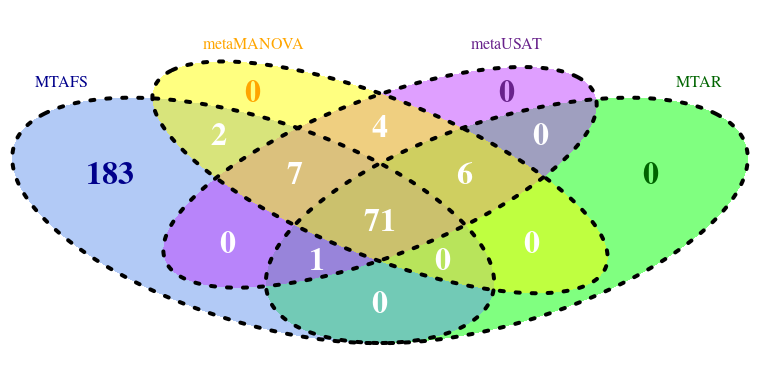}
    \caption{Venn diagram of number of significantly associated SNPs for Volume identified by different methods at $5 \times 10^{-8}$.}
    \label{fig:volume_venn}
\end{figure}

% competing gene heatmap
\begin{figure}[htbp]
    \centering
    \includegraphics[width=\linewidth]{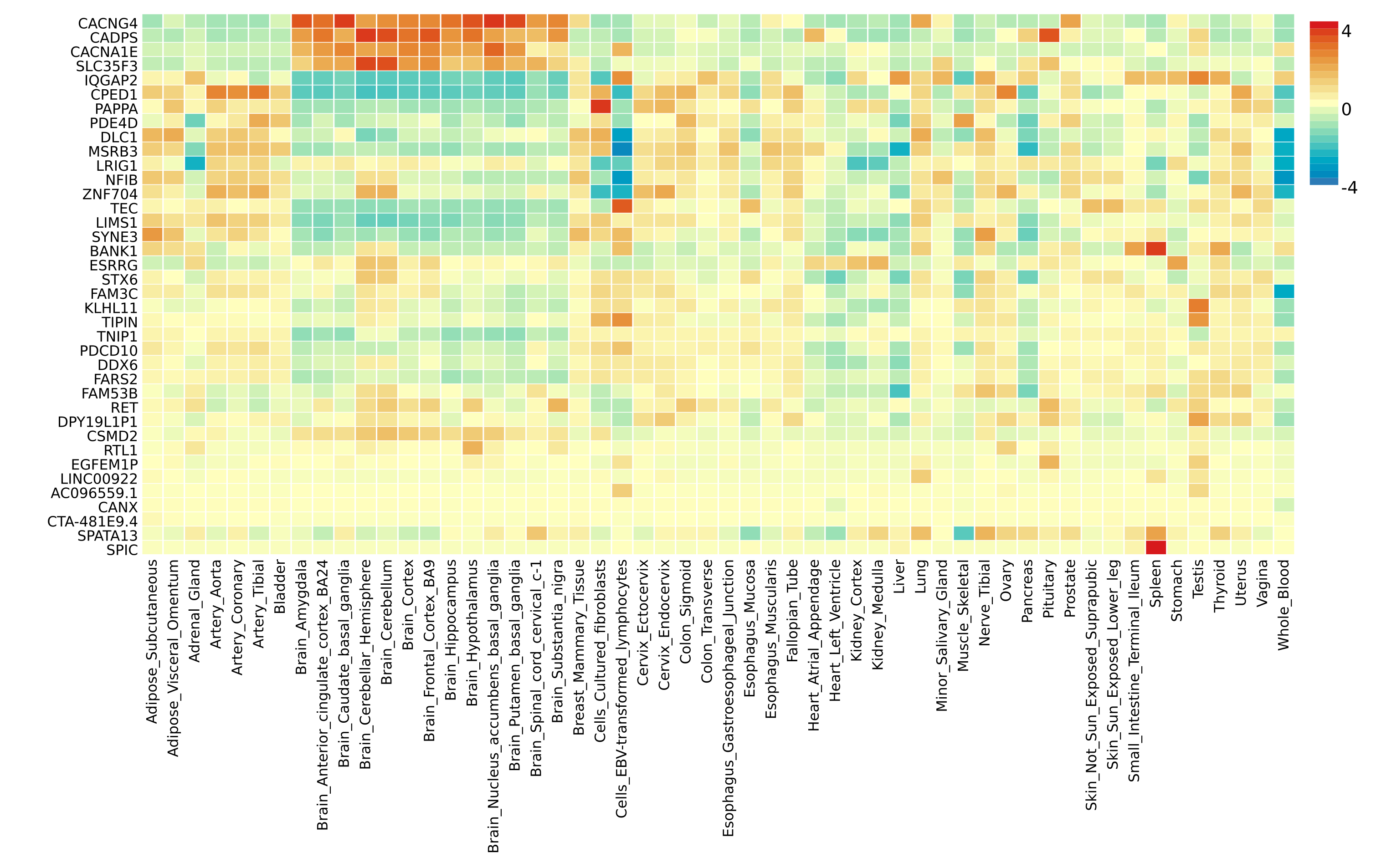}
    \caption{The expression heatmap of all genes identified by competing methods for volume. The red clusters have higher relative expression.}
    \label{fig:volume_other_heat}
\end{figure}

% competing gene expression
\begin{figure}[htbp]
    \centering
    \includegraphics[width=\linewidth]{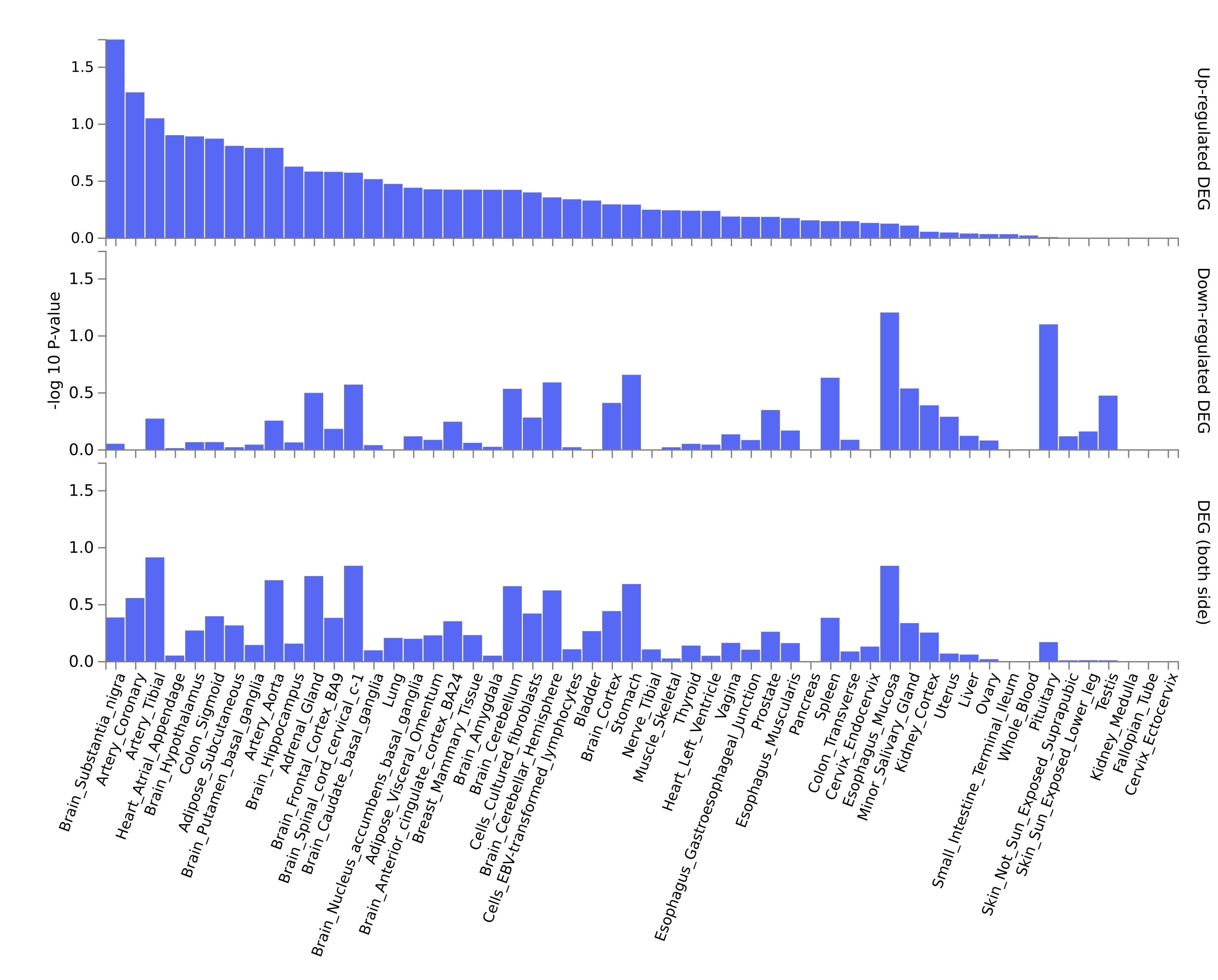}
    \caption{Tissue expression analysis for genes identified by competing methods for volume. Significant enrichment are in red with p-values less than 0.05 after Bonferroni correction.}
    \label{fig:volume_other_deg}
\end{figure}

% type1: area

\begin{table}[ht]
  \centering
  \begin{threeparttable}[htbp]
  \caption{Type 1 error$^a$ with Area trait correlation matrix}
  \label{tab:type1_area}
\begin{tabular}{cccccc}
\hline
& \multicolumn{5}{c}{Significance Levels}              \\ 
Methods & $5 \times 10^{-2}$ & $1 \times 10^{-2}$ & $1 \times 10^{-3}$ & $1 \times 10^{-4}$ & $1 \times 10^{-5}$ \\ \hline
metaMANOVA           & 1.05     & 1.07     & 1.09     & 1.08     & 1.21     \\
aMAT                 & 1.06     & 1.1      & 1.15     & 1.28     & 1.2      \\
MTAR                 & 0.99     & 1        & 1.04     & 1.08     & 0.97     \\
MTAFS                & 1.2      & 1.16     & 1.11     & 1        & 1.07 \\\hline
\end{tabular}
  \begin{tablenotes}
    \item[a] The values in the table are ratios of empirical Type I errors divided by the corresponding significance levels.

  \end{tablenotes}
 \end{threeparttable}
\end{table}

% area: Venn
\begin{figure}[H]
    \centering
    \includegraphics[width=0.7\linewidth]{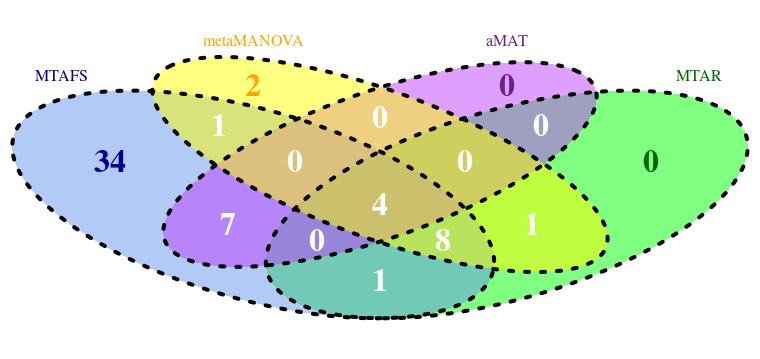}
    \caption{Venn diagram of number of significantly associated SNPs for Area identified by different methods at $5 \times 10^{-8}$.}
    \label{fig:area_venn}
\end{figure}

% area: competing heat expression
\begin{figure}[H]
    \centering
    \includegraphics[width=\linewidth]{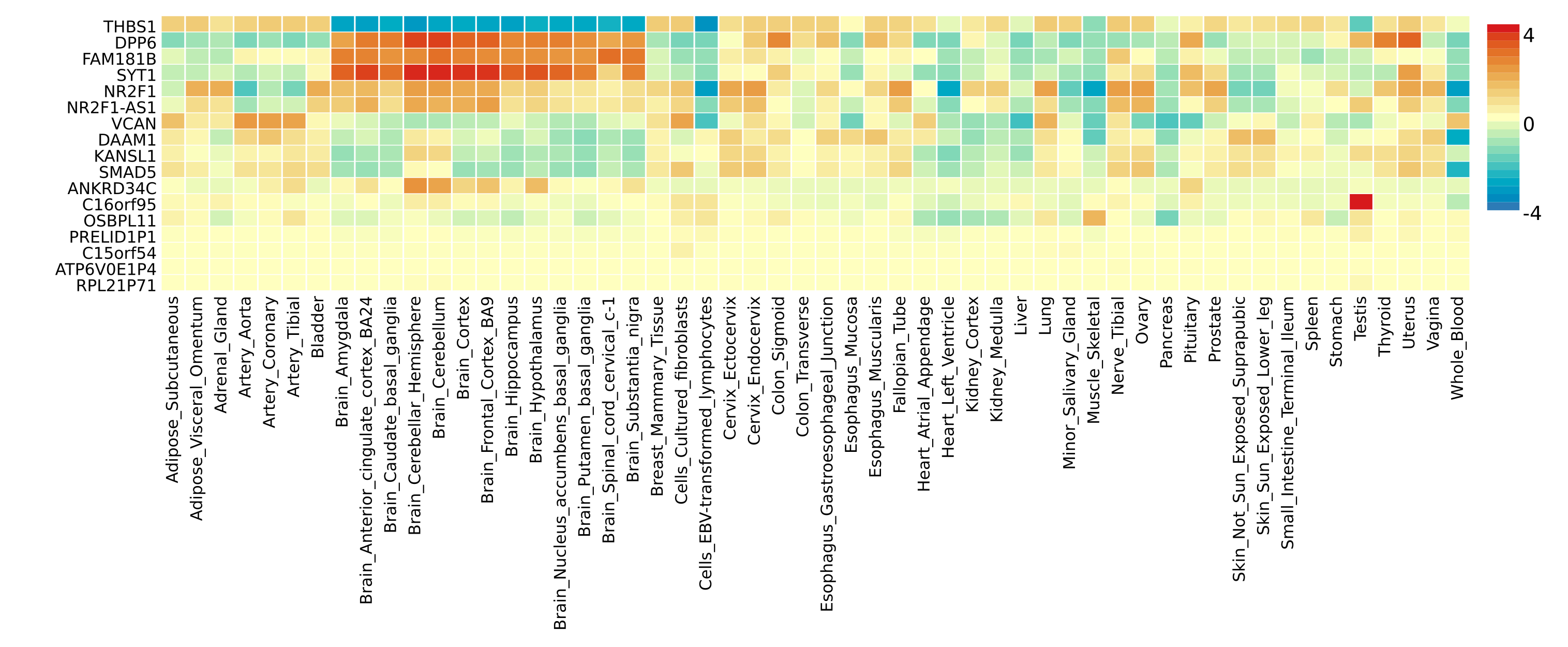}
    \caption{The expression heatmap of all genes identified by competing methods for area. The red clusters have higher relative expression.}
    \label{fig:area_other_heat}
\end{figure}

% area: competing gene expression
\begin{figure}[H]
    \centering
    \includegraphics[width=\linewidth]{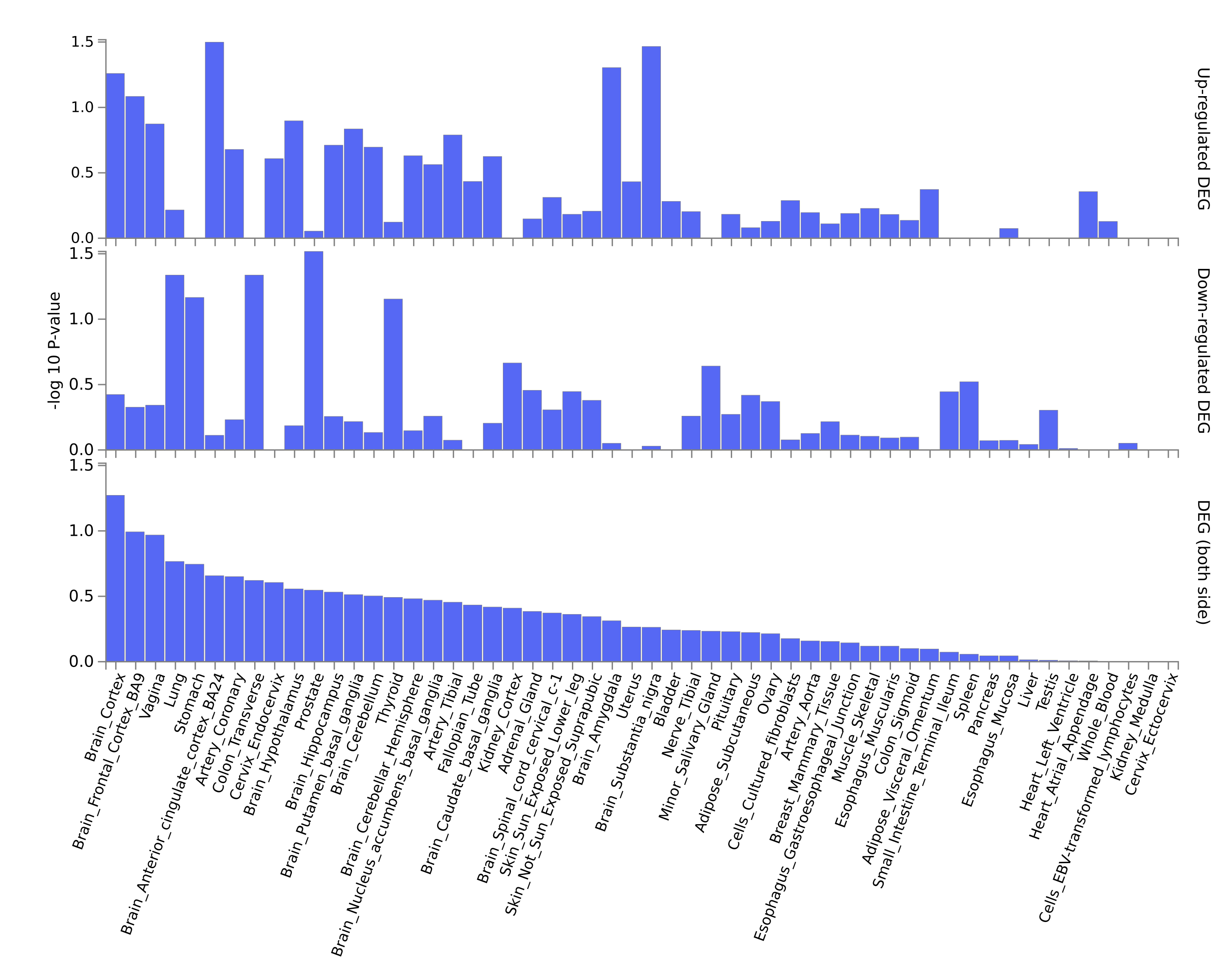}
    \caption{Tissue expression analysis for genes identified by competing methods for area. Significant enrichment are in red with p-values less than 0.05 after Bonferroni correction}
    \label{fig:area_other_deg}
\end{figure}

% computing time

\begin{table}[htbp]
  \centering
  \begin{threeparttable}[htbp]
  \caption{Computing time of different methods (in seconds)}
  \label{tab:time}
\begin{tabular}{cccccc}
                   & MTAFS & metaMANOVA & metaUSAT & aMAT & MTAR  \\ \hline
58 \textit{Volumetric} IDPs & 7200  & 8.4        & 9000     & 540  & 0.48  \\
212 \textit{Area} IDPs      & 600   & 60         & -        & 960  & 2     \\
UKCOR1, M1, Power  & 8.39  & 0.01       & 269      & 1.55 & 0.054 \\
UKCOR1, M1, Type 1 & 13.93 & 0.01       & 12.7     & 1.51 & 0.08  \\ \hline
\end{tabular}
  \begin{tablenotes}
    \item Notes: (1) For 212 \textit{Area} IDPs, MTAFS used 60 cores. (2) Power represents the power analysis which had 1000 SNPs and the effect size was 1. (3) Type 1 represents type 1 error analysis which had 1000 SNPs.

  \end{tablenotes}
 \end{threeparttable}
\end{table}

\end{document}